\documentclass[12pt,a4paper, titlepage, reqno]{article}
\RequirePackage{rotating}
\usepackage[margin=1in]{geometry}
\usepackage[shortlabels]{enumitem}
\usepackage[dvipsnames]{xcolor}
\usepackage{tocloft}
\usepackage[bookmarks, colorlinks, breaklinks]{hyperref} 
\hypersetup{
	linkcolor=blue,
	citecolor=blue,
	filecolor=black,
	urlcolor=blue}
\usepackage[most]{tcolorbox}
\usepackage{mathrsfs}
\usepackage{soul}
\usepackage{cleveref}
\crefname{equation}{}{}
\usepackage{amsmath,epsfig,array}
\numberwithin{equation}{section}
\usepackage{graphicx}
\usepackage{amssymb}
\usepackage{amsfonts,amstext,latexsym,xspace}

\newtheorem{definition}{Definition}[section]
\tcolorboxenvironment{definition}{enhanced jigsaw,colback={white!99!white}, colframe=black,boxrule=1pt, arc=0pt,outer arc=0pt}

\tcolorboxenvironment{conjecture}{enhanced jigsaw,colback={white!99!white}, colframe=black,boxrule=1pt, arc=0pt,outer arc=0pt}
\newtheorem{theorem}{Theorem}[section]
\newtheorem{remark}{Remark}[section]

\newcommand{\ut}{\underline{\tau}}
\newcommand{\N}{{\mathcal N}}

\newcommand{\eq}{\begin{equation}}
\newcommand{\en}{\end{equation}}
\newcommand{\eqn}{\begin{eqnarray}}
\newcommand{\enn}{\end{eqnarray}}

\newcommand{\beq}{\begin{equation}}
\newcommand{\eeq}{\end{equation}}
\newcommand{\M}{\ensuremath{\mathcal{M}}}

\newcommand{\vin}{\rotatebox[origin=c]{-90}{$\in$}}

\newcommand{\0}{\ensuremath{E_{10}}\xspace}

\newcommand{\cN}{\mathcal{N}}

\newcommand{\Tr}{\mathrm{Tr}}

\newcommand{\conf}[1]{\textbf{Conf}\left(#1\right)}
\newcommand{\aut}[1]{\textbf{Aut}\left(#1\right)}
\renewcommand{\lor}[1]{\textbf{Lor}\left(#1\right)}
\newcommand{\qconf}[1]{\textbf{QConf}\left(#1\right)}
\newcommand{\str}[1]{\textbf{Str}\left(#1\right)}
\newcommand{\rot}[1]{\textbf{Rot}\left(#1\right)}
\newcommand{\mob}[1]{\textbf{M\"ob}\left(#1\right)}
\newcommand{\stro}[1]{\textbf{Str}_0\left( #1 \right)}

\newcommand{\pmat}[4]{\begin{pmatrix} #1 & #2 \\ #3 & #4
\end{pmatrix}
}

\begin{document}

\begin{titlepage}
\vspace*{-2cm}
\begin{center}
 {\Large \bf {Octonionic Magical Supergravity, Niemeier \\[0.3cm] Lattices, and  Exceptional \& Hilbert Modular Forms}}
\end{center}
\vspace{0.2cm}
\begin{center}
 {\large Murat G\"unaydin$^{1,2}$ \& Abhiram Kidambi$^{3}$} \\[0.5cm]
\textit{$^1$ Stanford Institute of Theoretical Physics, Stanford University,\\ Palo Alto , CA 94305, \\[0.2cm] $^2$ Institute for Gravitation and the Cosmos, Physics Department,
Penn  State University,
University Park, PA 16802, USA, \\[0.2cm]
$^3$ Kavli IPMU, University of Tokyo, \\Kashiwanoha 5-1-5, 277-0813 Kashiwa, Chiba, Japan
 }\\[0.5cm]{ \large \textbf{\texttt{ \href{mailto:mgunaydin@psu.edu}{mgunaydin@psu.edu}, \href{mailto:abhiram.kidambi@ipmu.jp}{abhiram.kidambi@ipmu.jp}}} } \\
[0.5cm]

 \textit{ \large \color{blue} Dedicated to Hermann Nicolai on the occasion of his  70$^{\,th}$ birthday.} \\[0.5cm]
 \textbf{Abstract}
 \end{center}
We study the quantum degeneracies of BPS black holes of  octonionic magical supergravity in five dimensions that is defined by the exceptional Jordan algebra. We define the quantum degeneracy purely number theoretically as the number of distinct states in the charge space with a given set of invariant labels of the discrete U-duality group. We argue that the quantum degeneracies of  spherically symmetric stationary BPS black holes of octonionic magical supergravity in five dimensions  are given by the Fourier coefficients of the modular forms of the arithmetic subgroup of the exceptional group $E_{7(-25)}$ that had been proposed as a spectrum generating extension of its U-duality group $E_{6(-26)}$. The arithmetic group $E_{7(-25)}(\mathbb{Z})$ acts as the discrete conformal group of the exceptional Jordan algebra $J_3^{\mathbb{O}}(\mathcal{R})$ over the integral octonions of Coxeter $\mathcal{R}$. The charges of the black holes take values in the lattice defined by  $J_3^{\mathbb{O}}(\mathcal{R})$. The quantum degeneracies of charge states of rank one and rank two BPS black holes (zero area) are given by the Fourier coefficients of singular modular forms $E_4(Z)$ of weight 4 and $E_8(Z)=(E_4(Z))^2$ of weight 8 of $E_{7(-25)}(\mathbb{Z})$. The quantum degeneracies of charge states of rank 3 BPS (large) black holes and their automorphic properties will be studied elsewhere.
Following the work of N. Elkies and B. Gross on the embeddings of cubic rings $A$  into the exceptional Jordan algebra and their actions on the 24 dimensional orthogonal quadratic subspace of $J_3^{\mathbb{O}}(\mathcal{R})$, we show that the degeneracies of  charge states of  rank one black holes described by such embeddings are given by the Fourier coefficients of the Hilbert modular forms of $SL(2,A)$ which is a discrete subgroup of $SL(2,\mathbb{Z})^3$. If the discriminant of the cubic ring is $D=p^2$ with $p$ a prime number then the isotropic lines in the 24 dimensional quadratic space define a pair of  Niemeier lattices which can be taken as charge lattices of some BPS black holes. For $p=7$ they are the Leech lattice with no roots and the lattice $A_6^4$ with 168 root vectors. We also review the current status of the searches for the M/superstring theoretic origins of the octonionic magical supergravity. 

\end{titlepage}
\tableofcontents

\section{Introduction}
\label{sec:intro}
U-duality orbits of extremal, spherically symmetric, stationary black holes in $\mathcal N=2$ Maxwell-Einstein supergravity theories (MESGT)  defined by Euclidean Jordan algebras of degree three \cite{Gunaydin:1983rk,Gunaydin:1983bi,Gunaydin:1984ak,Gunaydin:1986fg}, as well as in maximal $\mathcal N=8$ supergravity in five and four dimensions were first classified in \cite{Ferrara:1997uz}.
They were later studied in greater detail in \cite{Ferrara:2006xx,Bellucci:2006xz} and further refined and extended in \cite{Cerchiai:2010xv,Borsten:2011ai,Borsten:2011nq}. $5d$ MESGTs with symmetric target spaces $G/H$ such that $G$ is a global symmetry of the Lagrangian  are uniquely defined by an underlying Euclidean Jordan algebra $J$ of degree three \cite{Gunaydin:1983bi,Gunaydin:1984ak}. The invariance group of the cubic norm of $J$, which is simply the Lorentz group of the Jordan algebra $J$, is also the U-duality group of the corresponding $5d$ MESGT.  Since the vector fields in these theories are in one-to-one correspondence with the elements of the Jordan algebra $J$, the charges  
of stationary, spherically symmetric, extremal black holes can be represented by elements of the Jordan algebra with integral entries. The entropy of the black holes is then given by the cubic norm of the corresponding charge matrices  that is invariant under the Lorentz group of $J$. 
For small black holes (i.e., black holes with vanishing  area ) the cubic norm vanishes. 
In addition to invariance under the Lorentz group, the norms of charge matrices representing small black holes with vanishing area are invariant under the special conformal transformations, as well as under scaling of the corresponding elements of the Jordan algebra $J$. This is similar to the invariance of light-like vectors under special conformal transformations in $4d$  Minkowski spacetime whose coordinates can be represented  by Pauli matrices, which in turn form a Jordan algebra under the symmetric Jordan product. Special conformal transformations act on large black holes with non-vanishing entropy and change their entropy. Therefore, the conformal groups $\conf{J}$ of underlying Jordan algebras were proposed as spectrum generating symmetry groups of black hole solutions of $\mathcal N=2$ MESGTs \cite{Ferrara:1997uz,Gunaydin:2000xr,Gunaydin:2005gd,Gunaydin:2004ku}. The U-duality groups of the corresponding   four dimensional supergravity theories obtained by dimensional reduction  are isomorphic to the respective conformal groups $\conf{J}$ of $J$.  The conformal group $\conf{J}$ acts linearly on the electric and magnetic charges of the $4d$ theory, and acts non-linearly on the complex scalar fields which parameterize the upper half-plane of the Jordan algebras $J$ \cite{Gunaydin:1983bi}.  

 This proposal led to the natural question as to whether the
$3d$ U-duality groups, $G_3$, could  be interpreted as spectrum generating groups of corresponding $4d$ supergravity theories whose scalar fields correspond to symmetric spaces  \cite{Gunaydin:2000xr}. This investigation resulted in   the discovery of novel geometric realizations of the U-duality groups of $3d$ supergravity theories as quasiconformal groups  $\qconf{J}$ that extend the $4d$ U-duality groups \cite{Gunaydin:2000xr}. These quasi-conformal
groups act  on the vector spaces of Freudenthal triple systems (FTS) $\mathcal{F}(J)$ associated with Jordan algebras $J$ of degree 3   extended by an extra singlet coordinate such that they
leave invariant a generalized light-cone with respect to a quartic
distance function. These quasiconformal extensions of $4d$ 
U-duality groups  were then  proposed as  spectrum generating symmetry groups of the $4d$ supergravity theories 
\cite{Gunaydin:2000xr,Gunaydin:2004ku,Gunaydin:2003qm,Gunaydin:2005gd,Gunaydin:2005mx,Gunaydin:2007bg}.
A concrete framework for realization  of this proposal for  spherically
symmetric stationary BPS black holes of $4d$ supergravity
theories was given  in
\cite{Gunaydin:2005mx,Gunaydin:2007bg,Gunaydin:2007qq}. This framework uses the fact that the attractor equations of  these black holes are equivalent to the equations of geodesic motion of a fiducial particle on the scalar manifold of the supergravity theory obtained by reduction on a time-like circle whose isometry group  is the U-duality group of the corresponding three dimensional supergravity.

Extremal black holes with non-vanishing entropy exhibit attractor phenomena \cite{Ferrara:1995ih,Ferrara:1996um} and the study of their connection to arithmetic was initiated by Moore \cite{Moore:1998pn,Moore:2004fg}. A major result of Moore is establishing the connection between the numbers of attractor black holes with a given area in $K3 \times T^2$ compactification of type II superstrings to four dimensions and the
class numbers of binary quadratic forms with negative
discriminant. The corresponding low energy supergravity theories have $\mathcal N=4$ supersymmetry. More recently it was pointed out that the relation 
 between attractors and arithmetic can be extended to  black holes in $\mathcal N=2$ supergravity and string models whose  equivalence classes involve more general forms under the action of arithmetic subgroups of the U-duality groups \cite{Gunaydin:2019xxl}. Most prominent examples involve ``magical" supergravities defined by Euclidean Jordan algebras of degree three and their number theoretic counterparts are directly related to the work of Bhargava \cite{bhargava1,  bhargava2}.

It is well known that continuous U-duality groups of supergravity theories are broken down to their arithmetic subgroups when they are embedded into M/superstring theories. In this paper we study the role of arithmetic subgroups of spectrum generating extensions of U-duality  groups of $5d$ supergravity theories  defined by Euclidean Jordan algebras of degree three. We will focus mainly on the octonionic magical supergravity whose continuous U-duality group in $5d$ is $E_{6(-26)}$ with the maximal compact subgroup $F_4$ and its spectrum generating conformal group is $E_{7(-25)}$ with the maximal compact subgroup $E_6\times U(1)$. The main reason for our choice is its connections to some deep mathematical structures as well as its potential relevance to physics \cite{Gunaydin:1983bi,Gunaydin:1984be}. Even though it is not yet known whether the quantum completion of the octonionic magical supergravity theory is a superstring theory or a novel phase of M-theory \textit{we assume that, at the quantum level, its continuous U-duality group is broken down to its maximal arithmetic subgroup}.   Furthermore the other magical supergravity theories can be obtained by truncation of the octonionic theory and it can also be truncated to a MESGT with 10 vector multiplets that belong to the infinite generic Jordan family of MESGTs defined by reducible Jordan algebras of degree three.  This MESGT with 10 vector multiplets in $5d$ describes the vector multiplet sector of the FHSV model\cite{Ferrara:1995yx}. 

{\it We should  stress that we define the quantum degeneracy of  charge states of BPS black holes as a purely number theoretic quantity, and it is not the same as the physical degeneracy of microstates of stringy black holes (see section 12 for further discussion of this point). 
Charge states are represented by matrices that are elements of the underlying exceptional Jordan algebra. The charge states with a given set of invariant labels transform in a representation of a finite subgroup of the maximal compact subgroup $F_4$ of the continuous U-duality group $E_{6(-26)}$.  }

 Small BPS black holes of rank one have one non-zero label  given by the linear trace form of the charge matrix and rank two BPS black holes have non-vanishing quadratic spur form as well as linear trace form. Charge matrices of large BPS black holes correspond to rank three elements of the Jordan algebra with non-vanishing cubic form in addition to non-vanishing spur and trace forms. BPS condition forces the non-vanishing labels to be positive.  This constrains the elements of the underlying Jordan algebra  that represent the charge states to lie in the exceptional cone defined over the exceptional Jordan algebra.

\subsection{Outline of the Paper}
In \autoref{sec:5dn=2}, we review the $5d, \ \mathcal{N}= 2$ MESGTs coupled to an arbitrary number of vector multiplets. These theories are uniquely determined by a symmetric tensor $C_{IJK}$ of rank 3.
MESGTs with symmetric scalar manifolds $G/H$ such that $G$ is a global symmetry of their Lagrangians are in one-to-one correspondence with Euclidean Jordan algebras of degree three. 
In \autoref{Jordansymmetries}, we review the symmetries of Euclidean  Jordan algebras. In \autoref{sec:magicalsupergravities}, we review the magical supergravity theories that are defined by {\it simple } Euclidean Jordan algebras of degree three which are realized by $3\times 3$ Hermitian matrices over the four division algebras. The largest magical supergravity is defined by the exceptional Jordan algebra of $3\times 3$ Hermitian  matrices over the octonions, which we call \textit{octonionic magical supergravity}.
 The \autoref{sec:embedding} discusses the question of embedding of the magical supergravity theories into  M/superstring theory. 
In \autoref{5dorbits}, we review the orbits of the spherically symmetric stationary extremal  black hole solutions of the $5d$ octonionic magical supergravity under the action of its U-duality group $E_{6(-26)}$.
\Cref{sec:specgen} reviews the proposal that the conformal groups of the Jordan algebras that underlie the MESGTs must act as their spectrum generating symmetry groups.  The relation between these spectrum generating symmetries and U-duality groups of dimensionally reduced theories is also  explained. In \autoref{sec:exceptionalcone}, we review the construction of the exceptional cone over the exceptional Jordan algebra following N. Elkies and B. Gross \cite{MR1411589} and introduce the concept of polarizations that will play a key role in distinguishing between different orbits of the quantum  BPS black holes under the discrete U duality group $E_{6(-26)}(\mathbb{Z})$. \Cref{modularform} reviews exceptional modular forms defined over the exceptional tube domain and their relation to the exceptional Jordan algebra over integral octonions $\mathcal{R}$. In particular, section 9.1 reviews the derivation of the degeneracy of rank one elements with a given trace form obtained by N. Elkies and B. Gross in \cite{MR1411589} and  section 9.3 gives a brief discussion of the relations between cubic rings and binary cubic forms. 
In \autoref{Springer}, we provide a brief interlude into the Springer decomposition of Jordan algebras of degree three.
In \autoref{sec:embeddings}, following \cite{MR1411589,MR1845183}, we review the embeddings of cubic rings into $J_3^{\mathbb{O}}(\mathcal{R})$ and their action on the 24 dimensional orthocomplement with a natural quadratic form. We discuss the theta functions of the Niemeier lattices that are defined by the isotropic lines in the 24 dimensional orthocomplement of  the cubic rings inside $J_3^{\mathbb{O}}(\mathcal{R})$. The signature of a Niemeier lattice, as we shall see, is determined by the number of  root vectors. These classical theta functions arise from Hilbert modular forms over the exceptional domain that were studied in \cite{MR1216126}.   In \autoref{sec:diffrankbhs}, we show how the Fourier coefficients of singular modular forms over the exceptional tube domain\footnote{We refer to modular forms over the exceptional tube domain as  exceptional modular forms for brevity.} of weight $4$ i.e., $E_4(Z)$ and of weight $8$ i.e., $E_8(Z)=E_4(Z)^2$ of Kim \cite{MR1216126} describe the degeneracies of charge states of rank one and rank two BPS black holes. We also discuss the Hilbert modular forms that describe the quantum degeneracies of charge states of rank one BPS black holes  whose charge lattices   are given by Niemeier lattices.  We then review  a reconstruction of the  singular modular forms $E_4(Z)$ and $E_8(Z)$ using Fourier Jacobi expansion over the upper half-plane of integral octonions as well as over the upper half-plane of the Jordan algebra of $2\times 2$ Hermitian matrices over the integral octonions
following \cite{MR1195510,MR2478255}.  Charge states of large BPS black holes are described by rank three elements of the exceptional Jordan algebra over the integral octonions that lie in the exceptional cone. In connection with the discussion of large BPS black holes we emphasize again that our number theoretic definition of quantum degeneracy  is not to be  confused with the degeneracy of microscopic states of large stringy BPS black holes.  Quantum degeneracies of the charge states of  rank three BPS black holes are related to higher powers $E_4(Z)^n$ of the singular modular form $E_4(Z)$ for  $n>2$.  The study of the relation between higher powers of $E_4(Z)$ and modular forms of higher weight over the exceptional domain studied by mathematicians and the quantum degeneracies of charge states of rank 3 BPS black holes that exhibit attractor phenomena is left to future studies. 
Finally, in \autoref{sec:cy}, we use  various results and interpretations to make an educated guess regarding the properties of a Calabi-Yau threefold that can embed the octonionic magical supergravity into M-theory/string theory. Although the Borcea-Voisin threefolds can be ruled out there are intricate subtleties that one must consider before realizing this Calabi-Yau as a variation of Hodge structure (VHS) or as a hypersurface in a toric variety. The issues pertaining to both are explained. We conclude the main part of the paper with discussions and comments of future directions of work in \autoref{sec:discussions}. \\
In \autoref{app:thetas}, we list the theta functions of all the Niemeier lattices  as well as the theta function of integral octonions $\mathcal{R}$  for the sake of the reader. We then provide an introduction to modular forms over the exceptional domain and their Fourier coefficients in \autoref{app:exceptionalmf}. Relevant information regarding the discrete subgroups of exceptional groups and their lattices is provided in \autoref{app:discgrps}. In \autoref{commutative}, we review the commutative subrings of the exceptional Jordan algebra. Finally, we provide a brief introduction to the theory of Hilbert modular forms in \autoref{HMF}.

\section{$5d$, $\mathcal{N}=2$ Maxwell-Einstein Supergravity Theories  and Jordan Algebras}
\label{sec:5dn=2}
\setcounter{equation}{0}
$\mathcal{N}=2$ MESGTs in $5d$ that describe
 the coupling of an arbitrary number $(n_V-1)$
of $\mathcal{N}=2$ vector multiplets to $\mathcal{N}=2$ supergravity were constructed
in \cite{Gunaydin:1983rk,Gunaydin:1983bi,Gunaydin:1984ak,Gunaydin:1986fg}.
The bosonic parts of their Lagrangians have a very simple form given by
\begin{eqnarray}\label{Lagrange}
e^{-1}\mathcal{L}_\textrm{bosonic} &=& -\frac{1}{2} R
-\frac{1}{4}{\stackrel{\circ}{a}}_{IJ} F_{\mu\nu}^{I} F^{J\mu\nu}-
 \frac{1}{2} g_{xy}(\partial_{\mu}\varphi^{x})
(\partial^{\mu} \varphi^{y})+\nonumber \\
 && + \frac{e^{-1}}{6\sqrt{6}} C_{IJK} \varepsilon^{\mu\nu\rho\sigma\lambda}
 F_{\mu\nu}^{I}F_{\rho\sigma}^{J}A_{\lambda}^{K},
\end{eqnarray}
where $e$ is the determinant of the f\"{u}nfbein  and $R$ is the
scalar curvature of $5d$ spacetime.  $F_{\mu\nu}^{I}$ denote the field strengths
of the vector fields $A_{\mu}^{I}$ including the graviphoton. $g_{xy}$ is  the metric of the
scalar manifold $\M_5$.  ${\stackrel{\circ}{a}}_{IJ}$ is the ``metric"  appearing in the kinetic energy term of the vector fields that
depends on the scalar fields $\varphi^{x}$. The range of indices are
\begin{eqnarray*}
I&=& 1,\ldots, n_V\\
a&=& 1,\ldots, (n_V-1)\\
x&=& 1,\ldots, (n_V-1)\\
\mu, \nu,... &=& 0,1,2,3,4.
\end{eqnarray*}

The
$\mathcal{N}=2$ MESGTs  in five dimensions have the remarkable feature that they are uniquely determined by the constant tensor $C_{IJK}$ describing the cubic couplings of vector fields
\cite{Gunaydin:1983bi}.  In particular, it was shown that the scalar manifold $\mathcal{M}_5$ can be interpreted as a hypersurface in an $n_V$ dimensional ambient space $\mathcal{C}_{n_V}$ whose metric $a_{IJ}(h)$ is determined by $C_{IJK}$ as follows \cite{Gunaydin:1983bi} :
\begin{equation}\label{aij}
  a_{IJ}(h):=-\frac{1}{3}\frac{\partial}{\partial h^{I}}
  \frac{\partial}{\partial h^{J}} \ln \mathcal{V}(h)\ ,
\end{equation}
where $\mathcal{V}(h)$ is a cubic polynomial in $n_V$ real variables
$h^{I}$ $(I=1,\ldots,n_V)$,
\begin{equation}
 \mathcal{V}(h):=C_{IJK} h^{I} h^{J} h^{K}\ .
\end{equation}

The  $(n_V-1)$-dimensional scalar  manifold, $\mathcal{M}_5$, of scalar fields
$\varphi^{x}$ is then simply the hypersurface in this ambient space defined by the constraint \cite{Gunaydin:1983bi}
\begin{equation}\label{hyper1}
{\mathcal V} (h)=C_{IJK}h^{I}h^{J}h^{K}=1 \ .
\end{equation}
 The ambient space
$\mathcal{C}_{n_V}$ is the domain of positivity (positive cone) as required by the positivity of the kinetic energy terms of scalars and vectors. The metric $g_{xy}$ of the scalar manifold is simply the pullback of
(\ref{aij}) to $\mathcal{M}_5$
\begin{equation}
g_{xy}(\varphi)= h^{I}_{x}  h^{J}_{y} a_{IJ}|_{\mathcal{V}=1} ~,
\end{equation}
where $ h^{I}_{x} =- \sqrt{\frac{3}{2}}
\frac{\partial}{\partial\phi^{x}} h^{I} $ and the ``metric''
${\stackrel{\circ}{a}}_{IJ}(\varphi)$ of the  kinetic energy term of the
vector fields is simply the restriction of the ambient metric
$a_{IJ}$ to the hypersurface $\mathcal{M}_5$:
\begin{equation*}
{\stackrel{\circ}{a}}_{IJ}(\varphi)=a_{IJ}|_{{\mathcal V}=1} \ .
\end{equation*}
The Riemann tensor of the scalar manifold $\mathcal{M}_5$ takes on a very simple form
\begin{equation}
  K_{xyzu}= \frac{4}{3} \left( g_{x[u} g_{z]y} + {T_{x[u}}^{w} T_{z]yw} \right) ~,
\end{equation}
where $T_{xyz}$ is  the pullback of the symmetric tensor $C_{IJK}$
\begin{equation}
   T_{xyz}= h^{I}_{x }h^{J}_{y} h^{K}_{z} C_{IJK}=- \left(\frac{3}{2}\right)^{3/2}  h^{I}_{,x }h^{J}_{,y} h^{K}_{,z} C_{IJK} ~.
\end{equation}
Conversely we have
\eq
C_{IJK} = \frac{5}{2} h_I h_J h_K -\frac{3}{2} \stackrel{\circ}{a}_{(IJ} h_{K)} + T_{xyz} h^x_I h^y_J h^z_K ~.
\en
Hence, if the $T-$tensor is covariantly constant i.e.,
$
T_{xyz ; w} = 0 $  we have
\begin{equation}
 K_{xyzu ; w} =0
\end{equation}
i.e.,  the  scalar manifold
is a locally symmetric space. Remarkably, the covariant constancy of
$T_{xyz}$ implies the ``adjoint
identity'' \cite{Gunaydin:1983bi}:
\begin{equation}
   C^{IJK} C_{J(MN} C_{PQ)K} = \delta^{I}_{(M} C_{NPQ)}
\end{equation}
and conversely\footnote{
Note that the indices are raised by the inverse
${\stackrel{\circ}{a}}{}^{IJ}$ of ${\stackrel{\circ}{a}}{}_{IJ}$.} and the  $\mathcal{N}=2$ MESGT's that
satisfy the adjoint identity are in one-to-one correspondence with
 Euclidean Jordan algebras of degree 3
\cite{Gunaydin:1983bi}. This correspondence follows from the identification of cubic norms defined by the $C_{IJK}$ tensor with the norms of degree three Jordan algebras. Furthermore,
cubic forms that satisfy the adjoint identity are also in one-to-one correspondence with
Legendre invariant cubic forms studied in
\cite{MR2094111}.

Scalar manifolds of $\mathcal{N}=2$ MESGTs defined by Euclidean Jordan
algebras of degree three are symmetric spaces of the form
\begin{equation}
    \mathcal{M} = \frac{\textbf{Str}_0 \left(J\right)}{ \aut{J}}~,
\end{equation}
where $\textbf{Str}_0\left(J\right)$ and $\aut{J}$
are the reduced structure group and automorphism
 group of the Jordan algebra $J$, respectively. Following established convention, we will refer to the reduced structure and automorphism groups as Lorentz and rotation groups of the underlying Jordan algebra $J$, respectively.  Their vector
fields including the graviphoton  are in one-to-one correspondence with
elements of $J$ and transform  linearly
under $\stro{J}$.


\section{\label{Jordansymmetries}  Rotation, Lorentz  and Conformal groups
of Generalized Spacetimes Coordinatized by Jordan Algebras}

Generalized spacetimes coordinatized by Jordan algebras were first introduced in  \cite{Gunaydin:1975mp}. For the $4d$ Minkowski spacetime, the underlying Jordan algebra $J_2^{\mathbb C}$ is generated by Pauli matrices including the identity matrix with the Jordan product defined as 1/2 the anticommutator i.e., $\displaystyle\frac 12 \lbrace \cdot, \cdot \rbrace$.
Then the rotation $SU(2)$, Lorentz $SL(2,\mathbb{C})$ and
conformal group $SU(2,2)$ in four dimensions are  simply  the automorphism,
reduced structure and M\"{o}bius (linear fractional) groups of the
Jordan algebra $J_2^{\mathbb
C}$, respectively \cite{Gunaydin:1975mp,Gunaydin:1979df}.   For   generalized spacetimes coordinatized by Jordan  algebras $J$, their
rotation $\rot{J}$, Lorentz $\lor{J}$ and conformal $\conf{J}$ groups correspond to the
automorphism $\aut{J}$, reduced structure $\stro{J}$ and M\"obius $\mob{J}$ groups of $J$, respectively
\cite{Gunaydin:1975mp,Gunaydin:1979df,Gunaydin:1989dq,Gunaydin:1992zh}. The norm of a coordinate vector remains invariant under the action of the Lorentz group $\lor J$.
Light-like coordinate vectors $X$ in these generalized spacetimes have  vanishing norms $\mathcal{N}(X)=0$.\footnote{We use $\mathcal N (\ast)$ to denote the norm of an element and $\mathcal N = \cdots$ to denote the number of supersymmetries.} They remain light-like under the action of generalized special conformal transformations and light-like separations between any two vectors $X,Y$ with respect to the norm $\mathcal{N}$, $\mathcal{N}(X-Y)=0$ are left invariant under the full conformal group $\conf{J}$ of the Jordan algebra.

The conformal groups of  spacetimes defined by Euclidean
Jordan algebras  all admit positive energy unitary
representations \cite{Gunaydin:1999jb}. They were shown to describe causal spacetimes  with a unitary time evolution as in $4d$ Minkowski spacetime \cite{Mack:2004pv}.
For these spacetimes, the maximal
compact subgroups of their conformal groups are simply the compact real forms of their Lorentz groups times
dilatations.

The conformal group $\conf{J}$  of a Jordan algebra $J$ is
generated by translations $T_{\mathbf{a}}$, $\mathfrak{a} \in J$, special conformal generators $K_{\mathbf{a}}$, dilatations and Lorentz transformations  $M_{\mathbf{a}\mathbf{b}}$ ($\mathbf{a,b} \in J$)
\cite{Gunaydin:1975mp,Gunaydin:1989dq,Gunaydin:1992zh}.  Its  Lie algebra $\mathfrak{conf}(J)$ admits    a 3-grading
with respect to the generator $D$ of dilatations.

Given a basis $e_I$ and a conjugate basis $\Tilde{e}^I$  of a  Jordan algebra $J$,   one can expand a general element
$\mathbf{x} \in J$ as \begin{align*} \mathbf{x} = e_I q^I=\tilde{e}^I q_I ~.\end{align*}
The generators of $\mathfrak{conf}(J)$ act as differential
operators  on the ``coordinates'' $q^I$ which can be twisted
by a unitary character $\lambda$, and take the form
\begin{align}
\begin{split}
  T_I & = \frac{\partial}{\partial q^I}  \\
  R^I_J& = - \Lambda^{IK}_{JL} q^L \frac{\partial}{\partial q^K}  - \lambda  \delta^I_J \\
  K^I & = \frac{1}{2} \Lambda^{IK}_{JL}  q^J q^L  \frac{\partial}{\partial q^K} + \lambda q^I ~, \end{split}
\end{align}
where $  \Lambda_{KL}^{IJ}$ are the structure constants of the Jordan triple product  $(e_K,e^I,e_L) $ defined as
\eq
(e_K,e^I,e_L) =\Lambda_{KL}^{IJ} e_J = ( e_K \circ e^I )\circ e_L + ( e_L \circ e^I ) \circ e_K - ( e_K \circ e_L ) \circ e^I ~.
\en
Here, $\circ$ denotes the Jordan product i.e. $\displaystyle X \circ Y = \frac 12 \{X,Y\}$.
The generators of $\mathfrak{conf}(J)$ satisfy the commutation relations
\begin{eqnarray}
& [ T_I, K^J ] =& - R^J_I \\
&[ R^J_I , T_K ] = &\Lambda_{IK}^{JL} T_L \\
&[R^J_I , K^K ] = &- \Lambda_{IK}^{JL} K^L ~.
\end{eqnarray}
The generators of the rotation  subgroup are simply
\begin{equation}
A_{IJ} = R_I^J -R_J^I ~,
\end{equation}
and the generator of scaling transformations is proportional to $R^I_I$.
 For Jordan algebras of degree 3, the tensor $ \Lambda_{KL}^{IJ}$ can be expressed in terms of the $C $ tensor as follows:
\begin{equation}
   \Lambda_{KL}^{IJ} = \delta_K^I \delta_L^J + \delta_L^I \delta^J_K - \frac{4}{3} C^{IJM} C_{KLM} ~.
\end{equation}

For discrete arithmetic subgroups of the conformal groups of Jordan algebras we must work with their global actions. Such an action was given by Koecher \cite{MR214630} who showed that the linear fractional  (conformal)  group $\conf{J}$ action on an element $X\in J$ can always be represented as follows:
\eq
\conf{J} : X  \longrightarrow \xi (X) =  W  \cdot \mathfrak{t}_A \cdot \mathfrak{j} \cdot \mathfrak{t}_B \cdot \mathfrak{j} \cdot \mathfrak{t}_C  \,  (X) =  W ( A - [ B - (X+C)^{-1} ]^{-1} ) ~,
\en
where $A,B,C \in J$.  The operator $\mathfrak{t}_A$ represents translation by $A$
\eq
\mathfrak{t}_A (X) = X + A~. \en
 $\mathfrak{j}$  represents inversion
\eq \mathfrak{j} (X) = - X^{-1} ~, \en
and $W$ is an element of the structure group $\str{J}$ which is the direct product of the reduced structure (Lorentz)  group  and dilatations of $J$. \\

Of relevance to us is
the case of the  non-linear action of the arithmetic subgroup of the conformal group $E_{7(-25)}$ of the exceptional Jordan algebra $J_3^{\mathbb{O}}$ on the exceptional domain $\mathcal{D}$  in $\mathbb{C}^{27}$ corresponding to the upper half-plane of $J_3^{\mathbb{O}}$ which was studied later by Baily Jr. in \cite{MR269779,MR242775}. The subsequent work on modular forms defined over the exceptional domain $\mathcal{D}$ will play a major role in our work. 
\section{Magical Supergravity Theories}
\label{sec:magicalsupergravities}

Among the supergravity theories defined by Euclidean Jordan algebras of degree 3, four of them are distinguished by the fact that they are unified theories. Their underlying Jordan algebras are simple and are realized by $3\times 3$ Hermitian symmetric matrices $J_3^{\mathbb{A}}$  over the four division algebras $\mathbb{A}$, namely the real numbers $\mathbb{R}$, complex numbers $\mathbb{C}$, quaternions $\mathbb{H}$ and octonions $\mathbb{O}$. They are referred to as \textit{magical} supergravity theories.

We shall follow the conventions of \cite{Gunaydin:2009dq} in labelling the elements of $J_3^{\mathbb{A}}$. For $J_3^{\mathbb{R}} $ a general element $Q$  has the form
\begin{equation}
Q =\left( {\begin{array}{*{20}c}
    q_1 & q_6 & q_5  \\
   q_6 & q_2 & q_4  \\
   q_5 & q_4 & q_3  \\
\end{array}} \right) \in J_3^{\mathbb{R}}
\end{equation}
where $q_1,...,q_6$ are real numbers and its cubic norm is given by the determinant\footnote{We should note that for MESGTs defined by Jordan algebras of degree 3 the tensor $C_{IJK}$  is an invariant tensor of $\stro{J}$ and  $C^{IJK}=C_{IJK}$.}
\begin{equation}
       \mathcal{N}(Q) = C^{IJK} q_I q_J q_K =  \left\{ q_1 q_2 q_3 -
            \left[  q_1 (q_4)^2 + q_2 (q_5)^2 + q_3 (q_6)^2 \right]
         + 2 q_4 q_5 q_6 \right\}
\end{equation}

For the Jordan algebras $J_3^{\mathbb{A}}$, where $\mathbb{A}=
\mathbb{C}$, $\mathbb{H}$ or $\mathbb{O}$ coordinates $q_4, q_5$ and $q_6 $ become
elements of $\mathbb{A}$, which we will denote by capital letters
$Q_4, Q_5$ and $Q_6$. Thus for $Q \in J_3 ^{\mathbb{A}}$ we have
\begin{equation}
Q = \left( {\begin{array}{*{20}c}
    q_1 & Q_6 & \bar{Q}_5  \\
   \bar{Q}_6 & q_2 & Q_4  \\
    Q_5 & \bar{Q}_4 & q_3  \\
\end{array}} \right)
\end{equation}
which we will denote as $Q=J(q_1,q_2,q_3;Q_4,Q_5,Q_6)$
The cubic norm of $Q$ is given by the ``determinant":
\begin{equation}
    \mathcal{N}(J(q_1,q_2,q_3;Q_4,Q_5,Q_6)) = \{ q_1 q_2 q_3 -
         \left( q_1 |Q_4|^2 + q_2 |Q_5|^2 + q_3 |Q_6|^2 \right)
        + \Tr (Q_4 Q_5 Q_6) \}
\end{equation}
where $\Tr(X):= X + \bar{X}$ denotes twice the real part of $X \in \mathbb{A}$ and $|X|^2 = X \bar{X} $.
If we expand the elements $Q_4, Q_5 $ and $Q_6$ in terms of their real components, we find
\begin{eqnarray}
 Q_4 = q_4 +q_{(4+3A)} j_A  \nonumber \\
 \bar{Q}_4 = q_4 -q_{(4+3A)} j_A \nonumber \\
 Q_5 = q_5 +q_{(5+3A)} j_A \nonumber \\
 \bar{Q}_5 = q_5 -q_{(5+3A)} j_A \\
 Q_6 = q_6 +q_{(6+3A)} j_A \nonumber \\
 \bar{Q}_6 = q_6 -q_{(6+3A)} j_A \nonumber ~,
\end{eqnarray}
where the index $A$ is summed over and using the fact that the imaginary units satisfy
\begin{equation}
 j_A j_B = - \delta_{AB} + \eta_{ABC} j_C ~.
\end{equation}
We can express the cubic norm as 
\begin{align}
\label{eq:cubicnorm}
\begin{split}
      \mathcal{N}(J(q_1,q_2,q_3;Q_4,Q_5,Q_6))& =    \big( q_1 q_2 q_3 -   q_1 [( q_4)^2  +q_{(4+3A)}q_{(4+3A)}] \\ &
       - q_2 [ (q_5)^2 +q_{(5+3A)}q_{(5+3A)} ] - q_3 [ (q_6)^2 +q_{(6+3A)}q_{(6+3A)} ]   \\ &
      + 2 [ q_4 q_5 q_6   - q_4 q_{(5+3A)} q_{(6+3A)} - q_5 q_{(4+3A)} q_{(6+3A)} -
	q_6 q_{(4+3A)}  q_{(5+3A)} ]  \\ &
      -2 \eta_{ABC}q_{(4+3A)} q_{(5+3B)} q_{(6+3C)} \big) ~.
      \end{split}
\end{align}
The indices $A,B,C$ take on the single value $1$ for complex numbers $\mathbb{C}$, range
from $1$ to $3$ for quaternions $\mathbb{H}$ and from $1$ to $7$ for octonions
$\mathbb{O}$. Note that $\eta_{ABC}$ vanishes for $\mathbb{C}$. For a real quaternion $X \in \mathbb{H}$ we have
\begin{eqnarray}
X&=&X_0 + X_1 j_1 + X_2 j_2 + X_3 j_3 \nonumber \\
\bar{X}& = &X_0 - X_1 j_1 - X_2 j_2 - X_3 j_3 \\
X \bar{X}& =& X_0^2 + X_1^2 + X_2^2 +X_3^2 ~, \nonumber
\end{eqnarray}
where the imaginary units $j_i$ satisfy
\begin{equation}
j_i j_j = - \delta_{ij} + \epsilon_{ijk} j_k ~.
\end{equation}
For a real  octonion $X\in \mathbb{O}$, we have
\begin{eqnarray}
X &=& X_0 + X_1 j_1 + X_2 j_2 + X_3 j_3  + X_4 j_4 + X_5 j_5  +X_6 j_6 + X_7 j_7 \nonumber \\
\bar{X} &=& X_0 - X_1 j_1 - X_2 j_2 - X_3 j_3  - X_4 j_4 - X_5 j_5  - X_6 j_6 - X_7 j_7 \\
X \bar{X} &= &X_0^2 + \sum_{A=1}^7 (X_A)^2 \nonumber ~.
\end{eqnarray}
The seven imaginary units of real  octonions satisfy
\begin{equation}
j_A j_B = - \delta_{AB} + \eta_{ABC} j_C ~,
\end{equation}
where $\eta_{ABC}$ is completely antisymmetric and, in the conventions of reference \cite{Gunaydin:1973rs}, take on the values
\begin{equation}
\eta_{ABC} = 1  \Leftrightarrow  (ABC)= (123), (471), (572), (673), (624), (435), (516) ~.
\end{equation}
The scalar manifolds of the $5d$  magical supergravity theories defined by the simple Euclidean Jordan algebras of degree three are the following symmetric spaces

\begin{equation}
    \begin{array}{cc}
        \mathcal{M}( J_3^\mathbb{R}) =\phantom{gis }  & \dfrac{{SL}(3, \mathbb{R})}
        {{SO}(3)} \\[12pt]
         \mathcal{M}(J_3^\mathbb{C} ) = \phantom{ges } & \dfrac{{SL}(3, \mathbb{C})}{{SU}(3)}
    \end{array}
    \phantom{ and also }
    \begin{array}{cc}
         \mathcal{M}(J_3^\mathbb{H}) =\phantom{ges }  & \dfrac{{SU}^\ast(6)}{{USp}(6)} \\[12pt]
         \mathcal{M}(J_3^\mathbb{O}) =\phantom{ges } & \dfrac{{E}_{6(-26)}}{{F}_4} ~.
    \end{array}
\end{equation}

The magical supergravity theories can be truncated to
theories belonging to the so-called generic Jordan family. To achieve this, one simply
restricts  the elements of $J_3^\mathbb{A}$
to  its  non-simple subalgebra $J = \mathbb{R} \oplus J_2^\mathbb{A}$. The U-duality  symmetry groups of the truncated theories are as follows:
\begin{equation}
\begin{split}
    J = \mathbb{R} \oplus J_2^\mathbb{R}  &: {SO}(1,1) \times {SO}\left(2,1\right) \subset {SL}\left(3, \mathbb{R}\right) \cr
    J = \mathbb{R} \oplus J_2^\mathbb{C}  &: {SO}(1,1) \times {SO}\left(3,1\right) \subset {SL}\left(3, \mathbb{C}\right) \cr
     J = \mathbb{R} \oplus J_2^\mathbb{H}  &: {SO}(1,1) \times {SO}\left(5,1\right) \subset {SU}^\ast(6) \cr
     J = \mathbb{R} \oplus J_2^\mathbb{O}  &: {SO}(1,1) \times {SO}\left(9,1\right) \subset {E}_{6(-26)} ~.
\end{split}
\end{equation}

The truncation of the octonionic magical supergravity defined by $\mathbb{R} \oplus J_2^\mathbb{O} $ is simply the $5d$ supergravity  that reduces to the Maxwell-Einstein  sector of the FHSV model in $4d$ describing the coupling of 10 vector multiplets to $\mathcal{N}=2$ supergravity. The full FHSV model has 12 hypermultiplets coupled to this Maxwell-Einstein supergravity theory~\cite{Ferrara:1995yx}.  The cubic form describing the Maxwell-Einstein sector of FHSV model in five dimensions can be obtained from the cubic norm \eqref{eq:cubicnorm} of the octonionic magical supergravity by setting two out the of the three octonions $Q_4, Q_5$ and $Q_6$ equal to zero. Setting $Q_5=Q_6=0$  the cubic norm of the FHSV model is given by the determinant of
\begin{equation}
J(q_1,q_2,q_3;Q_4,0,0)  = \left( {\begin{array}{*{20}c}
    q_1 & 0 & 0 \\
  0 & q_2 & Q_4  \\
  0 & \bar{Q}_4 & q_3  \\
\end{array}} \right)~,
\end{equation}
which is simply
\begin{equation}
  \mathcal{N}(J(q_1,q_2,q_3;Q_4,0,0)) =q_1 \{  q_2 q_3 -
         |Q_4|^2
         \}~.
\end{equation}
Let $q_1=X$ and $q_2=(Y_0+Y_1) $ , $q_3= (Y_0-Y_1)$ , $q_4= Y_2  $ and $q_{4+3A}=Y_{2+A}$ where $A=1,2,..,7$. Then we have
\begin{align*} \mathcal{N}(J(q_1,q_2,q_3;Q_4,0,0)) = X \left( (Y_0 )^2 - \sum_{i=1}^9 (Y_i)^2 \right) \end{align*}
that is invariant under $SO(9,1)\times SO(1,1)$ which is the global symmetry group of the Maxwell-Einstein sector of the FHSV model in five dimensions \cite{Ferrara:1995yx}.

\section{ M/Superstring Theory Embedding of Octonionic Magical Supergravity}
\label{sec:embedding}
The $\mathcal{N}=2$ supersymmetric octonionic magical supergravity  and the maximal $\mathcal N=8$ supergravity share certain common features.  They are both unified theories and have the exceptional groups of the $E$ series as U-duality groups in five , four and three dimensions. However their real  forms are different. Furthermore they have the same number of vector fields in five and four dimensions.  They have a common sector which is the $\mathcal{N}=2$ quaternionic magical supergravity defined by the Jordan algebra $J_3^{\mathbb{H}}$. Whether there exists a larger theory that can be truncated to both $\mathcal N=8$ supergravity and octonionic magical supergravity was posed as an open problem in \cite{Gunaydin:1984be}. After the discovery of Green-Schwarz anomaly cancellation mechanism in string theory \cite{Green:1984sg}, this question evolved into the question whether octonionic magical supergravity can arise as a low energy effective theory of superstring theory compactified on some \textit{exceptional}\footnote{We use the term exceptional here only to describe Calabi-Yau manifolds that could embed exceptional supergravities into M-theory/string theory.} Calabi-Yau (CY) manifold \cite{Gunaydin:1986fg}. A necessary condition for this is that the intersection numbers of the Calabi-Yau manifold must coincide with the ones given by $C-$tensor corresponding to the cubic norm of the underlying exceptional Jordan algebra $J_3^{\mathbb{O}}$.  Pure octonionic magical supergravity without hypermultiplets would require a rigid CY manifold.  We should note that the supergravity theory with $\mathcal N=2$ supersymmetry and 15 vector multiplets in $4d$ as obtained by Sen and Vafa \cite{Sen:1995ff} via the dual pair method from type II string theory describes the quaternionic magical supergravity without any hypermultiplets as was first pointed out by one of the current authors in \cite{talkparis}. Quaternionic magical supergravity is the largest common sector of the octonionic magical supergravity and the maximal supergravity. It was later realized that there exists an anomaly free supergravity theory in $6d$ which reduces to the octonionic magical supergravity theory coupled to 28 hypermultiplets in five dimensions \cite{gunaydinsezgin}. This led one of the authors to suggest that octonionic magical supergravity coupled to 28 hypermultiplets could  arise as  low energy effective theory of M/superstring theory on a self-mirror CY manifold~\cite{talkparis}.
 In $5d$, this theory would have the scalar manifold
\eq \mathcal{M}_V \times \mathcal{M}_H = \frac{E_{6(-26)}}{F_4} \times \frac{E_{8(-24)}}{E_7\times SU(2)}
\en as its moduli space.
In $4d$ this scalar manifold/moduli space would be the product manifold
\eq \mathcal{M}_V \times \mathcal{M}_H = \frac{E_{7(-25)}}{E_6\times U(1)} \times \frac{E_{8(-24)}}{E_7\times SU(2)}  \en  and in $3d$  the moduli space would be a doubly exceptional symmetric space
 \eq
 \mathcal{M}_V \times \mathcal{M}_H = \frac{E_{8(-24)}}{E_7\times SU(2)} \times \frac{E_{8(-24)}}{E_7\times SU(2)}
 \en
The FHSV model could be obtained as a truncation of this theory since moduli space of the FHSV model in the corresponding dimensions are
  \eqn
5d:~\mathcal{M}_V \times \mathcal{M}_H = \frac{SO(9,1) \times SO(1,1)}{SO(9) } \times \frac{SO(12,4)}{SO(12) \times SO(4)} \\
4d:~ \mathcal{M}_V \times \mathcal{M}_H = \frac{SO(10,2) \times SU(1,1)}{SO(10)\times U(1)\times U(1) } \times \frac{SO(12,4)}{SO(12) \times SO(4)} \\
3d:~ \mathcal{M}_V \times \mathcal{M}_H = \frac{SO(12,4)}{SO(12)\times SO(4) } \times \frac{SO(12,4)}{SO(12) \times SO(4)}~.
\enn	
In four dimensions, the scalar manifold of the vector fields of the octonionic magical supergravity is the symmetric space $E_{7(-25)}/E_6\times U(1)$. Remarkably on the mathematics side Benedict Gross posed it as an open problem whether this particular Hermitian symmetric space could arise as the moduli  space of variations of Hodge structures of a CY manifold \cite{MR1258484} with a number theoretic counterpart related to N\'eron-Severi groups posed in \cite{MR1369418}.

A candidate CY manifold would therefore have to reproduce the above. We however remark here that searching for this candidate CY manifold is not straightforward for reasons that are explained in \autoref{sec:cy}.

\section{Orbits of Extremal Black Holes of $5d$ Octonionic Magical Supergravity \label{5dorbits} }
The orbits of the extremal black holes of $5d$ $\mathcal{N}=2$ MESGTs defined by Jordan algebras of degree three were first classified in \cite{Ferrara:1997uz} and studied in further detail in \cite{Ferrara:2006xx,Cerchiai:2010xv}.
Here we shall review the orbits of the octonionic magical supergravity defined by the exceptional  Jordan algebra $J_3^{\mathbb{O}}$.
We should first  note that in five dimensions, asymptotically flat dyonic solutions do not exist; they are either purely electric black holes with charges charges $q_I$ or their magnetic duals which describe black strings with purely magnetic charges $p^I$.Therefore we shall restrict our study to extremal black hole solutions that are asymptotically flat, static, and spherically symmetric\cite{Ferrara:1997uz,Ferrara:2006xx,Cerchiai:2010xv}. The near horizon geometry of such black holes are of the form $AdS_2 \times S^3$ and their magnetic duals have the geometry $AdS_3\times S^2$.

The attractor mechanism  for $5d$, $\mathcal N=2$ MESGTs in an extremal black hole background is described by the
positive definite effective  black hole potential   \cite{Ferrara:1996um,Ferrara:1997tw}
\begin{equation}
V(\phi, q) = q_I {\stackrel{\circ}{a}}^{IJ} q_J ~,
\end{equation}
where ${\stackrel{\circ}{a}}^{IJ}$ is the inverse of the metric
${\stackrel{\circ}{a}}_{IJ}$ of the kinetic energy term of the
vector fields, and $q_I$ is 
the  $(n+1)$
dimensional charge vector \begin{align*} q_I = \int_{S^3} H_I = \int_{S^3}
\stackrel{\circ}{a}_{IJ} *F^J \hspace{1cm} (I=0,1,...n) ~.\end{align*} 
 The metric ${\stackrel{\circ}{a}}_{IJ}$ is related to
the metric $g_{xy}$ of the scalar manifold as follows
\begin{align}
\begin{split}
{\stackrel{\circ}{a}}_{IJ} = h_I h_J + \frac{3}{2} h_{I,x} h_{J,y}
g^{xy} \label{eq:vectormetric} \\
{\stackrel{\circ}{a}}^{IJ} =h^Ih^J + \frac{3}{2} h^I_{,x}
h^J_{,y} g^{xy} \end{split} \end{align}
and
\begin{equation}
g_{xy}= \frac{3}{2} h_{I,x} h_{J,y}{\stackrel{\circ}{a}}^{IJ} ~.\en
In terms of the central charge $ Z= q_Ih^I$
  the potential takes the form  \eq V(q,\phi) = Z^2 + \frac{3}{2} g^{xy} \partial_x Z
\partial_y Z \label{eq:N2} ~, \en
where $ \partial_x Z = q_I h^I_{,x} =\sqrt{2/3} \, q_I \,  h^I_x  $.
Using the identity  \eq h^I_{,x;y} =\frac{2}{3}( g_{xy} h^I -\sqrt{\frac{3}{2}}
 T_{xyz} g^{zw} h^I_{,w} )~, \en
the critical points of the potential are determined by the solutions of
 \eq \partial_x V= 2 ( 2Z\partial_x Z - \sqrt{3/2}
T_{xyz} g^{yy'}g^{zz'} \partial_{y'}Z \partial_{z'} Z ) =0 ~.
\label{eq:derpotential}\en
The BPS critical points are then given by  \eq
Z_x \equiv \partial_x Z=0 \en
and the non-BPS critical points are given by the equation  \eq 2 Z \partial_x Z = 
\sqrt{\frac{3}{2}} T_{xyz} \partial^y Z \partial^z Z = \sqrt{\frac{3}{2}} T_{xyz}  Z^y  \,  Z^z ~,
\label{eq:nonbps} \en where
\begin{align*}Z^x\equiv  \partial^x Z = g^{xy}  Z_y ~.\end{align*}

Using the identity 
 \eq
q_I = h_I Z -\frac{3}{2} h_{I,x} \partial^x Z \en that follows from \eqref{eq:vectormetric}, one finds that for BPS critical points  $\partial_x Z=0$ one has \eq
q_I = h_I Z ~, \en and for non-BPS critical points  $\partial_x Z \neq 0$ one has \eq q_I = h_I Z
- (3/2)^{3/2} \frac{1}{2Z} h_{I, x} T^{xyz}
 Z_y   Z_z ~. \en

The BPS orbit
with non-vanishing entropy  given by
$Z_x=0 $ is simply the symmetric space $ \displaystyle
    \frac{\stro J}{\aut J} $.
 \autoref{tablebps} lists these moduli spaces for theories defined by Euclidean Jordan algebras of degree three.  The corresponding  black hole  potentials take the value 
\eq V|_{Z_x=0}=Z^2 \en
at the horizon in terms of the central charge. 
\begin{table}
\centering
\begin{tabular}{|c|c|}  \hline
$J$ &$  \mathcal{O}_{BPS}= \stro J/\aut J $ \\ \hline $J_3^\mathbb{R}$ &
$
SL(3,\mathbb{R})/ SO(3) $ \\ \hline
$J_3^\mathbb{C}$   & $ SL(3,\mathbb{C})/ SU(3)  $ \\ \hline

$ J_3^\mathbb{H}$ & $ SU^*(6) / USp(6)  $ \\ \hline

$J_3^\mathbb{O}$  & $ E_{6(-26)} /F_4  $  \\ \hline
$\mathbb{R}\oplus \Gamma_{(1,n-1)}$ & $ SO(n-1,1)\times
SO(1,1) / SO(n-1) $ \\ \hline
\end{tabular}
\caption{ \label{tablebps} Orbits of spherically symmetric stationary BPS black holes  with non-vanishing entropy in $5D$  MESGTs defined by Euclidean Jordan algebras $J$  of degree three.  }
\end{table}

For the non-BPS  black hole solutions with non-vanishing entropy  given by \eqref{eq:nonbps}, the orbits are of the form 
\eq G/\tilde{H}= \stro J/ \widetilde{\aut J)} ~, \en 
where $ \widetilde{\aut J}$ is a non-compact real form of the automorphism group of $J$ and are listed in  \autoref{tablenonbps}.
For the non-BPS extremal black holes with non-vanishing entropy one finds that the black hole potential at the corresponding critical point takes the value
\eq 
V|_\text{non-BPS} = 9 Z^2 ~.
\en
\begin{table}
\centering
\begin{tabular}{|c|c|c|}  \hline
$ J$ & $ \mathcal{O}_{non-BPS}= \stro J/ \widetilde{\aut J)}  $& $ \tilde{K} \subset   \widetilde{\aut J
}$ \\ \hline $J_3^{\mathbb{R}}$ & $ SL(3,\mathbb{R}) / SO(2,1) $  & SO(2)\\ \hline
$J_3^{\mathbb{C}}$ & $SL(3,\mathbb{C}) / SU(2,1)$ &  $SU(2)\times U(1)$\\ \hline
$J_3^{\mathbb{H}}$ & $SU^*(6) / USp(4,2)$ &  $USp(4)\times USp(2)$\\ \hline
$J_3^{\mathbb{O}}$ & $ E_{6(-26)} / F_{4(-20)}$ &  $SO(9)$ \\
 \hline
 $\mathbb{R}\oplus
\Gamma_{(1,n-1)}$
& $ SO(n-1,1) \times SO(1,1) / SO(n-2,1) $ &   $SO(n-2)$ \\ \hline
\end{tabular}
\caption{\label{tablenonbps} The
 orbits of non-BPS extremal black holes of  $N=2$ MESGT's with non-vanishing entropy. The first column lists the underlying Jordan algebras of degree 3 . The third column lists their maximal compact
subgroups $\tilde{K}$ of $ \widetilde{\aut J}$. These non-BPS black holes have moduli described by the symmetric spaces $ \widetilde{\aut J}/\tilde{K}$.}
\end{table}
Since the black hole potential is determined by the metric  ${\stackrel{\circ}{a}}^{IJ}$ of the kinetic energy term of the vector fields, it is positive definite and vanishes only when all the charges $q_I$ vanish. Hence, the attractor mechanism that leads to the criticality condition for the black hole potential is valid only for the black holes with non-zero entropy. So-called small black holes with vanishing entropy do not exhibit the attractor mechanism and it is generally believed that their description requires going beyond the supergravity approximation to their quantum completions. 

In  MESGTs describing the coupling of an arbitrary number of vectors multiplets to pure $\mathcal N=2$ supergravity, one has to distinguish the bare graviphoton $A^0_\mu$ field strength $F^0_{\mu\nu}$ from the physical or ``dressed"  graviphoton field strength. This is given by the linear combination $h_I F^I_{\mu\nu} $ since it is this combination that is related by supersymmetry to the gravitino $\Psi_{\mu}^i$ in the interacting theory
\eqn
\delta e_{\mu}^{m}&=& \frac{1}{2}{\bar{\varepsilon}}^{i}
\Gamma^{m}\Psi_{\mu i}\\
\delta \Psi_{\mu i} &=&\nabla_{\mu} (\omega)\varepsilon_{i}+\frac{i
}
{4\sqrt{6}}h_{I}(\Gamma_{\mu}^{\:\:\:\nu\rho}-4\delta_{\mu}^{\nu}
\Gamma^{\rho})F_{\nu\rho}^{I}\varepsilon_{i} \\
\delta A_{\mu}^{I}&=& -\frac{1}{2}h_{a}^{I}{\bar{\varepsilon}}^{i}
\Gamma_{\mu}\lambda_{i}^{a}+\frac{i\sqrt{6}}{4}h^{I}
{\bar{\Psi}}_{\mu}^{i}\varepsilon_{i}\\
\delta \lambda_{i}^{a}  &=& -\frac{i}{2}f_{x}^{a}
\Gamma^{\mu}(\partial_{\mu}\varphi^{x})\varepsilon_{i} +
\frac{1}{4}h_{I}^{a}
\Gamma^{\mu\nu}
\varepsilon_{i}F_{\mu\nu}^{I} \\
\delta \varphi^{x}&=&\frac{i}{2}f^{x}_{a}{\bar{\varepsilon}}^{i}
\lambda_{i}^{a},\label{trafo}
\enn 
where $ \lambda_{i}^{a}$ are the spinor fields in the vector multiplets and $f^{x}_{a}$ is  the $n$-bein on the scalar manifold.
Hence the central charge $Z= q_I h^I$ is simply the dressed charge associated with  the physical graviphoton. Similarly one can interpret $Z^x$ as dressed charges with respect to the dressed vector field strengths $h_{I}^x F^I_{\mu\nu}$. 

The $C-$tensor $C_{IJK}$ that defines the $\mathcal N=2$ MESGT uniquely is a constant tensor and is given by the intersection numbers of the Calabi-Yau threefolds for those theories that descend from M-theory. The tensor $C^{IJK}$ obtained by raising the indices by the metric $\stackrel{\circ}{a}^{IJ}$ 

\eq 
 C^{IJK} = \stackrel{\circ}{a}^{II'}\stackrel{\circ}{a}^{JJ'}\stackrel{\circ}{a}^{K K'} C_{I' J' K' }
 \en
depends on the scalar fields in general. However, for those theories defined by the Euclidean Jordan algebras $J$  of degree three, the $C-$tensor is an invariant tensor of the U-duality group $\stro{J}$ and one has \cite{Gunaydin:1983bi}
\eq
C^{IJK} =C_{IJK} ~.
\en
Given a black hole solution of the $5d$ $\mathcal{N}=2$
 MESGT defined by the Jordan algebra $J$ with (electric) charges $q_{I}$, we associate an element
 $Q=e^{I}q_{I}$ of $J$, where $e^{I}, \, I=1,2,...,n_V $ form  a
basis of $J$.
The entropy $S$ of an
extremal black hole solution is then determined by the cubic norm
$\mathcal{N}(Q)$ of $Q$.
 More specifically
\eq
S= \pi \sqrt{|\mathcal{N}(Q)|}
\en 
where $\mathcal{N}(Q) =C^{IJK} q_I q_J q_K $. Using the fact that $ C^{IJK} =C_{IJK}$ one can write the cubic norm in terms of the dressed charges $Z, Z_x$ as follows:
\eq
\mathcal{N}(Q) = C^{IJK} q_I q_J q_K = Z^3 - (3/2)^2 Z Z_x Z_y g^{xy} - (3/2)^{3/2} T_{xyz}Z^x Z^y Z^z ~.
\en
Here we should stress the fact that while the bare charges $q_I$ take integer values  the dressed charges need not be integrally charged.

Specializing to the case of exceptional supergravity, the orbits of extremal black hole solutions under the action of U-duality group $E_{6(-26)}$ fall into three categories depending on the rank of the charge matrix $Q=J(q_1,q_2,q_3;Q_4,Q_5,Q_6)$\cite{Ferrara:1997uz,Ferrara:2006xx}. Firstly, we should note that by the action of the compact automorphism group $F_4$, any element $J\in J_3^{\mathbb{O}}$ can be  diagonalized:\footnote{ See \cite{Gunaydin:1978jq} and the references therein.}
\eq
F_4 : \qquad J \Rightarrow (\lambda_1 E_1 + \lambda_2 E_2 + \lambda_3 E_3 )
\en
where $\lambda_i$ are the eigenvalues of $J$ and  $E_i, \ (i=1,2,3) $ are  the irreducible idempotents of $J_3^{\mathbb{O}}$ defined as
\eq
E_1= J(1,0,0;0,0,0) , \qquad E_2=J(0,1,0;0,0,0) , \qquad E_3=J(0,0,1;0,0,0) ~.
\en
The cubic norm of $J$ is then simply given by $\cN (J) =\lambda_1 \lambda_2 \lambda_3$. The rank one elements can be brought to a multiple of an irreducible idempotent~\cite{Ferrara:1997uz}
\eq
\Lambda_i = \lambda E_i \qquad i=1,2,3 \qquad ,  \qquad \lambda \in \mathbb{R}~. \label{rank1}
\en
 The corresponding extremal black holes have vanishing entropy i.e., small black holes and their orbits are
\eq
\frac{E_{6(-26)}}{SO(9,1)\circledS T^{16} }~,
\en
where $SO(9,1)\circledS T^{16} $ represents the semi-direct product of $SO(9,1)$ with translations in its 16 dimensional (Majorana-Weyl)  spinor representation. 
They were called critical light-like orbits in \cite{Ferrara:1997uz} and describe 1/2 BPS black holes whose moduli spaces are \cite{Cerchiai:2010xv}
\eq \frac{SO(9,1)}{SO(9)}\circledS \mathbb{R}^{16}~ \label{rank1bps}\en
The rank 2 elements  can be brought to the form 
\begin{align*} S_{ij} =\lambda ( E_i+ E_j ) \qquad (i\neq j) \end{align*} or to the form
\begin{align*} A_{ij}= \lambda ( E_i -E_j) \end{align*} by the action of $E_{6(-26)}$\cite{Ferrara:1997uz,Ferrara:2006xx} and describe  black holes with vanishing entropy.  The orbits of black holes described by $S_{ij}$ are given by the  coset  space
\eq \frac{E_{6(-26)}}{SO(9)\circledS T^{16} } ~ \label{rank2bps} \en
They are 1/2 BPS black holes with moduli $\mathbb{R}^{16}$ \cite{Cerchiai:2010xv}. The black holes described by $A_{ij}$ are non-BPS and their orbits are given by the coset space
\eq
\frac{E_{6(-26)}}{SO(8,1)\circledS T^{16}} 
\en
with moduli spaces $\frac{SO(8,1)}{SO(8)} \circledS \mathbb{R}^{16} $. The orbits defined by $S_{ij}$ and $A_{ij}$ were called light-like orbits in \cite{Ferrara:1997uz}.
The elements of $J_3^{\mathbb{O}}$  with non-vanishing cubic norm ( rank 3)  can be brought to either of the following forms  by the action of $E_{6(-26)}$:
\eq  S_{ijk} = ( E_i +  E_j + \lambda E_k ) \en
or the form \eq  A_{ijk} = ( - E_i -  E_j + \lambda E_k )~, \en
where $i\neq j, j\neq k , i\neq k$.
The black holes whose charge matrix can be brought to the form $S_{ijk}$ with $\lambda >0$ describe 1/2 BPS  extremal black holes with non-vanishing entropy. They belong to the orbit
\eq
\frac{E_{6(-26)}}{F_4 } \label{rank3bps}
\en
and have no moduli. The extremal black holes described by charge matrices of the form $A_{ijk}$ with $\lambda >0$ are non-BPS  extremal black holes with orbits~\cite{Ferrara:1997uz}
\eq
\frac{E_{6(-26)}}{F_{4(-20)} }
\en
and moduli  spaces
\eq
\frac{F_{4(-20)}}{SO(9)}~.
\en

We should note that the black holes with vanishing entropy are commonly referred to as small black holes whereas those with non-vanishing entropy as large black holes in the literature and  we shall adopt this convention.

\section{ Conformal group $E_{7(-25)}$ of the Exceptional Jordan Algebra as Spectrum Generating Symmetry Group of  $5d$ Octonionic Magical Supergravity}
\label{sec:specgen}
 The U-duality group of $5d$ supergravity theory defined by a Euclidean Jordan algebra $J$ of degree 3  is simply the Lorentz group $\lor J$ of $J$ which is also the invariance group of the norm. Hence the entropy of a large  extremal black hole in these   supergravity theories given by the norm form of $J$  is invariant under the action of  $\stro J$. However, the small extremal black holes with vanishing entropy  have a larger symmetry, namely they remain light-like under the action of the semidirect product  group
 $\left( \lor J \times SO(1,1)\right)\rtimes S_J $, where $S_J$ represents the Abelian  group of special conformal transformations labelled by the elements of $J$.  Under the action of the full conformal group $\conf{J}$, reviewed in \autoref{Jordansymmetries},  that include translations $T_J$ as well as special conformal generators $S_J$, the light-like vectors do not remain light-like in general.
 Conformal group $\conf{J}$ changes the norm of a general
element $Q\in J$ representing the charges and hence the corresponding entropy of extremal  black hole
solution coordinatized by $Q$.
 Therefore conformal groups $\conf{J}$ of Jordan algebras
 were proposed as spectrum generating
symmetry groups of the solutions of the $5d$ supergravity theories defined by them
~\cite{Ferrara:1997uz,Gunaydin:2000xr,Gunaydin:2004ku,Gunaydin:2005gd,Gunaydin:2009pk}.

The conformal group $\conf{J}$ is isomorphic to the U-duality symmetry group
of the corresponding $4d$ supergravity theory  obtained by dimensional
reduction (R-map) from five dimensions.  The one-to-one correspondence between vector fields and the elements of the underlying Euclidean Jordan algebra $J$ of degree three  in five dimensions go over to the correspondence between  vector field strengths and their magnetic duals and  the Freudenthal triple system associated with the Jordan algebra $J$\cite{Gunaydin:1983bi,Gunaydin:2000xr,Ferrara:1997uz,Gunaydin:2005gd,Gunaydin:2005zz,Gunaydin:2009pk}.

Elements $X$ of a Freudenthal triple system (FTS) $\mathcal{F}(J) $ \cite
{MR0170974,MR0063358} associated with $J$ can be represented formally as a
$2\times 2$ ``matrix'':
\begin{equation}X= \left(
\begin{array}{ccc}
\alpha &  & \mathbf{x} \\
&  &  \\
\mathbf{y} &  & \beta
\end{array}
\right) \in \mathcal{F}(J)~,
\end{equation}
where $\alpha$, $\beta \in \mathbb{R}$ and $\mathbf{x}$,$\mathbf{y} \in J$.

Labelling the $4d$ graviphoton field strength and
its magnetic dual as $F_{\mu \nu }^{0}$ and
$\widetilde{F}_{0}^{\mu\nu}$, respectively, one has the correspondence
\begin{equation*}
\left(
  \begin{array}{ccc}
     F_{\mu \nu }^{0} &  & F_{\mu \nu }^{I} \\ &  &  \\
     \widetilde{F}_{I}^{\mu \nu } &  & \widetilde{F}_{0}^{\mu \nu }
  \end{array}
\right) \Longleftrightarrow
 \left(
    \begin{array}{ccc}
        e_0 &  & e_I \\
            &  &  \\
\tilde{e}^I &  & \tilde{e}^0
    \end{array}
 \right) \in \mathcal{F}(J),
\end{equation*}
where $e_I (\tilde{e}^I)$ are the basis elements of $J$ (its dual $\tilde{J}$).
Therefore, given a black hole solution with
electric and magnetic charges (fluxes) $\left(q_0, q_I,
p^{0},p^{I}\right)$ of the $4d$ MESGT defined by $J$, we can associate an element of the FTS
$\mathcal{F}\left( J \right) $ to it:
\begin{equation}
\left(
\begin{array}{ccc}
p^{0}e_0 &  & p^{I} e_I\\
&  &  \\
q_{I} \tilde{e}^I &  & q_{0}\tilde{e}^0
\end{array}
\right)
\in \mathcal{F}(J) ~.
\end{equation}
The U-duality group of the $4d$ MESGT is simply the
automorphism group of the FTS $\mathcal{F}(J)$. The FTS admits an
invariant symmetric quartic form and a skew-symmetric bilinear
form. The entropy of an extremal black hole solution  of the four dimensional theory with charges $ (p^0, p^I,q_0 ,q_I) $ is determined  by the quartic invariant $\mathcal{Q}_4 (q,p)$ of $\mathcal{F}(J)$. The orbits of extremal black holes of $4d$  $\mathcal N=2$ MESGT's with symmetric scalar manifolds were classified in \cite{Ferrara:1997uz,Bellucci:2006xz}.

The automorphism group $\aut{\mathcal{F}(J)}$ of the FTS  $\mathcal{F}(J)$ defined by $J$ is isomorphic to the conformal group $\conf{J}$ of $J$. There are two different actions of the conformal group $\conf{J}$ in the $4d$ theory. $\conf{J}$ acts linearly on the vector field strengths and their magnetic duals and non-linearly on the complex scalar fields. The manifold of complex scalar fields of the $4d$ theory can be mapped into the upper half plane of the corresponding Jordan algebra and the conformal group $\conf{J}$ acts holomorphically on the corresponding tube domain\cite{Gunaydin:1983rk,Gunaydin:1983bi}.

The proposal that the $4d$ U-duality group acts as a spectrum generating symmetry of the $5d$ supergravity raises the question whether the U-duality group of the  corresponding $3d$ supergravity can act as spectrum generating symmetry group  of the $4d$ theory. 
 This  question was first investigated in \cite{Gunaydin:2000xr} where 
it  was shown in that  the $3d$
U-duality groups of supergravity theories defined by Jordan algebras all have novel geometric realizations as quasi-conformal
groups. These quasiconformal groups act  non-linearly on the vector spaces of the corresponding  FTSs  extended by an extra singlet coordinate and 
leave light-like separations with respect to a quartic
distance function invariant. These quasiconformal actions of $3d$
U-duality groups  act  as  spectrum generating symmetry
groups of corresponding $4d$ supergravity theories
\cite{Gunaydin:2000xr,Gunaydin:2004ku,Gunaydin:2003qm,Gunaydin:2005gd,Gunaydin:2005mx,Gunaydin:2007bg,Gunaydin:2009dq}.
The quasiconformal groups defined over FTS's
$\mathcal{F}$ are denoted as
$\qconf{\mathcal{F}}$. When the corresponding  FTS is defined over a Jordan algebra $J$ of degree three they are denoted  either as $\qconf{\mathcal{F}}$ or simply as $\qconf{J}$. The construction given in \cite{Gunaydin:2000xr} is covariant under the $4d$ U-duality group of the corresponding supergravity. For
$\mathcal N =2$ MESGTs defined by Euclidean Jordan algebras of degree three,
quasiconformal realizations  of their $3d$ U-duality
groups  were given  in a
basis covariant with respect to their $6d$ duality groups  in \cite{Gunaydin:2005zz} and with respect to their $5d$ U-duality groups in \cite{Gunaydin:2009dq}. 

For the octonionic magical supergravity, the $3d$ U-duality group is the group $E_{8(-24)}$ with maximal compact subgroup $E_7 \times SU(2)$ and the corresponding FTS is 56 dimensional. Its quasiconformal realization as a spectrum generating symmetry group of black hole solutions  acts on a 57 dimensional space labelled 28 electric and 28 magnetic charges and the extra singlet coordinate was interpreted as entropy of the $4d$ blackholes in \cite{Gunaydin:2000xr}. 

\section{The Exceptional Cone}
\label{sec:exceptionalcone}
 In this section we shall review the exceptional cone defined by the exceptional Jordan algebra $J_3^{\mathbb{O}}$  following  the formulation given in \cite{MR1411589} where the lattices defined by elements of $J_3^{\mathbb{O}}$  over the integral octonions were studied. We shall denote the division algebra of octonions as $\mathbb{O}$ and, following  \cite{MR1411589}, label $J(a,b,c;x,y,z)$ as the elements  of $  J_3^{\mathbb{O}}$ of the form
\eq
\label{eq:Joctonions}
J=
\left(
  \begin{array}{ccc}
    a & z & \bar{y} \\
    \bar{z} & b & x \\
    y & \bar{x} & c \\
  \end{array}
\right)~,
\en
where $a,b,c \in \mathbb{R}$ and $x,y,z \in \mathbb{O}$. The exceptional cone $\mathcal{C}$ in $J_3^{\mathbb{O}}$ is defined by those elements $J$  which are positive semidefinite i.e., they satisfy the conditions
\begin{align}
\begin{split}
a,b,c \geq 0 \\
(bc - N(x)) \,  , \, ( ca -N(y))  \, , \, ( ab -N(z))  \geq 0 \\
\mathcal{N}(J)  \geq 0
\end{split}
\end{align}
and are denoted  as $J\geq 0$, where $N$ is the norm over the octonions.
The group of all invertible linear transformations of $J_3^{\mathbb{O}}$ that preserve the cubic norm is the reduced structure (Lorentz)  group $\stro{J_3^{\mathbb{O}}}=\lor{J_3^{\mathbb{O}}}$   which is $E_{6(-26)}$. The  group $E_{6(-26)}$ also leaves the symmetric trilinear form $(J_1,J_2,J_3)$, defined as
\eqn
(J_1,J_2,J_3) &:= & \mathcal{N}(J_1+J_2+J_3) - \mathcal{N}(J_1+J_2) - \mathcal{N}(J_2+J_3) -\mathcal{N}(J_1+J_3) \nonumber \\ && + \mathcal{N}(J_1) + \mathcal{N}(J_2) +  \mathcal{N}(J_3) ~,
\enn
invariant.
The cubic norm of $J$ is then related to the trilinear form as
\eq
\mathcal{N}(J) = \det (J)= \frac{1}{6} (J,J,J)
\en
The rank of an element $J \in J_3^{\mathbb{O}}$ is preserved by the action of $E_{6(-26)}$.
Furthermore the exceptional cone $\mathcal{C} \in J_3^{\mathbb{O}}$ is stabilized by $E_{6(-26)}$ which acts transitively on the set of elements $J\geq 0$ with unit cubic norm $\mathcal{N}(J)=1$. The stabilizer of an element in $\mathcal{C}$ with unit norm is simply the maximal compact subgroup $F_4$ of $E_{6(-26)}$.

The elements $E$ belonging to the exceptional cone $\mathcal{C}$ with $\mathcal{N}(E)=1 $ are called {\it polarizations}.  Given a polarization $E$ one can define a linear form that maps the elements of $J_3^{\mathbb{O}}$ into $\mathbb{R}$:
\eqn
T_E :& J_3^{\mathbb{O}} & \Longrightarrow \mathbb{R} \\
  T_E(J) &=& \frac{1}{2} (E,E,J) \qquad ,  \forall \,  J \, \, \in J_3^{\mathbb{O}}
 \enn
as well as a quadratic form
\eqn
R_E :& J_3^{\mathbb{O}} & \Longrightarrow \mathbb{R} \\
  R_E(J) &=& \frac{1}{2} (E,J,J) \qquad ,  \forall \,  J \, \, \in J_3^{\mathbb{O}}
 \enn
which is associated  with the bilinear form
\eq
(A,B)_E = R_E (A+B) -R_E(A) -R_E(B) =(E,A,B)
\en
The bilinear form $(A,B)_E$ has signature $(1,26)$. One can also define another bilinear form
$ \langle A,B \rangle_E$ \cite{MR1411589}, where
\eq
\langle A,B \rangle_E = T_E(A) T_E(B) - (A,B)_E
\en
and has signature (27,0). The linear form $T_E$ as well as the bilinear forms are invariant under the subgroup of the reduced structure group $E_{6(-26)}$ of $J_3^{\mathbb{O}}$ that leaves the polarization $E$ invariant.

The Jordan product $A\circ_E B$ depends on the choice of polarization $E$ and is defined by the identity~\cite{MR1411589}
\eqn
(2(A\circ_E B),C)_E &=& (A,B,C) + T_E(A) \langle B,C\rangle_E +T_E(B)  \langle C,A \rangle_E +T_E(C) \langle A,B \rangle_E \nonumber \\ &&  - T_E(A) T_E(B) T_E(C) \\
&=&  (A,B,C) - T_E(A) (B,C)_E -T_E(B)  (C,A)_E -T_E(C) ( A,B )_E \nonumber \\ && + 2 T_E(A) T_E(B) T_E(C)
\enn
Under the Jordan product $\circ_E$ the polarization $E$ acts as the identity element
\eqn
A\circ_E E &=& E \circ_E A = A \\
T_E (A\circ_E B) &=& \langle A,B \rangle_E \\
T_E(A\circ_E(B\circ_E C))&=& T_E ((A\circ_E B)\circ_E C ) ~.
\enn

The Jordan algebra defined by the product $\circ_E$ with the identity element $E$ is referred to as the \textit{isotope} of the Jordan algebra with respect to the product $\circ_I$, with the identity element $I$ given by the $3\times 3$ unit matrix. For a Jordan algebra $J$ with the Jordan product $\circ_I$ one  defines a Jordan triple product
\eq
\{ABC\} \equiv (A\circ_I B )\circ_I C + (C\circ_I B) \circ_I A - (A\circ_I C ) \circ_I B =\{CBA\} ~.
\en
Given an element $B$ of the Jordan algebra $J$ with an inverse $B^{-1}$ such that $B \circ_I B^{-1} = I $, one can define an isotope of $J$ with the Jordan product $\circ_B$ 
\eq
A\circ_B C \equiv \{A B^{-1} C \} = C\circ_B A 
\en 
with the identity element $B$ since
\eq
A\circ_B B = B \circ_B A = \{ A B^{-1} B \} = A ~.
\en
Hence the Jordan algebra with the product  $\circ_E$ is simply the isotope of the Jordan algebra with the product $\circ_I$ and one has
\eq
A \circ_E B = \{A E^{-1} B \}~,
\en
where $E^{-1}$ is the inverse of the polarization  $E$ 
with respect to the Jordan product $\circ_I$.

If the identity matrix $ I =(1,1,1;0,0,0)$ is chosen as the polarization then $T_I$ coincides with the ordinary matrix trace and the Jordan product is simply given by  1/2 the anticommutator
\eq
A \circ_I B = \frac{1}{2} ( A B + B A)
\en

The E-adjoint $A_E^{\#} $ of an element $A\in J_3^{\mathbb{O}}$ is defined as
\eq
A_E^{\#} = A\circ_E A - T_E(A) A + R_E(A) E~,
\en
and satisfies the following identities:
\eqn
E_E^{\#} = E \\
A\circ_E A_E^{\#} = \mathcal{N} (A) E \\
(A_E^{\#})_E^{\#}= \mathcal{N}(A) A \\
2 T_E(A\circ_E B_E^{\#}) = (A,B,B) ~.
\enn

\section{Exceptional Jordan Algebra over the  Integral Octonions and Exceptional Modular Forms \label{modularform}}
The real octonions with integer coefficients $\mathbb{O}(\mathbb{Z})$ form a  ring. However, as was shown by Coxeter, the ring $\mathbb{O}(\mathbb{Z})$ is not maximal. There is a maximal order $\mathcal{R}$ which has $\mathbb{O}(\mathbb{Z})$ as a subring \cite{MR19111}. It is generated by $\mathbb{O}(\mathbb{Z})$ and four additional octonions with all half-integer coefficients which can be chosen to be
\eqn
\begin{split}
\frac{1}{2}( 1+j_1 + j_2 + j_3) \\
\frac{1}{2} ( 1 + j_1 + j_6 + j_5) \\
\frac{1}{2} ( 1 + j_1 + j_4 + j_7) \\
\frac{1}{2} ( j_1 + j_2 + j_4 + j_6 )~.
\end{split}
\enn
The order $\mathcal{R}$ contains the element \cite{MR1411589}:
\eq
\beta = \frac{1}{2} ( -1 + j_1 +j_2 +j_3+j_4+j_5+j_6+j_7)
\en
which satisfies
\eqn
\begin{split}
\Tr(\beta)= -1 \\
\mathcal N(\beta)= \beta \bar{\beta} = 2 \\
\beta^2 + \beta +2 =0~.
\end{split}
\enn
Trace and norm  take on integral values on $\mathcal{R}$ and hence its elements  were  called integral Cayley numbers (octonions) by Coxeter \cite{MR19111}.
Following \cite{MR1411589},
we define a $\mathbb{Z}$ lattice $JL$ inside $J_3^{\mathbb{O}}$ by considering $3 \times 3$ Hermitian matrices over integral octonions:
\eq
 J(a,b,c;x,y,z) =
\left(
  \begin{array}{ccc}
    a & z & \bar{y} \\
    \bar{z} & b & x \\
    y & \bar{x} & c \\
  \end{array}
\right)
\en
  where $a,b,c \in \mathbb{Z}$ and $x,y,z \in \mathcal{R}$. The cubic norm of elements in $JL$ take on integral values and its invariance group is the discrete arithmetic subgroup $E_{6(-26)}(\mathbb{Z})$ of $E_{6(-26)}$ \cite{MR1369418}.
A  remarkable fact that was proven in \cite{MR1369418} is that while $E_{6(-26)}$ acts transitively on positive polarizations with determinant 1, its arithmetic subgroup  $E_{6(-26)}(\mathbb{Z})$  does not act transitively on the polarizations $E>0$ in $JL$ with determinant 1. There are precisely two orbits under the action of $E_{6(-26)}(\mathbb{Z})$ represented by the identity matrix $I=J(1,1,1,0,0,0)$  and the ``indecomposable" polarization $E_{ind}$
  \eq \label{indecomp}
  E_{ind}= J(2,2,2;\beta,\beta,\beta) =\left(
  \begin{array}{ccc}
    2 & \beta & \bar{\beta} \\
    \bar{\beta} & 2 & \beta \\
    \beta & \bar{\beta} & 2 \\
  \end{array}
\right)~,
  \en
 with $\mathcal{N}(E_{ind})=1$. There exist three rank 1 elements $A \in JL $ with respect to the identity polarization $I$  with $T_I(A)=1$. On the other hand  there are no rank 1 elements $A\in JL $  that satisfy  $T_{E_{ind}}(A)=1 $ \cite{MR1369418}.

\subsection{Integral Jordan Roots \label{Jordanroots}}
\label{sec:integraljordanroots}
Given a polarization $E$ we may define Jordan roots associated with $E$ as those elements $S$ in the exceptional cone $\mathcal{C}$  which have rank one and  satisfy \cite{MR1411589}
\eqn
\begin{split}
T(S) = 2 \\
(S,S)=0 \\
\langle S,S \rangle = (T(S))^2 =4 \\
S\circ S = 2 S ~.
\end{split}
\enn
The automorphism group $F_4$ that leaves  a given polarization $E$ invariant acts transitively on the set of Jordan roots associated with $E$.
The subgroup of $F_4$ that leaves a given Jordan root $S$ invariant is $\text{Spin}(9)$ and the symmetric space $F_4/\text{Spin}(9)$ can be identified with the Moufang plane.\footnote{ We should note that Jordan roots can be identified with the points in the octonionic projective plane (Moufang plane) and correspond to pure states in the octonionic quantum mechanics defined over $J_3^{\mathbb{O}}$\cite{Gunaydin:1978jq}.  The idempotents $P$ of $J_3^\mathbb{O}$ corresponding to pure states in octonionic quantum mechanics are normalized such that $P\circ_I P= P$.  }
This subgroup contains the involution $\tau_S $ of $J_3^{\mathbb{O}}$ defined by
\eq
\tau_S(A) = A - 2 (A\circ S ) + \langle A , S \rangle S ~.
\en

A root triple is defined as  three mutually orthogonal Jordan roots $S_1,S_2$ and $S_3$ that satisfy:
\eqn
2E = S_1 + S_2 + S_3 \\
\langle S_i , S_j \rangle =0 \qquad i \neq j \\
S_i \circ S_j =0  \qquad  i\neq j ~,
\enn
where $i,j=1,2,3$. The automorphism group $F_4$ acts transitively on the root triples associated with $E$ and the subgroup of $F_4$ that leaves invariant a given root triple is the semi-direct product group $\text{Spin}(8) \rtimes S_3$ where $S_3$ is the group of permutations of the root triple.

In the rest of this paper we shall adopt the convention to denote the indecomposable polarization $E_{ind}$   given in equation \ref{indecomp} simply as $E$. The linear  form $T_{E}$ defined by the polarization $E$  maps the lattice $JL$ into $\mathbb{Z}$. In \cite{MR1411589}, the possible values $T_{E}(A)$ for $A\geq 0$ in $JL$ with rank($A$)=1 were studied. All such elements satisfy $T_{E}(A)\geq 2$. A formula for the number $\mathfrak{N}(n)$ of such rank 1 elements $A$ such that $T_{E}(A)=n $ with $n\in \mathbb{N}$ turns out to depend on the values of the divisor sigma function \cite{MR1411589}
\eq
\sigma_{11} (n) =\sum_{d|n} d^{11}
\en
where $d|n$ indicates the sum over positive divisors of $n$ including 1, and the Ramanujan $\tau$ function defined by the $q-$series of the 24$^{th}$ power of the Dedekind eta function
\eq
\Delta= \eta(\tau)^{24} = q \prod_{m\geq 1} ( 1 - q^m)^{24}= \sum_{n\geq 1} \tau(n) q^n, \quad q := e^{2\pi i \tau}~.
\en
Let $c(A)$ denote the largest integer such that for any rank 1 element $A\in JL$  one still has $A/c(A)\in JL$.  Then  the number of linearly independent rank one elements with $T_E(A)=n$ is given by \cite{MR1411589}
\eq
\mathfrak{N}(n) = \sum_{T_{E}(A)=n ; \, \, rank(A)=1}  \left( \sum_{d|c(A)}d^3\right) =\frac{ 3\cdot 7 \cdot 13 }{691} \left( \sigma_{11} (n) - \tau(n) \right)~. \label{number_rankone}
\en
If $n=p$ is a prime number, one has $c(A)=1$ and the formula for $\mathfrak{N}(p)$ simplifies
\eq
\mathfrak{N}(p) = \frac{ 3\cdot 7 \cdot 13 }{691} \left( p^{11}  - \tau(p) +1 \right)~.
\en
We refer to $\mathfrak{N}(n)$ as the  multiplicity  of a rank one element with trace form $T_E(A)=n$.  Correspondingly we define the quantum degeneracy of a rank one black hole whose charge matrix $A$ satisfies $T_E(A)=q$  to be given by $\mathfrak{N}(q)$ with the adjective "quantum" deriving from the fact that the discrete U-duality group is expected to be a symmetry of the quantum completion of the  octonionic magical supergravity. 
The rank 1 elements with $T_{E}(A)=2 $ correspond to integral Jordan roots and  their number is $\mathfrak{N}(2)=819$ \cite{MR1411589} . Hence a BPS black hole whose charge matrix $A$ satisfies $T_E(A)=2$  can be in any one these 819 charge states. 

The proof of the formulas above giving the multiplicity $\mathfrak{N}(n)$ of rank 1 elements uses the theory of modular forms of weight 12 on the upper half-plane and   on the exceptional tube domain \cite{MR1216126}.
The space of modular forms of weight 12 for $SL(2,\mathbb{Z})$ is two dimensional spanned by $\Delta$ and the Eisenstein series $E_{12}$. They have the Fourier expansions
\begin{align}
\Delta(\tau) &= q\prod_{n \geq 1}(1-q^n)^{24}=q -24 q^2 +252 q^3 + \cdots
\\
\label{eq:eisensteinexample}
E_{12}(\tau) &= \zeta(-11)/2 + \sum_{m\geq 1} \sigma_{11}(m)\, q^m = \frac{691}{65520} + q +   2049 q^2 + \cdots ~.
\end{align}
The unique holomorphic modular $f(q)$  form of weight 12 for $SL(2,\mathbb{Z})$ whose Fourier series begins as $f(q) = 1 + 0 q + \cdots$ is given by
\eqn
f(q) &=& \frac{65520}{691} (E_{12} - \Delta ) = 1 + \frac{ 2^4 \cdot 3^2 \cdot 5 \cdot 7 \cdot 13 }{691}\sum_{n\geq 1} ( \sigma_{11} (n) - \tau (n) ) \nonumber \\ && = 1 + 0 q + 196560 q^2 + O(q)^3 ~. \label{qexpansion}
\enn
The modular form $f(q)$ is the theta function of the Leech lattice $ \Lambda \subset \mathbb{R}^{24} $ which is an even unimodular lattice with minimal norm $> 2$ \cite{MR1369418, conway2013sphere}.\footnote{See \autoref{app:thetas} and \autoref{tab:niemeiertheta} for relevant details on theta functions of Niemeier lattices.}

On the other hand, the ``upper half-plane" of the exceptional Jordan algebra $J_3^{\mathbb{O}}$ is spanned by elements of the form $ Z= ( X + i Y) $ where $ X $ is an arbitrary element of $J_3^{\mathbb{O}}$  and $Y>0$. This upper half-plane is in fact the \textit{exceptional tube domain} $\mathcal{D}$ of complex dimension 27. The conformal group $E_{7(-25)} $ of $J_3^{\mathbb{O}}$  acts holomorphically on the exceptional tube domain $\mathcal{D}$ and it was proposed as a spectrum generating symmetry group of extremal black holes of the octonionic magical supergravity in $5d$ \cite{Ferrara:1997uz,Gunaydin:2000xr,Gunaydin:2003qm,Gunaydin:2004ku,Gunaydin:2005gd}.

 Since the conformal group $\conf{J}$ includes translations $T_J$ by the elements of $J$ the Fourier coefficients of the modular forms of the arithmetic subgroup  $E_{7(-25)}(\mathbb{Z})$ of $E_{7(-25)}$ are expected to describe the degeneracies of charge states of  quantum extremal black holes of the octonionic magical supergravity theory whose bare charges are labelled by the elements of the exceptional Jordan algebra with integral coefficients. The above results show that this is the case for rank one charge states which act as building blocks of higher rank charge states as will be explained later in section \ref{sec:diffrankbhs}. In the next subsection we shall review the work of N. Elkies and B. Gross that establish the connections between the exceptional modular form of Kim\cite{MR1216126} on the exceptional tube domain, rank one elements of the exceptional Jordan algebra over the integral octonions and  the Leech lattice \cite{MR1411589}.
\subsection{Exceptional Modular Forms and Integral Jordan Roots}
\label{sec:exceptionalmf}
Let $F(Z)$ be  an holomorphic  function that maps  the exceptional tube domain $\mathcal{D}$ into complex numbers $\mathbb{C}$.
$F(Z)$ is a modular form of level 1 and weight $k$  of $E_{7(-25)}(\mathbb{Z}) $ if it satisfies the following conditions \cite{MR1216126}: 
\begin{enumerate}
    \item Invariance under translations by elements of $JL$
\eq F(Z+B) =F(Z) \ \ , \forall B \ \in JL
\en
\item Invariance under the action of $E_{6(-26)}(\mathbb{Z})$
\eq
F(g Z ) = F(Z) \qquad , \forall g \,\, \in E_{6(-26)}(\mathbb{Z})
\en
\item Under inversions, it satisfies the following identity
\eq
F(-Z^{-1})  = (\mathcal{N}(Z))^k F(Z)~,
\en
where the inversions are defined with respect to the identity polarization
\eq
Z^{-1} = \frac{Z_I^{\#}}{(\mathcal{N}(Z))}~.
\en
\end{enumerate} $F(Z)$ has a Fourier expansion of the form
\begin{align}
\label{eq:fourierexcep}
    F(Z) = \sum_{T \in J_3^\mathbb{O} \, , \, T\geq 0} a(T) e^{2\pi i \Tr(T\circ Z)}~. 
\end{align}

Given a holomorphic modular form $F(Z)$ of weight $k$ on the tube domain $\mathcal{D}$, then the function $f(\tau) = F(\tau E^{\#})$, where $\tau = x + iy$ is a complex number taking values in the upper half-plane and $E^\#$ is the adjoint of the indecomposable polarization $E$, is a holomorphic modular form of weight $3k$ of $SL(2,\mathbb{Z})\subset E_{7(-25)}(\mathbb{Z}) $. 
The  singular modular form $F(Z)=E_{4,0}(Z)$ of  weight 4  studied by Kim has the Fourier expansion
\eq
E_{4}(Z) = 1 +  240 \sum_{A\geq 0 \, \mathrm{in}\, JL \,; \,\mathrm{rank}(A)=1} \left( \sum_{d|c(A)} d^3 \right) e^{2\pi i  \Tr(A\circ_I Z)} ~.
\en
where the Jordan product $A\circ_I Z$ is with respect to the identity polarization $I$  of $J_3^{\mathbb{O}}$. Choosing $Z=E^{\#}$ and using the identity
\begin{align*} \Tr(A\circ_I E^{\#}) =\frac{1}{2} (A, E, E) = T(A)~, \end{align*}
one finds
\eq f(\tau) = F(\tau E^{\#}) = 1 + 240 \sum_{n\geq 1} \left( \sum_{T(A)=n } \left(\sum_{d|c(A)}d^3 \right) \right) q^n ~, \en
where $q =e^{2\pi i \tau}$ and $\mathrm{rank}(A)=1 $ with $A\geq 0$. Since $T(A) > 1$ for rank one elements $ A \in JL$ in the polarization $E$, the coefficient of $q$ is zero in the above series which is simply the $q$-expansion of the holomorphic form of weight 12 of $SL(2,\mathbb{Z})$ given in \eqref{qexpansion} and is also the theta function of the Leech lattice.

As was shown in  \cite{MR1411589}, the bilinear product $(A,B)$ is even and has discriminant 2 on $JL$. On the other hand, the pairing $\langle A,B \rangle $ is positive  definite and unimodular on $JL$. The polarization $E$ satisfies the identities  \cite{MR1411589}
\begin{align}
\begin{split}
\langle E,E \rangle &= 3 \\
\langle E,A \rangle &= \langle A,A \rangle \mod \ 2 ~,
\end{split}
\end{align}
 which implies that
 \eq
 2 ( A \circ_{E} B ) \in \, JL \qquad \forall \, \, A,B \in JL ~.
 \en
Involutions $\tau_S$ with respect  to Jordan roots $S$ map $JL$ into itself. The group $\aut{JL,\mathcal{N}, E}$  that leaves $JL, \mathcal{N}$ and the polarization $E$ invariant is a discrete subgroup of the compact group $F_4$ and, hence, is finite. It leaves invariant the submodule $\mathbb{Z} \, E$  and  acts faithfully on its orthogonal complement $JL_0$ of rank 26.

The subgroup $\Gamma$ of $\aut{JL,\mathcal{N}, E}$  generated by involutions with respect to 819 Jordan roots has order $211341312 $
and was shown to be isomorphic to the twisted Chevalley group $^3D_4(2)$ \cite{MR1411589}\footnote{ We use the conventions of the Atlas \cite{MR827219} in labelling finite groups.}.
The subgroup of $^3D_4(2)$ that leaves a given Jordan root $S$ invariant is  a  maximal parabolic subgroup isomorphic to  $2_+^{1+8} \cdot \, L_2(8)$. There are 2457 integral root triples and the group $\Gamma$ acts transitively on them. The subgroup of $\Gamma$ that leaves a given root triple invariant is the maximal parabolic subgroup  $2^{2+3+6} \cdot ( 7 \times L_2 (2))$. 
The group $^3D_4(2)$ acts transitively also on the set of rank one elements $A$ with $T(A)=3$ with the stabilizer being isomorphic to the maximal subgroup $L_2(2)\times L_2(8)$  \cite{MR1411589}.  
  
\subsection{ Cubic Rings and Binary Cubic Forms}
In \cite{MR2521487} it was shown that  the isomorphism classes of cubic rings $A$ over a local ring $R$ correspond to  the orbits of the action of $GL(2,R)$  on binary cubic polynomials over $R$. Given a binary cubic polynomial over $R$ of the form 
\eq
p(x,y)= a x^3 + b x^2 y + c x y^2 + d y^3 \label{binarycubic}
\en
with coefficients $a,b,c.d$ in $R$ and discriminant
\eq
\Delta (p) = b^2 c^2 + 18 a b c d  - 4 a c^3 - 4 d b^3 -27 a^2 d^2 \label{discriminant} 
\en
the twisted  action of $ g =\left(
                                 \begin{array}{cc}
                                   \alpha & \beta \\
                                   \gamma & \delta \\
                                 \end{array}
                               \right)
\in GL(2,R) $  on $(x,y)$ is defined as
\eq
g : \left(
            \begin{array}{c}
              x \\
              y\\
            \end{array}
          \right) \Longrightarrow \left(
                                    \begin{array}{c}
                                      x' \\
                                      y' \\
                                    \end{array}
                                  \right) \, = \, \frac{1}{(\alpha \delta - \beta \gamma)}
           \left(
                                    \begin{array}{cc}
                                      \alpha & \beta \\
                                      \gamma & \delta \\
                                    \end{array}
                                  \right) \, \, \left(
                                                  \begin{array}{c}
                                                    x \\
                                                    y \\
                                                  \end{array}
                                                \right) ~.
                                                \en

Under $GL(2,R)$, the discriminant $\Delta$ changes as follows  :
\eq \Delta(g \cdot p(x.y)) = (\det \,  g)^2 \Delta(p(x,y))~. \en

We should note that the discriminant $\Delta$ corresponds to the quartic invariant of an extremal black hole solution in $4d$ supergravity obtained by dimensional reduction of the pure $N=2$ supergravity in five dimensions whose electric and magnetic charges are related to $(a,b,c,d)$ if the ring $R$ is chosen to be the ring of integers $\mathbb{Z} $ \cite{Gunaydin:2007qq}. It is invariant under the $4d$ U-duality group $SL(2,R)$ and its relation to binary cubic forms and extremal black holes  were studied in \cite{Gunaydin:2019xxl}.

As was shown in \cite{MR2521487}, given a binary cubic form $p(x,y)$ and corresponding four dimensional $R-$module $M$, one can always define a cubic ring over $R$ with basis $(1,I,J)$ with multiplication rules
\eqn
 I \, J &=& -a d 1 \\
 I^2 &= &- a c 1 + b I - a J \\
 J^2 &=& - b d 1 + d I  - c J~.
 \enn
Such a basis  is referred to as a \textit{good basis}. The fact that the multiplication rules involve the constants of a binary cubic polynomial $p(x,y)$ over $M$ establishes a map from cubic rings to  binary cubic forms over $R$. The most general transformation from a good basis $(1,I,J)$ to another good basis $(1,I', J')$  has the form
\eq
\left(
  \begin{array}{c}
    1 \\
    I' \\
    J' \\
  \end{array}
\right) = \left(
            \begin{array}{ccc}
              1 & 0 & 0 \\
              u & \alpha & \beta \\
              v & \gamma & \delta \\
            \end{array}
          \right) \, \left(
                       \begin{array}{c}
                         1 \\
                         I \\
                         J \\
                       \end{array}
                     \right)~.
\en

\section{Springer Decomposition of Jordan Algebras of Degree Three \label{Springer}}
We now give a brief review of the Springer decomposition of  Jordan algebras $J_3^{\mathbb{A}}$ of degree three over a field $F$ with $\mathbb{A}$ representing a composition algebra following \cite{MR0138661,MR1632779}. We  will restrict ourselves to the case when $\mathbb{A}$ is the division algebra $\mathbb{O}$ of octonions. Let $A=F\times F \times F$ denote the subalgebra of diagonal matrices  
\eq 
\Lambda = \begin{pmatrix}
  \lambda_1 & 0 & 0 \\
0 & \lambda_2 & 0\\
  0 & 0 & \lambda_3 \\
\end{pmatrix}~,
\en
where $\lambda_i \in F $. We shall denote the matrix  $\Lambda$  simply as $\Lambda=(\lambda_1,\lambda_2,\lambda_3)\in A $. 
The orthogonal complement $L_c$  of $A$ with respect to the trace form  are given by the matrices
\eq 
\Omega = \begin{pmatrix}
  0 & \bar{o}_3 & o_2 \\
  o_3 & 0 & \bar{o}_1\\
  \bar{o}_2 & o_1 & 0 \\
\end{pmatrix}~,
\en
where $o_i \in \mathbb{O}$. We shall denote the matrix $\Omega$ as $(o_1,o_2,o_3)\in L_c $.  Springer defines the action `$\cdot$'  of the three dimensional subalgebra $A$ on the orthogonal complement $L_c$ as
\eq
\Lambda \cdot \Omega  \equiv - \Lambda \times \Omega = (\lambda_1 o_1, \lambda_2 o_2 , \lambda_3 o_3 ) \in L_c~,
\en 
where $\times$ is the Freudenthal product. 
Hence $L_c$ is an $A$ module under this action.
Representing  a general element of $J_3^{\mathbb{O}}$  as $ ( \Lambda, \Omega )$, we have
\eq 
(\Lambda, \Omega)^{\#} = ( \Lambda^{\#} - Q(\Omega) , \beta(\Omega) - \Lambda \cdot \Omega ) \in ( A \oplus L_c )~,
\en
where $ ( \Lambda^{\#} - Q(\Omega) )\in A $ and $( \beta(\Omega) - \Lambda \cdot \Omega ) \in L_c$. For our example we have
\eq 
Q( \Omega) = ( o_1 \bar{o}_1 , o_2 \bar{o}_2 , o_3 \bar{o}_3 )
\en 
and 
\eq
\beta(\Omega) = ( \bar{o}_2 \bar{o}_3 , \bar{o}_3 \bar{o}_1, \bar{o}_1 \bar{o}_2 )~.
\en
Therefore, the entire Jordan algebra $( \Delta , \Omega)$
can be viewed as a quadratic space over $A$ under the above action.

  \section{Embeddings of Cubic Rings in the Exceptional Jordan Algebra and Niemeier Lattices}
  \label{sec:embeddings}
  In their subsequent work \cite{MR1845183}, Elkies and Gross consider the embeddings of cubic rings into the $\mathbb{Z}$ lattice $JL$ of rank 27 defined by the $3\times 3$ Hermitian symmetric matrices over the Coxeter's ring of integral octonions $\mathcal{R}$.  Their work uses some of the results of earlier work by Gross and Gan \cite{MR1793601} on commutative subrings of certain non-associative rings which include the exceptional Jordan algebra which we summarize in \autoref{commutative}. The cubic rings $A$ considered are commutative rings which are  isomorphic to $\mathbb{Z}^3$ as additive groups and  their cubic norms are integral i.e.,\footnote{We use $\mathbf{N}$ to denote the norm in a cubic ring, as in \eqref{eq:cubicnormN}. However we also use this to denote norms of elements over a given field $F$. The usage should be clear from context.}
  \eq
  \label{eq:cubicnormN}
\mathbf{N} : A \rightarrow \mathbb{Z}~.
\en
An embedding of $A$ in $JL$ is a mapping $f$ such that
\eqn
\mathcal{N}(f(a)) = \mathbf{N} (a) \, \, \forall a \in A \\
f(1)= E~,
\enn
and such that $JL/f(A)$ is torsion-free. These conditions imply that
\eq
f(a\cdot b)= f(a)\circ_E  f(b)~,
\en
where $\circ_E$ denotes Jordan product with respect to the polarization $E$ and
\eq
\mathcal{N}(x E - f(a) ) =\mathbf{N}(x-a)~.
\en
Furthermore one has
\eqn
T(f(a))=\Tr(a) \\
\langle f(a),f(b)\rangle = T (f(a) \circ_E f(b) )= \Tr(a\cdot b)~,
\enn
where $\Tr$ is the trace form over the cubic ring $A$ and $\cdot$ refers to the product in $A$.

The cubic ring $A$ regarded as an integral lattice admits a dual lattice $ A^\vee \subset A \otimes \mathbb{Q}$ with respect to the bilinear form $\langle a, b\rangle = \Tr(a\cdot b) $ where $\mathbb{Q}$ denotes the rationals.
Discriminant $D$ of $A$ considered as a lattice is given by the order of the module $A^\vee /A$. Orthogonal complement of the image $f(A)$  of $A$ inside the 27 dimensional lattice $JL$ is a rank 24 even lattice $L_c$ such that
\eq
JL = f(A) \oplus L_c .
\en
Since the cubic ring has a unit which maps into the polarization vector $E$ the lattice $L_c$ is also a sublattice of $JL_0$ generated by traceless elements of $JL$. The lattice $JL_0$ is generated by vectors $A_0$ in $JL$ which satisfy the condition
\eq \langle A_0, E \rangle = T_E(A_0)=0~.
\en
It is an even lattice with a positive definite quadratic form $ q(v) \equiv \frac{1}{2} \langle v,v\rangle $ which maps the elements $v$ of $JL_0$ into $\mathbb{Z}$.
Since $\mathrm{det} JL_0=3$ it has index 3 inside its dual lattice $JL_0^{\vee}$.

The rank 24 sublattice $L_c=f(A)^{\bot}$ can be given an $A$-module structure as was shown by Springer \cite{MR0138661,MR1632779} and summarized in previous section. Elkies and Gross \cite{MR1845183} implement this decomposition using the adjoint map in the indecomposable polarization $E$\footnote{In the previous section we reviewed the Springer decomposition in the identity polarization for simplicity.}
\eqn B \rightarrow B^\# \\
B \circ_E B^\# = \mathcal{N}(B)  E~.
\enn
The quadratic form $q(v)$ on $JL_0$ is then  given by
\eq
q(v) = - \langle v^\# , E \rangle \label{quadratic}
\en for all $v\in JL_0$.  Furthermore they  define the $A$-module structure on the lattice $L_c$ with a positive definite quadratic map of $A$-modules
\eq
q_A : L_c \rightarrow A^\vee
\en
such that $ \Tr(q_A) = q $ on $L_c$ as defined in \eqref{quadratic}. They also define a quadratic map on the cubic ring $A$: $a \rightarrow a^\#$ such that $a \cdot a^\# = \mathbf{N}(a) $ and $f(a^\#) = f(a)^\# $ for a given embedding $f$.  In the polarization $E$, the Freudenthal product of two elements of the exceptional Jordan algebra  can be written in terms of the Jordan product as follows
\eq
B \times C =  (B+C)^\# - B^\# - C^\# = 2 (B\circ_E C ) - T(B) C - T(C) B + ( B,C ) E~.
\en
As reviewed in  section \ref{Springer} $L_c$ becomes an $A$ module under the action
\eq
a\cdot v = - f(a) \times v \, \quad , \forall a\in A~.
\en
That $a\cdot v $ lies in $L_c=f(A)^\bot $ follows from  the  identities:
\eq
\langle f(b), a \cdot v \rangle = - \langle f(b), f(a) \times v \rangle =-\langle f(b\times a), v\rangle =0~.
\en
One can also define $t_A : J \rightarrow A^\vee $  so  that $t_A(B)$ lies in $A^\vee$ in the decomposition $JL\subset A^\vee + L_c^\vee$\cite{MR1632779}.
We then have 
\eq
\Tr(t_A (B)) = \langle 1, t_A(B) \rangle = \langle E , B \rangle = T(B) \in \mathbb{Z} ~.
\en
The adjoint of an element $v\in L_c $ has a component in $A^\vee $ which can be used to define a quadratic map $q_A$:
\eq
v^\# = - q_A (v) + \beta (v)~,
\en
where $q_A(v) $ takes values in dual $A^\vee$ of $A$ and $\beta(v)$ takes values in the dual $L_c^\vee$ of $L_c$. One finds that
\eq \Tr(q_A(v)) = - T(v^\#) = -\langle v^\# , E\rangle = q(v)~. \en
Thus $L_c^\vee$ is also an $A$ module inside $L_c \otimes \mathbb{Q} $. Furthermore
\eq \beta(a\cdot v) = a^\# \cdot \beta(v) \in L_c^\vee \en and
\eq \mathcal{N} (v) = \langle v , \beta(v) \rangle _A ~, \label{norm}  \en where
\eq \langle v, w \rangle_A \equiv q_A(v+w) - q_A(v) - q_A (w)~. \en
Note that even though $\beta(v)$ lies in $L_c^\vee$,  the bilinear product \eqref{norm} takes values over the integers $\mathbb{Z}$ and
\eq \mathcal{N}( a\cdot v ) = \mathbf{N}(a) \cdot \mathcal{N}(v)~. \en
Given a totally real cubic ring $A$, an element of $A\otimes \mathbb{R}$ is said to be  \textit{totally positive} if each of its three $\mathbb{R}^3$ coordinates is non-negative and denoted as $(A\otimes\mathbb{R})_+ $ representing the self-dual cone of such elements \cite{MR1845183}. An embedding $ f : A \rightarrow JL $ maps totally positive $\alpha$ in $A$ to positive-semidefinite matrices $B=f(\alpha)$ in $JL$. Conversely if $B $ is positive-semidefinite then $\alpha = t_A (B) $ belongs to $A_+^\vee$. 

\subsection{ Hilbert Modular Forms and Cubic Rings}
Denoting the complex  upper half plane as $\mathcal{H}$, Elkies and Gross define a holomorphic function from $\mathcal{H}^3$ into $\mathbb{C}$ which has the convergent Fourier series
\eq
F(\underline{\tau}) = f(\tau_1,\tau_2,\tau_3) = \sum_{\alpha\in A_+^\vee} c(\alpha) e^{ 2\pi \imath (\alpha_1 \tau_1 + \alpha_2 \tau_2 +\alpha_3 \tau_3) }  \label{Hilbertform}
\en
where $c(0)=1$ and
\eq
c(\alpha) = 240 \sum_{\left[\begin{array}{c}S\in JL \\\text{rank}(S)=1 \\t_A(S)=\alpha \end{array}\right]} \left( \sum_{d|c(S)} d^3 \right)~,
\en
and show that it is a Hilbert modular form of weight $(4,4,4)$ for $SL(2,A)$ which is a discrete subgroup of $SL(2,\mathbb{R})^3$ i.e.,
\eq
F\left(\frac{a \underline{\tau} +b}{c \underline{\tau} + d } \right) = \mathbf{N} (c \underline{\tau} + d )^4 F(\underline{\tau})
\en
for all $\left(\begin{array}{cc}a& b \\c & d\end{array}\right)$ in $SL(2,A)$\footnote{See \autoref{HMF} for a review on Hilbert modular forms. }.
$c(S)$ is the largest positive integer that divides $S$ such that $S/c(S)\in JL$. 
When  $\alpha$ is primitive in $A^\vee$ then $c(S)=1$ and $c(a)$ simplifies to
\eq
c(a) = 240 \, \# \{ S : \text{rank} \, S=1, t_A(S)=2\}.
\en
Note that the sum is over rank one elements $S$ which  satisfy $S^\#=0$.  In general, a rank one element can be decomposed as
\eq S= \alpha + v, \en where $\alpha =t_A(S)$ and $ v\in L_c$.
Since
\eq S^\# = ( \alpha^\# - q_A(v)) + ( \beta(v) - \alpha \cdot v ) =0, \en
this implies $q_A(v)= \alpha^\# $ and $ \beta (v) = \alpha \cdot v $.

These results are proven using the singular form $F(Z)$ of Kim on the exceptional tube domain as seen in \autoref{sec:exceptionalmf} and \autoref{app:exceptionalmf}.\footnote{ Note that the work of Kim uses the identity matrix as the polarization. However any polarization $E$ determines an isomorphic discrete subgroup of the automorphisms of the tube domain \cite{MR1845183}.} Recall that $F(Z)$ has the Fourier expansion
\eq
F(Z) = 1 + 240 \sum_{S\geq 0, \text{rank}(S)=1} \left( \sum_{d|c(S)} d^3 \right) e^{2\pi i \langle S, Z \rangle}. \en

The function $F(\underline{\tau}) $ given in \eqref{Hilbertform} corresponds to the restriction of Kim's form $F(Z)$ to the sub-tube-domain
\eq
\mathcal{H}^3 = (A \otimes\mathbb{R}) + \imath (A \otimes\mathbb{R})_+ \en
which embeds into the exceptional tube domain $\mathcal{D}$ under the action of the embedding function of the cubic ring $A$ into the exceptional Jordan algebra over the integral octonions $\mathcal{R}$ . The function $F(\ut)$ satisfies
\eqn
\begin{split}
F(\ut + b ) = F(\ut) \, \, \, \forall \,\, b \, \in \,  A \\
F(\alpha^2 \cdot \ut) = F(\ut) \, \, \, \forall \,\, \alpha \, \in \, A^\vee \\
F\left(-\frac{1}{\ut}\right) = ( \mathbf{N}(\ut) )^4 F(\ut)~.
\end{split}
\enn
The corresponding matrices of $SL(2,A)$ are
\eq \left(\begin{array}{cc}1 & b\\0 & 1\end{array}\right) \, \, , \, \left(\begin{array}{cc}\alpha  & 0 \\0 & \alpha^{-1} \end{array}\right) \, \, , \, \left(\begin{array}{cc}0 & -1 \\1 & 0\end{array}\right)~, \en
where $b\in A$ and $\alpha \in A^\vee$.
 $F(\ut)$ has weight $(4,4,4)$ with respect to the discrete subgroup $SL(2,A)$ with Fourier expansion
\eqn
F(Z) &= &1 + 240 \sum_{S\geq 0, \text{rank}(S)=1} \left( \sum_{d|c(S)} d^3 \right) e^{2\pi\imath \Tr(t_A(S)\cdot \ut)  } \\
~ & =&  1 + 240 \sum_{\alpha \in A^\vee_+ , \alpha \neq 0} \left( \sum_{\text{rank}(S)=1, t_A(S)=\alpha} \left( \sum_{d|c(S)} d^3 \right) \right) e^{2\pi\imath \Tr (\alpha \cdot \ut )}. \label{modform}
 \enn

\subsection{ Cubic Rings with Discriminant $D=p^2$ and Niemeier Lattices}
\label{sec:gencubicring}

Recall that $L_c= f(A)^{\perp}$. We also have
\eq A \oplus L_c \subset JL \subset A^\vee \oplus L_c^\vee.
\en
Using the fact that the projections onto first and second components above 
\begin{align*} 
\alpha : J/(A \oplus L_c) \simeq A^\vee /A \quad , \quad 
\beta: J/(A \oplus L_c) \simeq L_c^\vee /L_c 
\end{align*}
define isomorphisms as finite Abelian groups, 
Elkies and Gross in \cite{MR1845183}
prove that 
\eq
\langle \gamma a , \gamma b \rangle  \equiv - \langle a, b\rangle  \quad ( \text{mod} \,  \mathbb{Z} ) \quad , \quad \forall \,  a,b \in A^\vee ~,
\en
where
\begin{align*} \gamma \equiv \beta \circ \alpha^{-1} \, : \, A^\vee/A \simeq L_c^\vee/L_c~. \end{align*}

Using these results, Elkies and Gross analyze, in particular,  the cases when the discriminant of the cubic ring $A$ is  $D=p^2$ with $p$ prime and the cubic ring $A$ is maximal. This analysis requires that $p \equiv 1 \mod 3$, $p\geq 7$, and $A$ consists of  integers in the cubic subfield  of the $p^{th}$ cyclotomic field\footnote{The $p^{th}$  cyclotomic field is generated by extending the rational numbers $\mathbb{Q}$ by $p$-th root of unity $\zeta$ such that $\zeta^p=1$. Galois group of a cyclotomic field is the multiplicative group $\mathbb{Z}_p$ of integers mod $p$. } \cite{MR1845183}. The quadratic space $ A^\vee/A$ is split over the integers mod $p$ and  has two isotropic lines $\bar{N}$ and $\bar{N}'$ which define unimodular lattices $N$ and $N'$ as sublattices in $A^\vee$. 
The two unimodular lattices $N$ and $N'$ both have rank 3 and  are isomorphic to $\mathbb{Z}^3$.
For the embeddings of these rings  $ f: \, \, A  \rightarrow JL $ they prove that there exist two even, integral unimodular lattices  $M$ and $M'$ of rank  24 which lie between $L_c $ and $L_c^\vee$ such that $M/L_c$ and $M'/L_c$ are the isotropic lines corresponding to  $N/A$ and $N'/A$, respectively. These are precisely the Niemeier lattices of rank 24.  

\subsubsection{Case $D=49$}

Among the examples of cubic ring embeddings studied by Elkies and Gross  with $D=p^2, \ p = 7$ is  the Dedekind domain $\mathbb{Z}[\cos(2\pi/7)] = \mathbb{Z}[\alpha]/(\alpha^3 + \alpha^2 - 2 \alpha -1) $. This corresponds to a particular binary cubic form given in \eqref{binarycubic} with 
coefficients 
\eq
a=b=1 \quad , \quad c=-2 \quad, \quad d=-1 
\en
Hence the discriminant given by \eqref{discriminant} is 49. In this case there are $2^9 3^4 13$ possible embeddings of the ring $f: A \rightarrow JL $ which are conjugate under the finite automorphism group
\eq \aut{JL,E} = \ ^3D_4 (2).3 \en
of order $2^12 3^5 7^2 13 $ \cite{MR1697445}. For a given embedding, the stability group is the finite group $7^2:2A_4$ of order $2^3 3 \cdot 7^2 $ . The normalizer of the stabilizer is the maximal subgroup $7^2 : 2 A_4\times 3$.  Their quotient is the cyclic group $\aut{A}=C_3$ of order 3.

One particular realization of the embedding with $D=49$ studied explicitly in \cite{MR1845183} is given by 
\eq
f(\alpha) =\left(\begin{array}{ccc}-1 & 1 & -\bar{\beta} \\1 & -1 & - \beta \\-\beta & - \bar{\beta}& -1\end{array}\right)
\en
with the identity $1$ of the cubic ring mapping into the polarization $E$

\eq
f(1) =E = \left(\begin{array}{ccc}2 & \beta & \bar{\beta} \\ \bar{\beta} & 2 &  \beta \\ \beta &  \bar{\beta}& 2\end{array}\right)~.
\en
The full image $f(A)$ of the cubic ring is a $\mathbb{Z}$-module parametrized by three integers $(f,p,r)$
\eq
f(A) =\left(\begin{array}{ccc}(f+p+r)  & (p-r + p \beta )  & (f -r -r \beta) \\ (p-r+p \bar{\beta}) & (f+p+r) & (f + r \beta ) \\ ( f - r - r \bar{\beta} )& ( f + r \bar{\beta} ) & (f+p+r) \end{array}\right)~. \label{gen49}
\en
Defining a mapping $t_A$ from $JL$ into the dual $A^\vee$ by requiring that $t_A(B)$ is the first component in the decomposition $JL\subset A^\vee + L_c^\vee$, one finds 
\eq
\Tr (t_A(B)) = \langle 1,t_A(B)\rangle = \langle E,B\rangle =T(B) \in \mathbb{Z}~.
\en
This implies that the embedding $f(A^\vee)$ of $A^\vee$ leads to matrices of the form \eqref{gen49} with the $(f,p,r)$ taking values in the rational numbers of the form $\mathbb{Z}/7$ such that $T(f(A))=(4f + 2p + r)$ takes integer values.
The trace form of the square of a  general element $f(A)$ is given by
\eq T(f(A)^2) = 10 f^2 + 10 f p + 6 p^2 -2 f r - 8 pr + 5 r^2~, \en
which take on values  $0,3,5,6, \cdots$.  The cubic norm of $f(A)$ is 
\eq 
\mathcal{N}(f(A)) = f^3 +p^3 +r^3 -2 p^2 r - p r^2 -f^2(2 p+ r) -f (p^2 +pr + 2 r^2) 
\en 
The number of roots $\lambda$ of the Niemeier lattice $M$ is equal to twice the number of Jordan roots $S \in JL$ that satisfy
\eq
t_A(S)= (1-n) \, \in \, A^\vee_+ ~,
\en
where $n$ is any short vector in $N$ with $\Tr(n)=1$ \cite{MR1845183}.
Conversely given a short vector $n\in N$ with $\Tr(n)=1$ and the corresponding totally positive element $ a=(1-n)$ in $A^\vee_+ $, one has
\eq
\# \{\mathrm{roots} \, \,  \lambda \in M \}=6 \cdot \# \{ \mathrm{Jordan \, roots }\,\,S \in JL\, ,  \,\,\,\mathrm{with} \,\,\, t_A(S)=a \}~.
\en
 Similar results hold for the lattices $M'$ and $N'$ with $\lambda, \, n \, , \, a$ replaced by  $\lambda', \, n' \, , \, a'$.
 Furthermore, the above numbers can be calculated using the Hilbert modular form $F(\ut ) $ of weight $(4,4,4)$ under $SL(2,A)$ \cite{MR1845183}.
The space of such forms is two dimensional and can be expanded in terms of the forms $E_4$ and $(E_2)^2$ where $E_k$ here is the weight-$(k,k,k)$ Hilbert modular form whose Fourier expansions can be written in the form
\eq
E_k = \frac{1}{2^3} \zeta_A(1-k) + \sum_{a>0\, , a\in A^\vee} \left( \sum_{c|(a)p^2} ( \mathbf{ N}c)^{k-1} \right) q^a ~,
\en
where $\zeta_A$ is the zeta function valued over the ideal class $A$.\footnote{For an ideal class $A$, its zeta function $\displaystyle \zeta_A(k) = \sum_{\substack{c \in A,\\ c \in \mathfrak O(A)}} (\textbf{N}c)^{-k}, $ where $\mathfrak O(A)$ is the ring of integers over $A$ and $\textbf{N}$ is the $\mathbb Q$-norm.}
Substituting the values of the zeta function for $k=2$ and $  k=4$ \cite{van2012hilbert, MR1845183}, one has
\eq
\begin{split}
E_2= -\frac{1}{2^3\cdot  3 \cdot 7} + \sum_{a>0\, , a\in A^\vee} \left( \sum_{c|(a)p^2} ( \mathbf{ N}c)  \right) q^a, \ \text{and} \\
E_4= -\frac{79}{2^4\cdot  3 \cdot 5 \cdot 7} + \sum_{a>0\, , a\in A^\vee} \left( \sum_{c|(a)p^2} ( \mathbf{ N}c)^3  \right) q^a ~.
\end{split}
\en
Under the action of $\aut{A}$ there is a unique orbit of elements $a>0$ in $A^\vee$ with $\Tr(a)=1$. They are represented by the squares $n^2$ of short vectors in $N$. Since the relevant space of modular forms is spanned by $(E_2)^2$ and $E_4$ one finds that there exists a unique modular form $F(\ut)$ with constant Fourier coefficient $c(0)=1$ and $c(n^2)=0$ which is 
\eq
\label{eq:uniqueHMF}
F(\ut)= 2^4 \cdot 3 \cdot 5\cdot7 \, E_2(\ut)^2 + 2^2 \cdot 5 E_4(\ut) ~.
\en
The corresponding form $F(\ut)$ coincides with the form given in \eqref{modform} since it satisfies the conditions $c(0)=1$ and $c(n^2)=0$.
Under the action of $\aut{A}$ on elements $a>0$ in $A^\vee$ with $\Tr(a)=2$, it was found {that \eqref{eq:uniqueHMF} has} five different orbits \cite{MR1845183}. {Two of these orbits as given below correspond to the theta functions of Niemeier lattices.} The Fourier coefficients $c(a)$ of $F(\ut)$ on these orbits were given in \cite[Table 1]{MR1845183} which we reproduce in \autoref{orbits}.
 Therefore we can read off the Fourier coefficients $c(a)$ for elements $a$ with $\Tr(a)=2$ :
\eq
c(a) = 240 \# \{S= \mathrm{Jordan \, roots \, of \, JL \, with }  \, t_A(S)=a ~. \}
\en
For the lattice $N$  from the third row of \autoref{orbits} one can read off  $6\cdot 28=168$ roots and for the lattice $N'$ from the fourth row of \autoref{orbits}, one reads off  $6\cdot 0 =0$ roots. Thus the corresponding Niemeier lattices are
\eq
N \simeq (A_6)^4 
\en
\eq
N' = \mathrm{Leech \,\, lattice}~,
\en
which are unique Niemeier lattices with 168 roots and Coxeter number $h = 7$, and no roots and Coxeter number $h = 0$, respectively. A table of theta functions of Niemeier lattices can be found in \autoref{app:thetas} in \autoref{tab:niemeiertheta}. In the below table, $\mathfrak p$ is the unique prime in the ring of integers in the $p^{th}$ cyclotomic field.
\begin{table}[htp]
\begin{center}
\begin{tabular}{|c|c|c|}
\hline 
$a>0 \, \, \mathrm{in}\,\,  A^\vee \, , \mathrm{\Tr}(a)=2$ & $(a) \,\mathfrak{p}^2$ & $ c(a) \, \mathrm{of}\, F(\ut) $ 
\\
\hline 
$2 \cdot n^2 $ & $(2)$ & $240\cdot 49 $ \\ \hline 
$(1-n)$ & $\mathfrak{p}$  & $ 240\cdot 28 $  \\ \hline
$( 1- n')$ & 1 & 0 \\ \hline
$(1-n^2) $ & $\mathrm{a \, prime \,  of  \, norm \, 13 }$ & $240\cdot 196 $ \\ \hline 
$(1 -2n + n^2) $ & 1 & 0 \\  \hline
\end{tabular}
\end{center}
\caption{Orbits of $\aut A$ on elements $a>0$ in $A^\vee$ with $\Tr(a)=2 $ and their Fourier coefficients $c(a)$ in $F(\ut) $\cite{MR1845183}}
\label{orbits}
\end{table}

\subsubsection {Case $D=16$} 
Although $D=16$ is not of the $p^2$ type as studied in the previous subsection, it nevertheless is an interesting case to study since it corresponds to the case where the two orbits that give rise to Niemeier lattice theta functions coincide, i.e., there is only one Niemeier lattice defined by the isotropic lines \cite{MR1845183}. \\ 
In this case, the embedding of a cubic ring in $JL$ studied in \cite{MR1845183} is the ring of triples of integers $(a,b,c)$ with $a\equiv b\equiv c \, (\mathrm{mod}\, 2 ) $ which has discriminant $D=16$. This embedding is also unique modulo the conjugacy by the automorphism group $\aut{JL, \mathcal{N}, E}$. A particular embedding maps the triples $(2,0,0), (0,2,0)$ and $(0,0,2)$ into the rank one elements $S_1,S_2$ and $S_3$ in the polarization $E$:
\eq 
f(2,0,0)= S_1, \,\,\, f(0,2,0) = S_2 \, , \,\, f(0,0,2) = S_3 ~,
\en
such that they satisfy
\eqn 
S_i^2  &=& 2 S_i \,\,\, i,j =1,2,3 \\
S_i S_j &=& 0 \,\,\, i\neq j \\
S_1 + S_2 + S_3 &=& 2E ~.
\enn
These rank 1 elements form a root triple. The group $\aut{JL, \mathcal{N}, E} =\, ^3D_4(3).3 $ of order $2^12 \cdot 3^5\cdot 7^2 \cdot 13 $  acts transitively on root triples. The subgroup that leaves invariant a particular set of root triples  is $2^{2+3+6}\cdot7\cdot3$ which means that there are $14742$ inequivalent embeddings $f: A \rightarrow JL$. 

For a particular embedding with $L_c=A^\bot$, one has $L_c^\vee/L_c \simeq A^\vee/A$ where $A^\vee$ is the subgroup of $((1/2) \mathbb{Z})^3$  formed by triples $(a,b,c)$ such that $(a+b+c)\in \mathbb{Z}$ . Hence $A^\vee /A \simeq (\mathbb{Z}/4 \mathbb{Z})^2$.
The corresponding Niemeier lattice $M$ between $L_c$ and $L_c^\vee$ turns out to be isomorphic to root system of $A_1^{24}$ with 48 roots vectors \cite{MR1845183}. To establish this fact Elkies and Gross, first determine the modular form $F(\ut)$ of weight $(4,4,4)$ for $SL(2,A)$. For this cubic form the relevant modular form turns out to be of weight $(4,4,4)$ with respect to the congruence subgroup $SL(2,A) =\Gamma(2)^3$ of $SL(2,\mathbb{Z})^3$.
 \section{Modular Forms of Spectrum Generating Symmetry $E_{7(-25)}$ and Quantum Degeneracies of  Charge States of BPS  Black Holes of the $5d$ Octonionic Magical Supergravity}
\label{sec:diffrankbhs}

The continuous U-duality group $G$ of any supergravity theory that arises as the  low energy effective theory of M-/superstring theory gets broken down to its discrete arithmetic subgroup $G(\mathbb{Z})$ by the stringy corrections \cite{Hull:1994ys}. 
Therefore, we shall assume that the quantum completion of the octonionic magical supergravity lies within M-/superstring theory framework or an extension thereof and  
its  continuous  U-duality group $E_{6(-26)}$ gets broken down to its arithmetic subgroup  $E_{6(-26)}(\mathbb{Z})$. The arithmetic subgroup $E_{6(-26)}(\mathbb{Z})$ was first studied by Benedict Gross in \cite{MR1369418}.

The orbits of extremal black hole solutions of the octonionic magical supergravity in $5d$ under the continuous U-duality group  $E_{6(-26)}$  were studied earlier and has been reviewed in previous sections. Here we will try to extend  those results to the orbits under the discrete arithmetic subgroup of $E_{6(-26)}$. If the octonionic magical supergravity can be obtained from a compactification of M-theory over a Calabi-Yau (CY) threefold then the exceptional cone $\mathcal{C}$  can be identified  with the K\"ahler cone of that CY manifold and the K\"ahler moduli get identified with the scalar fields of the octonionic magical supergravity. If we choose a basis $J_I$ of the 27 $(1,1)$ forms, the K\"ahler form $J$ can be expanded as
\eq J = h^I J_I~, \en
where $h^I$ are 27 functions of the 26 scalars of the octonionic magical supergravity that satisfy\footnote{ Here we are assumimg a definite value of the volume of the CY threefold which is associated with the universal hypermultiplet which will not play any role in the dicussion of extremal black holes.}
\eq C_{IJK} h^I h^J h^K =1 ~.\en
K\"ahler moduli are given by the volumes of the 2-cycles $\Omega^I$ of the CY manifold
\eq
h^I = \int_{\Omega^I} J~,
\en
and the intersection numbers $C_{IJK}$ are defined as
\begin{align}
\label{eq:intring}
C_{IJK} = \int_{Vol} J_I \wedge J_J \wedge J_K~.
\end{align}
The cohomology lattice of the CY threefold as well as the bare charge lattice of black holes in $5d$ are lattices with integer valued coordinates. On the other hand, the lattice $JL$ defined over the Coxeter order of integral octonions involve integer as well as half-integer coefficients.  We will identify the lattice $JL$ with the lattice of dressed charges and $h^I$ as the vector that determines the polarization. Note that $h^I$ depends on the scalar fields. If the bare charges are denoted as $q_I$, then $h^I q_I$ corresponds to the central charge  as explained in section \ref{5dorbits}. Recall that the physical (dressed)  graviphoton is given by the linear combination $h^I A_{I \mu}$, and in a given vacuum of the theory, the physical graviphoton  is given by $\langle h^I \rangle A_{I \mu}$ where $\langle h^I \rangle$ is the VEV of $h^I$. If we choose the identity polarization $\langle h^I \rangle J_I = I_3 $ then the graviphoton is the bare graviphoton of the theory \cite{Gunaydin:1983rk}. Under the action of the continuous U-duality group $E_{6(-26)}$, the identity polarization can be mapped into any other polarization. However, as we previously explained, under the action of the arithmetic subgroup  $E_{6(-26)}(\mathbb{Z})$ we have two distinct orbits namely the polarizations in the orbit of the identity polarization $I_3$ and the polarizations that are in the orbit of the indecomposable polarization $E$ \cite{MR1369418}.  Physically what that means is that at the quantum level we have two distinct families of vacua , separated possibly by some sort of a phase transition, that are not connected by the arithmetic subgroup of the continuous U-duality group of the classical theory. One could interpret the vacua in the orbit of the identity polarization as the perturbative vacua since the vacuum with the bare graviphoton belongs to it , and the family defined by the indecomposable polarization $E$ as the non-perturbative vacua.

\subsection{Rank 1 BPS Black Holes}
Black holes of supergravity described by rank one elements given in \eqref{rank1} of $J_3^{\mathbb{O}}$ have vanishing entropy (area) and their orbits, which were called critical light-like,  under the continuous duality group is
\begin{align*} \frac{E_{6(-26)}}{SO(9,1)\circledS T^{16}}~. \end{align*}
In addition to rank one condition, they are uniquely labelled by the trace (linear) form $T(A)$. Only those rank one elements $A$ with positive $T(A)$ lie in the exceptional cone.  
For quantum black holes described by rank one elements $A$  of the exceptional Jordan algebra over the integral octonions, $T(A)$ takes on integral values. We interpret the number of rank 1 elements with a given value of $T(A)$ as the degeneracy  of charge states of critical light-like ( small)  1/2 BPS   black holes with the  "quantum number" $T(A)$.
The number $\mathfrak{N}(n)$ of rank one elements $A$  in the positive cone with  $T(A)=n $  ($ n \in \mathbb{N} $) in the indecomposable polarization $E$  as obtained by Elkies and Gross  was given in  \eqref{number_rankone}\footnote{ Since the rank one elements with $T(A)=-n $ differ from those with $T(A)=n$ by an overall sign their countings coincide.}.  For $n=p$ prime it simplifies to
\begin{align*} \mathfrak{N}(p) = \frac{ 3\cdot 7 \cdot 13 }{691} \left( p^{11}  - \tau(p) +1 \right). \end{align*}

For $p=2$ corresponding to Jordan roots, we have $\mathfrak{N}(2)=819$. The group $^3D_4(2)$  acts transitively on the 819 Jordan roots and the stabilizer of a given Jordan root is the subgroup $2_+^{1+8} \cdot L_2(8)$.
According to \cite{MR1845183} the group $^3D_4(2)$ is also expected to act transitively on the set of $A\geq 0$ in $JL$ with rank($A$)=1  and $T(A)=3$ and  the stabilizer  is  a maximal subgroup of $^3D_4(2)$ isomorphic to $L_2(2) \times L_2(8)$. 

Degeneracies of  charge states of  critical light-like  1/2 BPS black holes with $T(A)=n$ with $n\geq 2$ are given by the Fourier coefficients of the singular modular form of weight 4 over the exceptional domain. As was summarized in \autoref{modularform} and reviewed further in \autoref{app:exceptionalmf}, the singular modular form of weight 4 over the exceptional domain $\mathcal{D}$ as studied in \cite{MR1216126,MR1437509} has the Fourier expansion
\eq
E_{4}(Z) = 1 +  240 \sum_{\substack{T\geq 0, \ T \in  JL , \\ \mathrm{rank}(T)=1}} \sigma_3(c(T))  e^{2\pi i  \Tr(T\circ_I Z)} \quad , \quad Z\in \mathcal{D} 
\label{singular4}\en
where 
\begin{align*} \sigma_k(m) := \sum_{d\in \mathbb{N}, d|m} d^k \quad , \quad m\in \mathbb{N}, 
\end{align*}
and 
\begin{align*}
	c(T) = \text{max}(r \in \mathbb N) \ \text{such that } \frac{1}{r} T \in JL ~. 
\end{align*}
Under the action of  the arithmetic subgroup $E_{6(-26)}(\mathbb{Z})$,  there are two orbits: one characterized by the identity polarization $I_3$ and the other orbit is characterized by the indecomposable polarization $E$. If we choose $Z= \tau E^{\#}$ where $\tau$ is the complex coordinate in the upper half-plane, then we obtain the modular form of weight 12 of   $SL(2,\mathbb{Z})$ discussed in \autoref{Jordanroots}. Since all rank one elements $A$  in the polarization $E$ have $T_E(A) >1$, the resulting modular form of weight twelve of $SL(2,\mathbb{Z})$ turns out to be the theta function of the Leech lattice. Its Fourier expansion coefficients count the number of distinct critical light-like BPS black holes with $T_E(A)=n$ with $n>1$
\begin{align*}
\mathfrak{N}(n) =\frac{ 3\cdot 7 \cdot 13 }{691} \left( \sigma_{11} (n) - \tau(n) \right)~. \label{number_rankone}
\end{align*}

If we choose $Z=\tau  I $ then the coefficients of the resulting modular form of $SL(2,\mathbb{Z})$ would count the number $\mathfrak{N}(n)$ of distinct BPS black holes with $T_I(A)=n$ and $n\geq 1$, where $\mathfrak{N}(n)$ is generated by the series
$$f(q):= 1 + 240 \sum_{r \geq 1} (\sigma_{11}(r) - \tau(r))q^r = 1 + \sum_{r\geq 1} \mathfrak{N}(r) q^r ~.$$

Under the continuous U-duality group $E_{6(-26)}$ action on the exceptional Jordan algebra, there is a single orbit corresponding to 1/2 BPS black holes with non-vanishing entropy, namely  $ E_{6(-26)}/F_4$ with the compact automorphism group $F_4$ acting as stabilizer. Any element $J$ of $J_3^{\mathbb{O}}$ can be brought to the diagonal form 
\eq 
J = \lambda_1 P_1 + \lambda_2 P_2 + \lambda_3 P_3~, \label{idenpol}
\en
where $P_i$ are the idempotents of rank one such that the identity polarization has the decomposition
\begin{align*} I = P_1 + P_2 + P_3~. \end{align*}
For the exceptional Jordan algebra over the Coxeter's order of integral octonions we have two distinct polarizations, namely $I$ and $E$, that belong to different orbits. If $J$ belongs to the orbit of the indecomposable polarization then the  analog of the decomposition \eqref{idenpol} for the indecomposable polarization $E$ is
\eq 
J = \mu_1 S_1 + \mu_2 S_2 +\mu_3 S_3 ~,
\en
where $S_i$ are the Jordan roots such that
\begin{align*} 2E =S_1 + S_2 + S_3 ~.\end{align*}
The Hilbert modular forms defined and studied by Elkies and Gross \cite{MR1845183} are related to this decomposition.  When extended to the exceptional domain  $\mathcal{D}$ the element $J$ goes over to 
\eq
Z = \tau_1 S_1 + \tau_2 S_2 + \tau_3 S_3
\en
When substituted into \eqref{singular4}, one obtains the Hilbert modular form of weight $(4,4,4)$ of $SL(2,\mathbb{Z})^3$. Modular form of weight 12 of $SL(2,\mathbb{Z})$ corresponds to setting 
\eq
\tau=\tau_1=\tau_2=\tau_3 
\en
and restricting to a diagonal subgroup of $SL(2,\mathbb Z)$.
For each embedding of cubic ring $A$ in $JL$, the rank 1 elements $S$ decompose as $ (\alpha + v )$ with $\alpha \in A$ and $v \in L_c$.
When the discriminant of the cubic ring is $p^2$ with $p$ prime, the charges of the critical light-like black holes in the orthogonal complement of $A$ take values in a Niemeier lattice and restriction of the $Z\in \mathcal{D}$  to the subdomain $\underline{\tau}\cdot\underline{S}$ leads to the Hilbert modular forms studied by Elkies and Gross \cite{MR1845183} and reviewed in \autoref{HMF} and \autoref{app:exceptionalmf}. For the choice $ p=7$ the Niemeier lattices are the Leech lattice and the lattice $A_6^4$ with $4\cdot 48 =168$ root vectors. These two lattices coresponding to the two isotropic lines lead to different modular forms. They are uniquely distinguished by the coefficient of the first order term which is given by the number of root vectors. For the Leech lattice this coefficient is zero. This is true in general for embeddings of cubic rings with discriminant $D=p^2$ with $p$ prime whose two isotropic lines define two different Niemeier lattices. For the embedding of cubic ring with $D=16$ studied by  Elkies and Gross and reviewed above the rank one black hole charges lie on the  Niemeier lattice $A_1^{24}$. 

\subsection{Rank 2 BPS Black Holes}
Rank 2 black holes that were called light-like in \cite{Ferrara:1997uz} 
are characterized the conditions that the element $A$ of the Jordan algebra representing the charges has vanishing norm and, hence, vanishing area : 
\begin{align*} \mathcal{N}(A) =0 \end{align*} and 
\begin{align*} A^\# \neq 0~. \end{align*}
They are characterized by trace ( linear) form $T(A)$ and spur ( quadratic) form 
$S(A) =T(A^\#)$ following the definition of McCrimmon \cite{mccrimmonbook}.  
The rank 2 elements  can be brought to the form 
\begin{align*} S_{ij} =\lambda ( P_i+ P_j ) \qquad (i\neq j) \end{align*} or to the form
\begin{align*} A_{ij}= \lambda ( P_i -P_j) \end{align*} under the action of $E_{6(-26)}$ \cite{Ferrara:1997uz,Ferrara:2006xx} as explained previously. 

$S_{ij}$ satisfies 
\eq T(S_{ij})= 2 \lambda \, \quad , \quad S(S_{ij})= 2 \lambda^2 \en 
and $A_{ij}$ satisfies 
\eq T(A_{ij})=0 \, \quad , \quad S(A_{ij})= -2\lambda^2 \en 

For $\lambda >0$, $S_{ij}$ lie in the exceptional cone and describe 1/2 BPS black holes\footnote{In this paper we will restrict ourselves to BPS black holes.}. Their orbits $S_{ij}$ are given by the coset space
\begin{align*} \frac{E_{6(-26)}}{SO(9)\circledS T^{16} }~. \end{align*}

Over the integral octonions $\mathcal{R}$   rank two elements $S_{ij}$ of  the exceptional  Jordan algebra in the exceptional cone can be written as a linear combination of two mutually orthogonal rank one elements 
\eq 
S_{ij} = S_i + S_j \en with positive trace forms $T_E(S_i)$ and $T_E(S_j)$. Then

\eq T_E(S_{ij})= T_E(S_i) + T_E(S_j)  \quad , \quad S(S_{ij})= T(S_{ij}^\# ) \en 

By squaring the exceptional singular modular form of weight 4, we get the exceptional  singular modular form of weight 8 \cite{MR1437509} whose rank two terms in its Fourier expansion involve such sums of two rank one elements in the exceptional cone. Rank two terms of the form $A_{ij} = S_i - S_j$  which correspond to charge states of non-BPS rank two extremal black holes do not appear in the expansion of $E^2_4(Z)$. \footnote{ Rank two element with both $T_E(S_1)$ and $T_E(S_2)$ negative differ by an overall sign from those with both of them positive and their quantum degeneracies will also be given by  the Fourier coefficients of $E^2_4(Z)$.}  
 Hence we expect the singular modular form of weight 8 of $E_{7(-25)}$ to describe the quantum degeneracies of charge states of rank 2 BPS black holes. In the formulation of Krieg \cite{MR1437509}, it takes the form
\eq
\label{eq:e8}
E_{8}(Z) =E_{4}(Z)^2 = \sum_{T\in JL \, , \, T\geq 0} \alpha(T) \,   e^{2\pi i  \Tr(T\circ_I Z)} \quad , \quad Z\in \mathcal{D} ,
\en
where 
\begin{align}
\label{eq:fouriereqweight8}
	\alpha(T) = \begin{cases}
		1 & \text{if } T = 0 \\ 
		480 \cdot \sigma_7(c(T)) & \text{if } \text{rank}(T) = 1 \\
		240 \cdot 480 \cdot {\displaystyle \sum_{d \in \mathbb N, d \vert c(T)}} d^7\sigma_3(c(T^\#/d^2)) & \text{if } \text{rank}(T) = 2 \\ 
		0 & \text{if } \text{rank}(T) = 3 
	\end{cases}~.
\end{align}
The Fourier coefficients $\alpha(T)$  in \eqref{eq:e8} and \eqref{eq:fouriereqweight8} are all rational integers. 
Coefficients of rank 2 elements $T$ with $T^\# \neq 0$ count the quantum degeneracies of charge states of rank 2 BPS black holes.

 Krieg's derivation of these results used  the Fourier-Jacobi expansion of $E_4(Z)$\cite{MR1195510} by decomposing the coordinates $Z$ of the exceptional domain over the integral octonions  as 
\eq Z= \left( \begin{array}{cc} Z_1 & W \\ W^\dagger & z_3 \end{array} \right) \en
where $Z_1$ lies in the upper half plane of the Jordan algebra of $2\times 2$ Hermitian matrices over the integral octonions $\mathcal{R}$ , $W$ is a $(2\times 1)$ matrix over complex integral octonions and $z_3$ is a complex variable in the upper half-plane.  We shall summarize this derivation following \cite{MR2478255}.  The Fourier-Jacobi expansion of $E_4(Z)$ takes the form
\eq E_4(Z)= f_4(Z_1) + \sum_{m=1}^\infty \phi_m (Z_1,W)  e^{2\pi i m z_3 } \en
where $f_4(Z_1)$ is a modular form of weight 4 on $\mathcal{H}_2$ and $\phi_m(Z_1,W)$ is a Jacobi form of weight  4 and index $m$ on $\mathcal{H}_2 \times \mathbb{O}_{\mathbb{C}}^2$. The coefficient $\alpha(T)$ is given by $$ \alpha(\left( \begin{array}{cc} T_1 & 0 \\ 0 & 0 \end{array} \right))=240 \left(\sum_{d|\epsilon(T_1)} d^3\right)$$  
if $det (T_1)\neq 0 $ and $T_1\neq 0$. 
When we square the modular form $E_4(Z)$ the Fourier coefficients $\beta(T)$ in its expansion 
\eq
E^2_4(Z) = \sum_{T\in JL} \beta(T) e^{2\pi  i (T,Z)}
\en
are of the form $\beta(T)= \alpha(T_1) \alpha(T_2) $ where $\alpha(T_i)$ are the Fourier coefficients of $E_4(Z)$
and hence clearly vanish unless the rank of $T=T_1+T_2 $ is less than or equal to two. 
Fourier-Jacobi expansion of $E^2_4(Z)$
\eq
E^2_4(Z)= g_8(Z_1) + \sum_{m=1}^\infty \psi_m(Z_1,W) e^{2\pi i m z_3} 
\en
then follows from that of $E_4(Z)$ and one has\cite{MR2478255}
\begin{align}
g_8(Z_1)= [f_4(Z_1)]^2=\sum_{T_1} b\left( \begin{array}{cc} T_1 & 0 \\ 0 & 0 \end{array} \right) e^{2\pi i (T_1,Z_1)} = \lim_{\lambda\rightarrow \infty} E^2_4 \left( \left( \begin{array}{cc} Z_1 & 0 \\ 0 & i \lambda \end{array} \right) \right)
\end{align}

where 
\eq
b\left(\left[\begin{array}{cc} T_1 &0 \\ 0 &0 \end{array} \right] \right) = \# \{h_1,h_2 \in \mathcal{R}^2 \, | \,  h_1 h_1^\dagger + h_2 h_2^\dagger = T_1 \} \en
If $T_1$ is of the form $\left(\begin{array}{cc} n & 0 \\ 0 & 0 \end{array} \right) $  where  $n\in \mathbb{N}$ then
\eq b\left(\begin{array}{cc} T_1 & 0 \\ 0 & 0 \end{array} \right)   = \#\{ o_1, o_2 \in \mathcal{R} \, | \, N(o_1)+N(o_2) =n\} = 480 \sum_{d|n} d^7 \en
and when $T_1= \left[ \begin{array}{cc} n & t \\ \bar{t}& 1\end{array}\right]$ then 
\eqn b\left(\begin{array}{cc} T_1 & 0 \\ 0 & 0 \end{array} \right)&=&  \# \{ o_1,o_2,o_3 \in \mathcal{R}\, | \, N(o_3)=1, N(o_1)+N(o_2) = n , o_1 \bar{o}_3 = t \} \nonumber \\
&& = 240 \cdot 480 \sum_{d| (n - N(t))} d^3 
\enn
where $N(t) = t \bar{t}$.

\subsection{Rank 3 Large BPS Black Holes} 
 $E_{4}(Z)$ and $E_8(Z)=(E_{4}(Z))^2$ are the only two singular modular forms over the  exceptional domain \cite{MR1437509}. The Fourier coefficients of non-singular higher weight forms were obtained by Kim and Yamauchi in \cite{kim2016cusp, kim2015ikeda} based on work by Karel \cite{karel1974fourier}.
  Using the notation of \cite{kim2016cusp} and \cite{MR1437509}, the Fourier coefficients in full generality are given  below. Consider a weight $  k \in 4 \mathbb Z$ modular form over the exceptional domain with the Fourier expansion
  \begin{align}
  	F(Z)_{k, \ k\geq 12, \ k\in 4\mathbb Z} =  \sum_{T \geq 0, T \in JL} \alpha_k(T) e^{2 \pi i \Tr(T\circ_I Z)} ~. 
  \end{align}
 The Fourier coefficients $\alpha_k(T)$ above are detailed in \autoref{app:exceptionalmf}. We only quote the main results here, and we refer the reader to the appendix for more details. For a higher weight exceptional modular form, the Fourier coefficients are given by
\begin{align}
	\label{eq:allfouriercoeffs}
		\alpha_k(T) = \begin{cases}
		1 & \text{if } T = 0 \\ 
		-\frac{2k}{B_{k}} \cdot \sigma_{k-1}(c(T)) & \text{if } \text{rank}(T) = 1 \\
		\frac{4k (k-4)}{B_k B_{k-4}} \cdot {\displaystyle \sum_{d \in \mathbb N, d \vert c(T)}} d^{k-1}\sigma_{k-5}(c(T^\#/d^2)) & \text{if } \text{rank}(T) = 2 \\ 
		 2^{15} \frac{k}{B_k} \cdot \frac{k-4}{B_{k-4}} \cdot \frac{k-8}{B_{k-8}}  \cdot \det(T)^{(k-9)/2} \cdot {\displaystyle \prod_{p \vert \det T}} f_T^p(p^{(k-9)/2}) & \text{if } \text{rank}(T) = 3 
	\end{cases}~,
\end{align} 
where $f_T^p$ is a monic Laurent polynomial that depends only on  $T$ and $p$, and mostly evaluates to unity \cite{kim2015ikeda, kim2016cusp}.

 Rank 3 extremal  black holes exhibit attractor phenomena and their analysis is more subtle. For rank three elements $T$, the entropy (area) of the large extremal black hole is given by  the squareroot of the cubic form $\det(T)$.\footnote{ We should note that the trilinear form in Kim and Yamauchi \cite{kim2016cusp} given by the determinental form  differs by a factor six from the one used by Elkies and Gross. In particular
  $ \det(J)= (J,J,J)_{KY} = \frac{1}{6} (J,J,J)_{EG}$. }
  A rank 3 element of the Jordan algebra element $J$ can be written as a linear combination of the three rank one elements. 
Over the reals $\mathbb{R}$ every element of the exceptional Jordan algebra with non-zero cubic form can be brought to a diagonal form under the action of the compact automorphism group $F_4$ of the exceptional Jordan algebra. Over the integral octonions of Coxeter, not all the elements can be brought to a diagonal form by the action of a finite subgroup of the compact group $F_4$.  This is due to the fact that the arithmetic group $E_{6(-26)}(\mathbb{Z})$ does not act transitively on positive polarizations $E$  in the exceptional cone with $det(E)=1$. It has two distinct orbits.  One is the identity polarization  $I=I_3$ which corresponds to the perturbative vacuum of the theory and the other one is the indecomposable polarization $E$
that corresponds to the non-perturbative vacuum of the octonionic magical supergravity at the quantum level. 
The little group of the identity polarization $I=I_3$ is $2^2\cdot O_8^+(2)\cdot S_3 $, whereas the little group of the indecomposable polarization $E$ is the finite group  $^3D_4(2)\cdot 3$.  We should stress that in this paper we are interested in the quantum degeneracies of charge states of BPS black holes whose charges lie in the positive cone of the exceptional Jordan algebra. The charge states of large extremal non-BPS black holes do not lie in the exceptional cone. The little groups of rank two and rank 3 elements corresponding to non-BPS extremal black holes are non-compact discrete subgroups of $E_{6(-26)}(\mathbb{Z})$ .

By squaring the singular modular form $E_4(Z)$  one obtains  a singular modular form $E_8(Z)$ whose Fourier coefficients give the degeneracies of rank two BPS black holes. Taking higher powers of $E_4(Z)$ does not lead to any new singular modular forms. 

When taking the third power of $E_4(Z)$ we will get three different types of terms. First terms of the form 
\begin{align} \sum_{T \geq 0, T \in JL} \alpha_k(T) e^{2 \pi i \Tr(T\circ_I Z)}\end{align}
that are relevant for charge states of  rank 1 BPS black holes, while terms  of the form 
\begin{align} \sum_{T > 0, T \in JL} \alpha_k(T) e^{2 \pi i \Tr(T\circ_I Z)}  \sum_{S > 0, S \in JL} \alpha_k(S) e^{2 \pi i \Tr(S\circ_I Z)} 
\end{align}
 are relevant to the degeneracies of charge states of rank 2 BPS black holes, and terms  of the form
\begin{align} \sum_{T > 0, T \in JL} \alpha_k(T) e^{2 \pi i \Tr(T\circ_I Z)}  \sum_{S > 0, S \in JL} \alpha_k(S) e^{2 \pi i \Tr(S\circ_I Z)}  \sum_{U > 0, U \in JL} \alpha_k(U) e^{2 \pi i \Tr(U\circ_I Z)} \end{align}
are relevant to  the degeneracies of charge states of rank 3 BPS black holes.
The Fourier coefficients of terms of the form \[  \alpha_k(S) \alpha_k(T) \alpha_k(U)e^{2 \pi i \Tr(T\circ_I Z)} e^{2 \pi i \Tr(U\circ_I Z)}e^{2 \pi i \Tr(S\circ_I Z)} = \alpha_k(S) \alpha_k(T) \alpha_k(U)e^{2 \pi i \Tr((S+T+U)\circ_I Z)} \] for $S,T,U$ all distinct describe the degeneracies of charge states of rank 3 BPS black holes since the cubic norm of $(S+T+U)$ is non-vanishing for such terms. Furthermore sums of rank one elements in the exceptional cone lie in the exceptional cone and are relevant for BPS black holes. 

 Higher powers $E_4^n(Z)$ for $n>2$ also contain terms of the  forms  ($Tr \sum_i^n T_i)\cdot Z$ which describe rank 3 as well as rank 2 and rank 1 elements in the exceptional cone. A rank 3 element with a given cubic norm can occur in $E_4^n(Z)$ for different values of $n$. Unlike the rank 1 and 2 cases norms of rank 3 elements are not invariant under discrete special conformal transformations. To our knowledge the relationship between  higher powers of $E_4(Z)$  and the higher weight $k$ modular forms or cusp forms  of \cite{karel1974fourier,kim2016cusp, kim2015ikeda} have not yet been studied by mathematicians. 
 
 At this point we should stress the difference between our definition of quantum degeneracy of a given charge state represented by an element of the exceptional Jordan algebra over integral octonions and the degeneracy of microstates that underlie extremal black holes solutions in string theory. For spherically symmetric large $5d$ extremal black holes in string theory the microscopic degeneracy $d_{micro}$ is related to the entropy $S_{string}$  in the limit of large charges via a formula of the form
 \eq S_{string}= ln\, (d_{micro} ) = \mathcal{N}_3 ( Q_i) \en
 where $\mathcal{N}_3 ( Q_i)$ is the cubic norm determined by the charges. 
 This shows that the microscopic degeneracy $d_{micro}$ grows exponentially as a function of the cubic norm defined by the charges.
 In our case quantum degeneracy of a rank 3 BPS black hole charge state $J$ depends not only on the cubic norm but also on the quadratic spur form as well as  the linear trace form.  This is a purely number theoretic calculation as in the case of rank one and rank two black holes. The only physical assumption we are making is that the quantum completion of the octonionic magical supergravity breaks the U-duality group to its maximal arithmetic subgroup  and the charges take values in the lattice defined by the exceptional Jordan algebra over the integral octonions. The BPS condition restricts the charge states to lie in the exceptional cone. 
 What makes the analysis of rank three BPS black holes harder is the non-uniqueness of  non-singular modular forms in contrast to unique singular modular forms describing the degeneracies of  rank one and rank two BPS black holes.
 
 From the physics point of view rank 3 BPS black holes are distinguished by the fact that they exhibit attractor phenomena.  We shall leave the investigation of quantum degeneracies of  charge states of large rank three BPS black holes to future investigations.

\section{The Geometric Embedding of Octonionic Magical Supergravity}
\label{sec:cy}
Consider $\mathcal D$, a bounded symmetric domain. It was shown by Deligne in \cite{deligne1979varietes} that $\mathcal D$ is the moduli space of Hodge structures (canonical). Consider a simply-connected, simple real algebraic group $G$ which has a transitive action on $\mathcal D$ and let $K$ be its maximally compact subgroup such that $\mathcal D = G/K$. The tube domain is then classified by pairs $(D,v)$ where $D$ is a connected Dynkin diagram and $v$ is a special vertex of $D$. For the case that we considere here, $\mathcal D $ is a tube domain \cite{MR1258484}. 
We are interested in the case where the group $E_7$ acts transitively on $\mathcal D$. In this case, we have $G=E_{7(-25)} \equiv E_{7,3}, \ K = U(1)\times E_6$. The positive cone is the exceptional cone $\mathcal{C}$ 
 defined over the exceptional Jordan algebra. This gives us $\displaystyle\mathcal D = \frac{E_{7,3}}{E_6 \times U(1)}$ with rank$(\mathcal D) = 3$, $\dim(\mathcal D) = 27$. The vertex $v$ then determines a fundamental irreducible representation of $E_{7,3}$ over $\mathbb R$, which is in fact just the 56 dimensional unique miniscule representation of $E_{7,3}$, which we denote as $V$. This fundamental representation gives rise to the canonical variation of Hodge structures on $\mathcal D$, in the sense of Deligne \cite{deligne1979varietes}. While we omit the details here, the variation of real Hodge structure $\mathcal V$ can be obtained as the tensor product of the unique miniscule representation of $E_{7,3}$ with an equivariant holomorphic vector bundle on $\mathcal{D}$. The Hodge structures here are all of weight 3. Now, given any tube domain $\mathcal D$, we may ask if and how the variation of Hodge structures $\mathcal V$ arises geometrically. This requires the existence of a reduction of the pair $(E_{7,3},V)$ onto the field $\mathbb Q$ from $\mathbb R$. Generically, the reduction of the variation of Hodge structure might arise from a sub-Hodge structure on a dimension 3 projective primitive cohomology \cite{MR1258484}. For the case of when $G = \text{Spin}(2,10)$, the positive cone associated to the domain is defined over $J_2^{\mathbb{O}}$. Note that the octonions have a unique reduction to $\mathbb Q$. In this case, the descent of the Hodge structure is of type $(1,10,1)$ and can be realized as the pull back of a suitable sub-Hodge structure of polarized $K3$ surfaces with an Enriques involution.\\ If one expects to realize the variation of Hodge structures from geometry, one therefore needs an analogous setup for the case when $G = E_{7,3}$. In the case of the $E_7$ tube domain, there are no variations of Hodge structure of abelian variety type \cite{ZhangPhD}, which makes the problem more intricate. The Hodge numbers of the descent onto $\mathbb Q$ in this case are expected to be $(1,27,27,1)$. If there is a Calabi-Yau threefold that satisfies these criteria, the Hodge numbers ensure consistency with the physical requirements with respect to the number of vector-multiplets and the hyper-multiplets of the dimensionally reduced octonionic magical supergravity. However, it is unlikely due to mirror symmetry that the descent of the variation of Hodge structure onto $\mathbb Q$ is the pullback of the entire cohomology of a smooth complex projective variety. Since the Hodge structures do not admit Picard-Lefschetz degenerations, it excludes most complete intersection Calabi-Yau manifolds (CICYs) in weighted projective spaces. In fact, such a Calabi-Yau does not exist in the current database of CICY threefolds available at \url{http://www-thphys.physics.ox.ac.uk/projects/CalabiYau/cicylist/}.\\ \\
One may also approach this question from a more bottom up approach where the Lie algebras associated to projective varieties are analyzed. This was done by Looijenga and Lunts in \cite{looijenga1997lie}, where it was demonstrated that classical Jordan algebras arise geometrically and that the $E_7$ algebra that we have discussed thus far arise topologically. This happens if there exists a 27 dimensional $K-$vector space (where $K$ is a field of characteristic zero) endowed with a cubic form $c:W^3 \rightarrow K$ that does not factor through the proper linear quotient of $W$ such that the cubic form only takes even integral values. If these conditions are satisfied, then there exists a closed oriented 6-manifold for which the integral cohomology ring is isomorphic to the integral algebra associated to the the vector space and the integral structure endowed on it \cite{jupp, wall}. This question is posed in  \cite{MR1258484}, \cite{MR1411589} and \cite{looijenga1997lie} in various avatars. We reiterate the question in a more unified manner here. Is there a (Calabi-Yau) threefold with a Picard group of rank 27 whose N\'eron-Severi\footnote{The N\'eron-Severi group of a CY manifold is the group of divisors of the CY manifold modulo algebraic equivalence. It is an Abelian group (and hence often referred to as a lattice). The rank of the N\'eron-Severi group is the Picard rank or number.} group contains the Leech lattice, and whose Lie algebra is the $E_7$ Lie algebra? \\ 
 If octonionic magical supergravity can be embedded canonically in M-theory (and F-theory) by a(n) (elliptically fibered) Calabi-Yau threefold, it must have Hodge numbers $h_{11} = h_{21}=27$ and its N\'eron-Severi lattice must be defined in terms of the exceptional Jordan algebra such that it admits the action of the $E_7$ group. The ample cone\footnote{For any projective variety $X$ with an inclusion, a line bundle is \textit{very ample} if it can be obtained by the pulling back the natural line bundle on $X$ via a closed immersion. A line bundle $\ell$ is said to be  \textit{ample} if $\exists\  n\in \mathbb Z_+$ such that $\underbrace{\ell\otimes \cdots \otimes \ell}_{n \ \text{times}}$ is very ample. The ample cone is then simply the convex cone in $\textbf{H}^2(X;\mathbb Q)$ generated by $c_1(\ell)$. The case in question here is when $X$ is a CY variety whose ample cone is the positive cone of the Jordan algebra.}  \cite{mukai1998abelian} corresponds then to the positive cone of the Jordan algebra. Furthermore, the intersection polynomial of this threefold is the cubic norm \eqref{eq:cubicnorm}, following the discussion in \autoref{sec:embedding}.
It is important to note that although there are physical reasons to search for a Calabi-Yau threefold with elliptic fibration (the reason here being that there is also an F-theory embedding of the model), there is no requirement a-priori that the threefold required  has to be Calabi-Yau. The possibility of the case where the manifold does not satisfy the Calabi-Yau conditions might have some important implications in string theory, the starkest one being that there might be an(other) additional phase(s) of M-theory that describes the quantum completion of octonionic magical supergravity.
\subsection{Search for Candidate Calabi--Yau Threefolds}
As was explained in section \ref{sec:embedding} magical supergravity theories were discovered before the so-called first string revolution. Whether the octonionic magical supergravity can be obtained from M/superstring theory 
by compactification on some exceptional Calabi-Yau manifold was posed as an open problem in \cite{Gunaydin:1986fg} shortly thereafter.
Since the quaternionic magical supergravity without any hypermultiplets can be obtained from superstring theory suggests that it might be possible to obtain the octonionic magical supergravity without  hyper multiplets  from M/superstring theory on some rigid Calabi-Yau manifold. However to this date no such rigid Calabi-Yau manifold has been found. The focus eventually shifted to look for a self-mirror Calabi-Yau manifold with $h_{11}=h_{12}=27$  after it was realized that there exists an anomaly free supergravity theory in six dimensions that reduce to the octonionic magical supergravity theory coupled to 28 hypermultiplets with the target space $E_{8(-24)}/E_7\times SU(2) $ in five and four dimensions.  

To search for  Calabi-Yau threefolds with $h_{11}= h_{21}= 27$ such that they have intersection numbers given by the cubic norm \eqref{eq:cubicnorm} of the exceptional Jordan algebra that defines the octonionic magical supergravity  we detail certain `experimental' approaches. 
The two most plausible approaches are the Borcea--Voisin threefolds which have been completely classified, and those Calabi-Yau manifolds that are realizable as hypersurfaces in a toric variety.
\subsubsection{Ruling out Borcea--Voisin Threefolds} The Borcea--Voisin (BV) threefolds are one of the best known examples of mirror pairs of Calabi--Yau threefolds that are constructed by acting with an involution on the product of a $K3$ surface and an elliptic curve \cite{borcea1,borcea2,voisin1}. These manifolds have also been studied in the context of F-theory compactification \cite{vafamorrison1,vafamorrison2}. \\[0.2cm] BV manifolds are a class of elliptically fibered threefolds with base $K3/\sigma$, where $\sigma$ is an anti-holomorphic involution on $K3$ that flips the sign of the holomorphic $2-$form as $\sigma: \omega_{2,0} \rightarrow - \omega_{2,0}$. From this involution, one may construct a CY threefold as
\begin{align}
 CY_3 \cong \frac{K3 \times T^2}{\sigma \times \hat \sigma}
 \end{align}
 where the involution $\hat \sigma$ acts on the torus coordinate $T$ as $\hat \sigma: T\rightarrow -T$ and the holomorphic $3-$form of the threefold is given by $\omega_{3,0} = \omega_{2,0} \wedge dT$. Such BV Calabi-Yau threefolds are determined by three integers $(r,a,\delta)$, where $\delta = 1,2$ represents the canonical class parity, $r \in [1,20]$ is the rank of the sublattice of $\textbf{H}^2(\text{K3}, \mathbb Z)$ that is invariant under the involution $\sigma$, and $a \in [1,11]$ is the rank of the N\'eron-Severi group of $\text{K3}/\sigma$ \cite{nikulin1986discrete}. Thus, given a triple $(r,a,\delta)$, one may determine the Hodge numbers of the Calabi-Yau threefolds in a straightforward manner:
 \begin{align}
 h_{11}(\text{K3}/\sigma) = r; \ \  h_{11}(\text{K3}\times T^2/\sigma') = 3 r - 2a + 5; \ \ h_{21} = 65 - 3r - 2a.
 \end{align}
 One may also construct mirror pairs of threefolds as
 \begin{align}
 (r,a,\delta) \xrightarrow {\text{Mirror}}  (20-r, a, \delta).
 \end{align}
Since we are interested in models with $h_{11} = h_{21} = 27$ for the threefold $\text{K3} \times T^2/\sigma'$, we have $ (r,a) = (10,4)$. \\ \\
 Motivated by the above considerations,  Bianchi and Ferrara reconsidered the string derivation of FHSV model \cite{Ferrara:1995yx} and investigated whether the  octonionic magical supergravity might also admit a string interpretation along the lines of Enriques model on a particular self-mirror Calabi-Yau of Borcea--Voisin  type \cite{Bianchi:2007va}. However, it was argued in \cite{Bianchi:2007va} that this particular Calabi--Yau threefold cannot realize the octonionic magical supergravity theory since in the six dimensional reduction of this theory  one of the $SO(8)$ (out of the rank 16 $SO(8)^4$) can be broken by the adjoint hypers. In the four dimensional reduction, a  restoration of symmetry amongst the 16 Cartan generators is not likely, despite the theory having the required number of vector- and hyper-multiplets \cite{Bianchi:2007va}. Here we also stress on another compelling reason why BV threefolds cannot realize octonionic magical supergravity that was not mentioned in \cite{Bianchi:2007va}. The moduli spaces of variation of Hodge structures of BV threefolds have a direct product form \cite{borcea2} as in generic Jordan families like the FHSV model \cite{Ferrara:1995yx} and hence cannot include the moduli space of octonionic magical supergravity \cite{mukai1998abelian}. The N\'eron-Severi group of the FHSV model is given by the reducible Jordan algebra $J_2^\mathbb O\oplus \mathbb R$ whose conformal group is $SO(10,2)\times SU(1,1) \subset E_{7(-25)}$. 

\subsubsection{CY Threefolds as Hypersurfaces in Toric Varieties}
Since the $(r = 10, a = 4)$ BV threefold, despite yielding the right Hodge numbers and matter content, does not respect the symmetry of the Cartan subgroup, we consider turning to determining the threefold that results in octonionic magical supergravity using more brute force techniques. By this, we mean searching for threefolds that are hypersurfaces in toric varieties. For a review of how this construction works, we refer the reader to \cite{Altman2015}. The construction realized by Kreuzer--Skarke \cite{Kreuzer2002, Kreuzer2002a} makes use of an important condition for a hypersurface in a toric variety to be Calabi-Yau viz., the lattice polytope is reflexive \cite{Batyrev1994}. Inequivalent reflexive polyhedra yield different Calabi-Yau manifolds. Classification of inequivalent polytopes is therefore a necessary problem and is a problem in combinatorics. This has been done in the Kreuzer-Skarke (KS) database aka \texttt{PALP} \cite{Kreuzer2004}. \\[0.3cm] However it is not clear from the work of KS as to how many distinct Calabi--Yau manifolds actually emerge from this. This is due to the fact that different triangulations of simplices of a given polytope can in principle give rise to different Calabi-Yau manifolds. This means that there are quite likely more Calabi-Yau manifolds than there are reflexive polytopes.
The Kreuzer-Skarke database (KSD) is a construction and classification of all reflexive polytopes for dimension  $D \geq 4$. In $D=4$, there are  473,800,776 reflexive polytopes. \\ \\
We focus our attention to the case of $h_{11} = h_{21} = 27$. These Hodge numbers represent the case where there are a maximal number of reflexive polyhedra in four dimensional toric varieties \cite{Ashmore:2011yw}. \\ \\ To ascertain the correct Calabi-Yau threefold, we start with a reflexive polyhedron in $D=4$. For this reflexive polyhedron, we then obtain all possible FRS triangulations and compute the Mori cone and the Stanley-Reisner ideal. We can then compute the triple intersection polynomial for all the distinct Calabi-Yau manifolds that can be obtained from inequivalent triangulations (flops) and compare with \eqref{eq:cubicnorm}. \texttt{PALP} performs triangulations on polytopes corresponding to small Hodge numbers. A more extended catalogue of Calabi-Yau threefolds with computations up to and including $h_{11} = 7$ has been done in \cite{Altman2015} and \cite{Crino:2022zjk}.\footnote{We thank Andreas Schachner for informing of the reference \cite{Crino:2022zjk}.} However, it is difficult to extend the computation of distinct intersection numbers to much beyond this due to the rapid growth of number of polytopes and their triangulations, and increased CPU usage in computing the Gr\"obner basis \cite{He:2018jtw}. \\[0.3cm] The first bottle neck here lies in triangulating these polyhedra and computing the topological quantities in the same way as \cite{Altman2015}. The number of vertices of the Newton polytope and its dual scale with $h_{11}$ i.e. all Calabi--Yau threefolds that are constructed from triangulations of hypersurfaces in toric varieties have convex reflexive Newton polytopes whose number of vertices and faces increase with increasing $h_{11}$. The number of triangulations for a higher dimensional convex polytope of $n$ points is the $(n-2)^{th}$ Catalan number \cite{santosicm} as 
\begin{align}
C_n = \frac{1}{n-1} \begin{pmatrix}2n - 4 \\n -2 \end{pmatrix}.
\end{align}
However, we stress here that this is only a coarse argument for lower estimate for the number of triangulations since the dependence of polytope parameters has only been observed with respect to $h_{11}$ and its precise scaling with respect to $h_{22}$ is unknown.
For example, in considering $h_{11} = 7$, the number of points is still reasonable (13 for the polytope and 12 for the dual polytope). This gives us $O(10^5)$ triangulations. For the case of $h_{11} = 27$, the number is factorially much larger. A search would have to run through all triangulation configurations for all possible toric threefold constructions with $h_{11} = 27, h_{21}= 27$. In the space of toric threefolds, there is a sharp peak in the number of threefolds for precisely these Hodge numbers, with 910,113 Calabi-Yau threefolds \cite{Ashmore:2011yw}.\footnote{Instead of using TOPCOM \url{https://www.wm.uni-bayreuth.de/de/team/rambau_joerg/TOPCOM/index.html} to triangulate the convex polytopes, one could employ parallelization techniques based on \url{https://polymake.org/doku.php/mptopcom} i.e., modified TOPCOM algorithms (Called MPTOPCOM).  However, the space of $h_{11}=h_{21}=27$ threefolds is still quite extensive.} \\[0.2cm] If the octonionic magical supergravity threefold coupled to 28 hypermultiplets can indeed be obtained from a self-mirror Calabi-Yau manifold that is an  hypersurface in a toric variety, we expect the exact search to be a numerically and computationally challenging task. A possible avenue is to pursue this problem as one on the interface of computational complex geometry and machine learning as in \cite{Bao:2021ofk,Berglund:2021ztg,He:2018jtw, Demirtas:2020dbm}. However, due to the mathematical relevance of the problem, it is our hope that there exist symmetry based arguments as of yet unclear to us regarding the existence of a Calabi--Yau threefold satisfying all the physical and mathematical requirements.\footnote{Following the appearance of first version of this paper on the arXiv, there has been some remarkable advancements in the computational study of triangulations and constructions of toric CY manifolds by the authors of CYTools (\url{https://cy.tools/}) \cite{Demirtas:2020dbm}. It is therefore work left for the immediate future to exploit these tools to make more concrete statements regarding the existence of the relevant CY manifold discussed in this section. Furthermore, one of the authors (AK) thanks Liam McAllister, Richard Nally, Andreas Schachner, Jakob Moritz and Naomi Gendler for extensive discussions and explanations regarding the CYTools software, as well as hospitality in Cornell.} 

\section{Discussions \& Conclusions}
\label{sec:conclusions}
A summary of our  main results was given in the introduction (\Cref{sec:intro}). In this section, we would like to point out how our work can be further developed and extended in various directions.
\label{sec:discussions}

The most pertinent extension is the precise determination of the Fourier coefficients of different rank elements of the exceptional Jordan algebra over integral octonions that lie in the exceptional cone  in  the decomposition of  higher powers $E_4^n$ ($ n\geq 3$). These decompositions can then be used to determine the quantum degeneracies of charge states of rank three (large) BPS black holes. 

Secondly, the other magical supergravity theories can be obtained from the octonionic magical supergravity theory by truncation. Their spectrum generating conformal symmetry groups are $SO^*(12),SU(3,3)$ and $Sp(6,\mathbb{R})$. By restricting the conformal group $E_{7(-25)}$ to these subgroups one can obtain the corresponding results for the other magical supergravity theories.

Another obvious question is if and  how  our work can be extended to rotating black holes and black rings in five dimensional MESGTs and their relation to $4d/5d$ uplifts. A well studied example of the $4d/5d$ uplift is for the case of $\mathcal{N}=4$ compactification of string theory. Extending our results to the quantum degeneracies of $\mathcal{N}=4$ supergravity theories in $5d$ whose underlying Jordan algebras are not Euclidean will be an important exercise which is currently being investigated \cite{MGAK}.

As explained in the introduction (\Cref{sec:intro}), quasiconformal groups associated with Jordan algebras of degree three were proposed as spectrum generating symmetry groups of $4d$ supergravity theories. For the octonionic magical supergravity this group is $E_{8(-24)}$ which acts nonlinearly on a 57 dimensional space coordinatized by the Freudenthal triple system associated with $J_3^{\mathbb{O}} $ extended by a singlet coordinate. Extension of our work to the spectra of $4d$ octonionic magical supergravity requires the extension of the nonlinear conformal action of $E_{7(-25)}(\mathbb{Z})$ on the exceptional Jordan algebra over the integral octonions to the nonlinear action of $E_{8(-24)}(\mathbb{Z})$ on the 57 dimensional space coordinatized by the exceptional Freudental triple system over integral octonions extended by $\mathbb{Z}$ which will be the subject of a separate study \cite{MG}. 

Another direction along which our work can be extended is to maximal supergravity in $5d$ whose U-duality group is $E_{6(6)}$ and its spectrum generating conformal group is $E_{7(7)}$. In this case, the exceptional Jordan algebra over the real octonions is replaced by the exceptional Jordan algebra over the split octonions which is not Euclidean\cite{Ferrara:1997uz,Gunaydin:2000xr,Gunaydin:2009zza}.

There is a sharp peak in the number of self-mirror  Calabi-Yau manifolds with $h_{11}= h_{21} = 27$, many of which are elliptically fibered. It is also known that there is only one  anomaly free $6d$ supergravity theory that reduces to the octonionic magical supergravity in five dimensions. What, if any, are the other anomaly free supergravity theories in 6d that descend from F-theory on these self-mirror  CY manifolds? Does the peak of the number of CY manifolds at $h_{11} = h_{21}= 27$ correspond to something physical in terms of a large family of anomaly free theories that descend from F/M/string theory?
\section*{Acknowledgements}

We would like to thank Shamit Kachru and Brandon Rayhaun for invaluable discussions and collaboration during the early stages of this project. We would like to thank Atish Dabholkar, Mehmet Demirtas, Xenia de la Ossa, James Gray, Olaf Lechtenfeld, Liam McAllister, Sameer Murthy, Richard Nally, Matthias Sch\"utt, and Harald Skarke for helpful discussions.  We would also especially like to thank Aloys Krieg for correspondence and for bringing to our attention some references relevant to this work. AK acknowledges previous sources of financial support that were crucial towards the completion of this work: The Riemann Fellowship endowed by the Riemann Center for Geometry and Physics, The Austrian Marshall Plan Foundation, the Stanford Institute for Theoretical Physics, FWF W1252-N27 (Austrian Academy of Sciences). MG would like to express his gratitude to the Stanford Institute of Theoretical Physics for its hospitality during his visits there including a sabbatical stay when  much of this work was carried out. AK would like to acknowledge prior affiliations that also provided a fruitful environment for the completion of this work: Institute for Theoretical Physics (TU Vienna), Stanford Institute for Theoretical Physics, and the Institute for Theoretical Physics (Leibniz University of Hannover)/Riemann Center for Geometry and Physics, and would like to thank the Abdus Salam ICTP for hospitality during the final stages of the project.
\appendix

\section{Theta Functions of Niemeier Lattices}
\label{app:thetas}
The theta function of degree 1 of a Niemeier lattice is given by the expression 
\begin{align}
    \theta(\tau) = E_4(\tau)^3 + (24 h - 720) \Delta ~,
\end{align}
where $h$ is the Coxeter number of the Niemeier lattice. From this we can write down the $q-$series of the theta function (where $q:= e^{2\pi i \tau}$) as in the \autoref{tab:niemeiertheta}. The $n-$th coefficient in $q-$series of the theta function of a lattice defines how many vectors of norm $2n$ there are in the lattice. For example, consider the case of the Leech lattice whose $q^1$ coefficient is 0, implying that there are no vectors of norm 2 i.e., no roots in the Leech lattice.
\begin{sidewaystable}[ht!]
\begin{tabular}{|c|c|c|}
\hline 
$h$ & Niemeier Lattice & Theta Function of Niemeier Lattice of Degree 1 \\ \hline
 0 & Leech & $1+196560 q^2+16773120 q^3+398034000 q^4+4629381120 q^5+34417656000 q^6+O\left(q^7\right)$ \\
 2 &  $A_{1}^{24}$ & $1+48 q+195408 q^2+16785216 q^3+397963344 q^4+4629612960 q^5+34417365696 q^6+O\left(q^7\right)$\\
 3 & $A_3^{12}$ & $1+72 q+194832 q^2+16791264 q^3+397928016 q^4+4629728880 q^5+34417220544 q^6+O\left(q^7\right)$\\
 4 & $A_3^8$ & $1+96 q+194256 q^2+16797312 q^3+397892688 q^4+4629844800 q^5+34417075392 q^6+O\left(q^7\right)$\\
 5 & $A_4^6$ & $1+120 q+193680 q^2+16803360 q^3+397857360 q^4+4629960720 q^5+34416930240 q^6+O\left(q^7\right)$\\
 6 & $A_5^4 D_4, \ D_4^6$ & $1+144 q+193104 q^2+16809408 q^3+397822032 q^4+4630076640 q^5+34416785088 q^6+O\left(q^7\right)$\\
 7 & $A_6^4$ & $1+168 q+192528 q^2+16815456 q^3+397786704 q^4+4630192560 q^5+34416639936 q^6+O\left(q^7\right)$\\
 8 & $A_7^2 D_5^2$ & $1+192 q+191952 q^2+16821504 q^3+397751376 q^4+4630308480 q^5+34416494784 q^6+O\left(q^7\right)$\\
 9 & $A_8^3$ & $1+216 q+191376 q^2+16827552 q^3+397716048 q^4+4630424400 q^5+34416349632 q^6+O\left(q^7\right)$\\
 10 & $A_9^2 D_6, \ D_6^4 $ & $1+240 q+190800 q^2+16833600 q^3+397680720 q^4+4630540320 q^5+34416204480 q^6+O\left(q^7\right)$\\
 12 & $A_{11}D_7E_6 , \ E_6^4$& $1+288 q+189648 q^2+16845696 q^3+397610064 q^4+4630772160 q^5+34415914176 q^6+O\left(q^7\right)$\\
 13 & $A_{12}^2$ & $1+312 q+189072 q^2+16851744 q^3+397574736 q^4+4630888080 q^5+34415769024 q^6+O\left(q^7\right)$\\
 14 & $D_8^3$ & $1+336 q+188496 q^2+16857792 q^3+397539408 q^4+4631004000 q^5+34415623872 q^6+O\left(q^7\right)$\\
 16 & $A_{15}D_9$ & $1+384 q+187344 q^2+16869888 q^3+397468752 q^4+4631235840 q^5+34415333568 q^6+O\left(q^7\right)$\\
 18 & $A_{17}E_7, D_{10}E_7^2 $& $1+432 q+186192 q^2+16881984 q^3+397398096 q^4+4631467680 q^5+34415043264 q^6+O\left(q^7\right)$\\
 22 & $D_{12}^2 $& $1+528 q+183888 q^2+16906176 q^3+397256784 q^4+4631931360 q^5+34414462656 q^6+O\left(q^7\right)$\\
 25 & $A_{24}$ & $1+600 q+182160 q^2+16924320 q^3+397150800 q^4+4632279120 q^5+34414027200 q^6+O\left(q^7\right)$\\
 30 & $D_{16}E_8, \ E_8^3$ & $1+720 q+179280 q^2+16954560 q^3+396974160 q^4+4632858720 q^5+34413301440 q^6+O\left(q^7\right)$\\
 46 & $D_{24}$ & $1+1104 q+170064 q^2+17051328 q^3+396408912 q^4+4634713440 q^5+34410979008 q^6+O\left(q^7\right)$\\
 \hline
\end{tabular}\\
 \caption{Coxeter numbers $h$ of Niemeier lattices and their associated classical theta functions of degree 1}
 \label{tab:niemeiertheta}
\end{sidewaystable}
\section{ Theta function of integral octonions $\mathcal{R}$ of Coxeter}
Let $Tr$  and $\mathcal{N}$ denote integral valued  trace and quadratic norm form of integral octonions $\mathcal{R}$ defined as 
\eq
Tr(t)= t +\bar{t} \, , \quad \, \mathcal{N}(t) = t \, \bar{t} \quad , \quad \forall \, t \in \mathcal{R} 
\en 
For $ z \in \mathcal{H}$ the theta series for integral octonions $\mathcal{R}$ is defined as \cite{MR2478255,MR1195510}
\eq
\theta(z)= \sum_{t\in \mathcal{R}} e^{2\pi \, i\, \mathcal{N}(t)z }
\en
This theta function is a modular form of weight 4 and can be expressed as a normalized Eisenstein series
\eq \theta(z) = E_4(z) = 1 + 240 \sum_{n=1}^{\infty} \left( \sum_{d|n} d^3 \right) e^{2\pi i n z} \en 
where the factor $(240\sum_{d|n} d^3 )$ counts the number of solutions to the equation 
\eq \mathcal{N}(t)= n \, , \quad t\in \mathcal{R} \en
\section{Modular Forms on the 27 Dimensional Exceptional Domain and Theta Functions}
\label{app:exceptionalmf}

Following the pioneering work of Baily Jr. on the the exceptional arithmetic subgroup of $E_{7(-25)}$ and its Eisentein series \cite{MR269779}, 
Resnikoff showed that the non-constant singular modular forms defined over the exceptional domain given by  the exceptional Jordan algebra are exactly of weight 4 and 8 \cite{MR412495}. They were first obtained by Kim by analytic continuation of non-holomorphic Eisenstein series \cite{MR1216126}. 
The exceptional tube domain $\mathcal{D}$ is simply the ``upper half-plane" of the exceptional Jordan algebra $J_3^{\mathbb{O}}$: 
\eq
\mathcal{D} = \{ Z= X +i Y ; X, Y \in J_3^{\mathbb{O}} \textrm{ with} \,  Y>0 \}~.
\en
An exceptional modular form of weight $k$ of $E_{7(-25)}(\mathbb{Z}) $ on $\mathcal{D}$ is a holomorphic function $F(Z)$ that satisfies
the conditions: \begin{enumerate}
    \item \eq F(Z+B) =F(Z) \qquad \, \forall \,\, B \,\, \in L
\en
\item \eq
F(g Z ) = F(Z) \qquad  \forall \,\, g \,\, \in E_{6(-26)}(\mathbb{Z})
\en
\item \eq
F(-Z^{-1})  = (\mathcal{N}(Z))^k F(Z)~,
\en
\end{enumerate}
where
\eq
Z^{-1} = \frac{Z_I^{\#}}{(\mathcal{N}(Z))}~.
\en
Such modular forms have  absolutely convergent Fourier expansions of the form
\eq 
F(Z) = \sum_{T\in L, \,\, T\geq 0} a(T) e^{2\pi i (T,Z)} ~,
\en
and they are called \textit{singular} if $a(T)=0$ for all $T>0$. 

The classical theta series associated with the complex upper half-plane 
\eq
\theta(z,u)= \sum_{n\in \mathbb{Z}}  e^{(\pi i \cdot z n^2 + 2\pi i \cdot n u )}
\en
where $z\in \mathcal{H}$ and $u \in \mathbb{C}$ has been generalized to theta series associated with the upper half-planes of all formally real Jordan algebras \cite{MR412495,MR486675} that do not have the exceptional Jordan algebra as a direct summand. Further generalizations of classical theta series were discussed recently and it was shown that the exceptional tube domain does not admit a theta function \cite{dorfmeistertheta}. The case of the singular modular forms studied by Kim \cite{MR1216126} cannot be identified with the theta series associated with the exceptional Jordan  algebra. However, it has been noted  that the singular modular forms of Kim \cite{MR1216126} can be obtained from theta series on the $10-$dimensional boundary component identified with the Jordan algebra of the $2\times 2$ Hermitian matrices over $\mathbb O$ \cite{MR1437509}.
A simpler construction of the singular modular forms of Kim was obtained by Krieg \cite{MR1437509} using the theta series on the upper half-plane of the Jordan algebra of Hermitian $2\times 2$ matrices over the integral octonions $\mathcal{R}$ of Coxeter which he refers to as the Cayley half-plane of degree two. The modular forms in question arise from the theta series on the Cayley half-plane of degree two via the use of Fourier-Jacobi expansion. 

As was summarized in \autoref{modularform} the singular modular form of degree 4 over the exceptional domain $\mathcal{D}$ has the Fourier expansion
\begin{align*}
E_{4}(Z) = 1 +  240 \sum_{T\geq 0, \ T \in \, JL \,; \,\mathrm{rank}(T)=1} \sigma_3(c(T))  e^{2\pi i  \Tr(T\circ_I Z)} \quad , \quad Z\in \mathcal{D} ~,
\end{align*}
where 
\begin{align*} \sigma_k(m) := \sum_{d\in \mathbb{N}, d|m} d^k \quad , \quad m\in \mathbb{N}, 
\end{align*}
and 
\begin{align*}
	c(T) = \text{max}(r \in \mathbb N) \ \text{such that } \frac{1}{r} T \in JL ~. 
\end{align*}
Squaring the modular form of weight 4 leads to the singular modular form of weight 8. In the formulation of Krieg it takes the form \cite{MR1437509}
\eq
E_{8}(Z) =E_{4}(Z)^2 = \sum_{T\in JL \, , \, T\geq 0} \alpha(T) \,   e^{2\pi i  \Tr(T\circ_I Z)} \quad , \quad Z\in \mathcal{D} ~,
\en
where 
\begin{align*}
	\alpha(T) = \begin{cases}
		1 & \text{if } T = 0 \\ 
		480 \cdot \sigma_7(c(T)) & \text{if } \text{rank}(T) = 1 \\
		240 \cdot 480 \cdot {\displaystyle \sum_{d \in \mathbb N, d \vert c(T)}} d^7\sigma_3(c(T^\#/d^2)) & \text{if } \text{rank}(T) = 2 \\ 
		0 & \text{if } \text{rank}(T) = 3 
	\end{cases}~.
\end{align*}

  $E_{4}(Z)$ and $E_8(Z)$ are the only  singular modular forms on the  exceptional domain \cite{MR1437509}.  The general computation of the Fourier coefficients of higher modular forms was given in \cite{kim2016cusp, kim2015ikeda}. Using the notation of \cite{kim2016cusp} and \cite{MR1437509}, the Fourier coefficients in full generality are given as below. Consider a weight $ k \in 4 \mathbb Z$ modular form over the exceptional domain with the Fourier expansion
  \begin{align}
  	F(Z)_k  = \sum_{T \geq 0, T \in JL} \alpha_k(T) e^{2 \pi i \Tr(T\circ_I Z)} ~. 
  \end{align}
 The Fourier coefficients $\alpha_k(T)$ above are  
\begin{align}
	\label{eq:allfouriercoeffs}
		\alpha_k(T) = \begin{cases}
		1 & \text{if } T = 0 \\ 
		-\frac{2k}{B_{k}} \cdot \sigma_{k-1}(c(T)) & \text{if } \text{rank}(T) = 1 \\
		\frac{4k (k-4)}{B_k B_{k-4}} \cdot {\displaystyle \sum_{d \in \mathbb N, d \vert c(T)}} d^{k-1}\sigma_{k-5}(c(T^\#/d^2)) & \text{if } \text{rank}(T) = 2 \\ 
		2^{15} \frac{k}{B_k} \cdot \frac{k-4}{B_{k-4}} \cdot \frac{k-8}{B_{k-8}}  \cdot \det(T)^{(k-9)/2} \cdot {\displaystyle \prod_{p \vert \det T}} f_T^p(p^{\frac{k-9}{2}}) & \text{if } \text{rank}(T) = 3 
	\end{cases}~,
\end{align} 
where $f_T^p$ above is a monic Laurent polynomial of degree $d = \text{ord}_p (\det(T))$ such that the polynomial satisfies the functional equation $f_T^p(X) = f_T^p(X^{-1})$, and $B_k$ is the $k^{th}-$Bernoulli number. These polynomials depend only on $T$ and $p$ and are mostly identically equal to 1 except for a finite few cases of $p$. Following \cite{kim2016cusp}, the modular form
$\tilde F(Z)_k = \Xi_k \times F(Z)_k$, where $\Xi_k$ is the numerator of$ \left(\prod_{n=0}^{\text{rank}(T)-1} B_{k-4n}\right)$, has coefficients valued in $\mathbb Z$.
As can be seen, for the case of weight $4$, the rank 2 and 3 coefficients are zero, while for the case of weight $8$, the rank 3 coefficients are zero.

  \section{Discrete subgroups of exceptional groups.}
  \label{app:discgrps}
The groups defined over the integers $\mathbb{Z}$ were studied by Gross\cite{MR1369418}. In this appendix we will summarize some of these discrete groups relevant to our work.
The automorphism group $\aut{\mathcal{R}}$ of integral octonions  is a certain form of the exceptional group $G_2$ . This group is finite since it describes a compact group over the reals $\mathbb{R}$. It is of order $2^6\cdot 3^3 \cdot 7$ and is denoted as $G_{2}(2)$ \cite{MR827219}. The order of integral octonions $\mathcal{R}$ is invariant under octonion conjugation and the trace form $\Tr(x) := x + \bar{x}$ takes on integer values over it.
The symmetric bilinear $\langle x, y \rangle $ defined as
\begin{align*} \langle x, y \rangle := \Tr (\bar{x} y) \qquad x, y \in \mathcal{R} \end{align*} is even, positive-definite and has determinant 1. Hence, it defines a unimodular lattice $M$ over $\mathcal{R}$ which is isomorphic to the $E_8$ root lattice. The sublattice $M_0$ orthogonal to the identity $1$ is isomorphic to the $E_7$ root lattice and has determinant $2$\cite{MR1369418}. $\aut{\mathcal{R}}(\mathbb{Z}) $ has a seven dimensional representation on $M_0$ with determinant 2. It also leaves the trilinear form defined as
\begin{align*} \Tr(xyz)=\Tr(x(yz))=\Tr((xy)z)\qquad x, y, z \in \mathcal{R} \end{align*}
invariant. On the sublattice $M_0$ this form is completely antisymmetric.

The arithmetic subgroup of $E_{6(-26)}(\mathbb{Z})$ of the U-duality group of octonionic magical supergravity in five dimensions is defined as the group of transformations acting on the exceptional Jordan algebra $J_3^\mathbb{O}$  defined over the Coxeter order of integral  octonions $\mathcal{R}$ that leave the cubic norm invariant. It has two inequivalent 27 dimensional representations one acting on $J_3^\mathbb{O}$  and another one on its dual $J_3^{\mathbb{O}^\vee}$.

Groups of type $F_4$ over the integers $\mathbb{Z}$ can be obtained as subgroups of  $E_{6(-26)}(\mathbb{Z})$ as invariance groups of elements $E$ with unit norm.  If one chooses
\begin{align*} E=E_0 = \left(
             \begin{array}{ccc}
               0 & 0 & -1 \\
               0 & -1 & 0 \\
               -1 & 0 & 0 \\
             \end{array}
           \right)~, \end{align*}
         its invariance group is $F_{4(-20)}(\mathbb{Z}) $. On the other hand if one chooses $E$ to be positive definite the invariance group is finite over $\mathbb{Z}$ since the corresponding group over $\mathbb{R}$ is compact real form of $F_4$. For example if $E$ is the $3\times 3$ identity matrix $I_3$  its invariance group over $\mathbb{Z}$ has order $2^{15}\cdot 3^6\cdot 5^2\cdot 7$ and is isomorphic to the finite  group
         $2^2\cdot O_8^+(2)\cdot S_3 $ in the notation of \cite{MR827219}. The corresponding lattice $JL_0$ of rank 26 and determinant 3 is the root lattice
         \eq
         JL_0  \cong E_8 \oplus E_8 \oplus E_8 \oplus A_2
         \en
         which has 726 root vectors $v$ satisfying $\langle v,v \rangle =2$. The corresponding dual  lattice is
          \eq
         JL_0^\vee  \cong E_8 \oplus E_8 \oplus E_8 \oplus A_2^\vee
         \en
         which has 6 short vectors $\langle v,v \rangle =2/3$.

        The invariance group of  the  polarization $E=E_2 = J(2,2,2;-\bar{\beta} , -1, \-\bar{\beta})$ has order $2^{12}\cdot 3^5 \cdot 7^2 \cdot 13$ and is isomorphic to the group $^3D_4(2)\cdot 3$ \cite{MR1369418}. The corresponding lattice $JL_0$ does not have any roots and has 117936 short vectors that satisfy $\langle v,v \rangle =4$. It is the unique even lattice of rank 26 and determinant 3 with no roots as was shown independently by Borcherds \cite{borcherds1999leech} and Elkies \cite{elkies1996mock}.
    \section{ Commutative Subrings of the Exceptional Jordan Algebra \label{commutative}}
    Gross and Gan studied  the commutative subrings of integral exceptional Jordan algebra as well as  its subalgebra $J_2^{\mathbb{O}}(\mathcal{R})$ generated by $2\times 2$ Hermitian matrices over the integral octonions $\mathcal{R}$ \cite{MR1793601}. To this end they first determine the number $N(A,R)$ of different ways the ring $A$ of integers in an imaginary quadratic field $K$ of discriminant $D$ can be embedded into  the Coxeter's ring $\mathcal{R} $ of integral octonions. They prove that it is given by
    \eq
    N(A,R) = \frac{L(\epsilon_A, -2 )}{\zeta(-5)}= -252 \cdot L(\epsilon_A, -2) ~,
    \en 
    where 
    \eq
    \epsilon_A : ( \mathbb{Z}/D\mathbb{Z)^\times}\longrightarrow \langle \pm 1 \rangle 
    \en
    is the odd quadratic Dirichlet character associated with $K$. 
    
  The  $2\times 2$ Hermitian matrices  of the form
   \eq 
   X = \left(\begin{array}{cc}a & x \\ \bar{x} & b\end{array}\right) ~,
   \en
   where $a,b \in \mathbb{Z}$ and $x\in \mathcal{R}$  form a Jordan algebra under the product
   \eq X \circ Y =\frac{1}{2} ( XY + YX)~. \en Under addition, such matrices form an Abelian  group of rank 10 which we shall label as $ JL_2 $. The determinant 
   \eq \det(X) = ab - {N}(x) \en defines a quadratic form of signature $(1,9)$ and discriminant $-1$.
   
   If $A$ is the ring of integral elements in a quadratic algebra $K$ with discriminant $D$ then  the number of ways it can be embedded in $J_2$ is given by the formula
   \eq
   N(A,J_2) = \frac{L(\epsilon_A, -3)}{\zeta(-7)}= 240 \cdot L(\epsilon_A, -3) 
   \en
   where 
   \eq
   \epsilon_A : (\mathbb{Z}/D\mathbb{Z} )^\times \longrightarrow \langle \pm 1 \rangle 
   \en
   is the even quadratic character of $K$ and $ L(\epsilon_A, -3) $ is the Dirichlet $L-$function \cite{MR1793601}. 
   
   Most interestingly from our point of view, Gross and Gan study the embedding of the ring of integral elements in an \'etale cubic algebra $K$ over $\mathbb{Q}$ with discriminant $D$ into the $JL$ considered as an Abelian group generated by $3\times 3$ Hermitian matrices over the integral octonions $\mathcal{R}$. In this case $A$ is either $\mathbb{Z}^3$ when $D=1$ , $\mathbb{Z}\bigoplus B $ where $ B$ is the ring of integers in a quadratic field or $A$ is the ring of integers in a totally real cubic field. They denote the Jordan algebra defined over $JL$ with the polarization $I$  or $E$  as the identity element as $J_I$ or $J_{E}$, respectively, and derive the formula
   \eq
   91 N(A,J_I) + 600 N(A,J_{E}) = 2^7 \cdot 3^3 \cdot 5^2 \cdot 7 \cdot 13 \cdot L(V_A, -3) ~,
   \en
  where $N(A,J_I) $ and $N(A,J_{E})$ denote the number of embeddings into $J_I$ and $J_{E}$, respectively. They stress that they do not have  formulas for  
    $N(A,J_I) $ and $N(A,J_{E})$  separately in general. However when  $A$ is not an integral domain    which is the case when $A=\mathbb{Z}\bigoplus B $ then $N(A,J_{E}) =0$ and one has 
    \eq
    N(A,J_I) = 2^7\cdot 3^3\cdot 5^2 \cdot L(V_A, -3)~.
    \en   
\section{Hilbert Modular Forms \label{HMF}}
This appendix is devoted to summarizing the relevant background information on Hilbert Modular Forms (HMFs). Standard references on HMFs include \cite{van2012hilbert,freitag1990hilbert,bruinier2008hilbert}. We begin with the definition of Hilbert modular group. \\ 
Consider the upper half plane ($\mathbb{H}$) on which $SL(2,\mathbb{R})$ acts via fractional linear transformations.
Consider a positive integer $n \in \mathbb{Z}_{+}$. We consider $n$ copies of the upper half plane which we denote by $\mathbb{H}^n$. Let $F$ be a totally real number field of degree $n$ over $\mathbb{Q}$ such that $F$ admits $n$ distinct embeddings into $\mathbb{R}$ i.e., 
\begin{align}
\begin{array}{@{}c@{\;}c@{\;}c@{\;}c@{\;}c@{}}
 F \hookrightarrow R  \Rightarrow & \alpha & \longmapsto & a^{(i)}  \\
& \vin && \vin  \\
& F && \mathbb{R}^{n}
\end{array}, \ a^{(i)} = \left( a^1,\cdots, a^n \right) ~.
\end{align}
The group $SL(2,F) = \left\{ \begin{pmatrix} a & b\\c & d\end{pmatrix} \bigg \vert a, b, c, b \in F, \ ad - bc > 0  \right\}\footnote{Or rather, $GL(2,F)$.} $ can be embedded into $SL(2,\mathbb{R})^n$ $n$ times by means of the embedding $F \hookrightarrow \mathbb{R}$.
Consider now the ring of integers in $F$, which is usually denoted as $\mathcal O_F$.
$\mathcal O_F$ is a \textit{Dedekind domain} i.e., it is an integral domain that is an integrally closed  Noetherian ring\footnote{It satisfies the ascending chain condition on both left and right ideals.} in which every non-trivial prime ideal is maximal i.e., there are no other ideals between the prime ideal and $\mathcal O_F$.
\begin{definition}[Hilbert Modular Group]
The Hilbert Modular Group for $F$ is the group $\Gamma_F = PSL(2,\mathcal O_F) = SL(2, \mathcal O_F)/\left\{ +1,-1 \right\} \subset PGL^+(2,F)$. 
\end{definition}
\begin{remark}
The classical modular group is the special case when $F = \mathbb{Q}$ and $n = 1$.
\end{remark}
\begin{definition}[Hilbert modular variety]
A Hilbert modular variety of degree $n$ is an algebraic variety obtained by quotienting $n$ copies of the upper half plane ($\mathbb{H}^n$) by the Hilbert modular group $\Gamma_F$. 
\end{definition}
The group $\Gamma_F$ acts on $\mathbb{H}^n$ as 
\begin{align}
    \Gamma_F \ni \begin{pmatrix}
a & b \\ c & d
    \end{pmatrix} (\tau_1, \cdots, \tau_n) = \left( \frac{a_1 \tau_1 +b_1}{c_1 \tau_1 + d_1}, \cdots, \frac{a_n \tau_n +b_n}{c_n \tau_n + d_n} \right). 
\end{align}
For the sake of convenience, we denote $\tau^k := \left( \tau_1, \cdots, \tau_k \right) $. 
\begin{definition}[Hilbert Modular Form]
A Hilbert Modular Form (HMF) of weight $ k = (k_1, \cdots , k_n)$ and degree $n$ is a function $f : \mathbb{H}^n \rightarrow \mathbb{C}$ such that 
\begin{align}
    f(\gamma \tau^n) = \left( \prod_{j = 1}^n (c_j \tau_j + d_j)^{k_j} \right) f(\tau^n), \ \forall \begin{pmatrix}
a & b \\ c & d
    \end{pmatrix} \in \Gamma_F,
\end{align}
and the function $f$ is regular at the cusps of $\Gamma_F$. The HMF is holomorphic if it is holomorphic on $\mathbb{H}^n$. Similar definition for meromorphic HMFs holds. If for a HMF of weight $k = (k_1, \cdots, k_n)$ such that $k_1 = k_2 = \cdots = k_n$, then the HMF is simiply said to have \textit{parallel} weight $k$ and degree $n$.
\end{definition}
\subsection{Fourier expansions of Hilbert Modular Forms}
Holomorphic HMFs admit a Fourier expansion at the cusp at $\infty$. Let use consider a $\mathbb{Z}-$module $M \subset F$ and let $V \subset \mathcal{O}_F^*$, where as usual $\mathcal{O}_F^*$ is the group of units in $\mathcal{O}_F$. $V$ has an action on $M$ in the following sense. Let $G(M,V) = \left\{ \pmat{e}{\mu}{0}{e^{-1}} \bigg \vert \mu \in M, \ e \in V \right\} $ be a finite index subgroup of $\Gamma_\infty$, the stabilizer of $\infty$. This defines the analog of the ``T--transform'' as for classical modular forms for the case of HMFs. This gives us the periodic structure of $f$ as $f(\tau_k + \mu) = f(\tau_k), \ \forall \mu\in M$. This periodic property allows us to write the convergent Fourier expansion of a holomorphic HMF as 
\begin{align}
    f(\tau_n) = \sum_{\nu \in M^\vee} a_\nu e^{2 \pi i \Tr(\nu \tau_n)},
\end{align}
where $\displaystyle M^\vee = \left\{ \lambda \in F  \ \vert \ \Tr(\mu \lambda) \in \mathbb{Z} \ \forall \mu \in M \right\} $ is the dual lattice to $M$ w.r.t the trace norm of $F$. The Fourier expansion of holomorphic HMFs can be expressed at its cusp as 
\begin{align}
    f(\tau_n) = a_0 + \sum_{\substack{\nu \in M^\vee,\\ \nu > 0}} a_\nu e^{2 \pi i \Tr(\nu \tau_n)}~.
\end{align}
Holomorphic HMFs for which $a_0$ as above is zero are \textit{Hilbert Cusp Forms}. 
\begin{theorem}[G\"otzky--Koecher principle]
All non-parallel HMFs are cusp forms. 
\end{theorem}
We will not prove this theorem here, although we do point the reader to any of the standard references on HMFs for a proof of the G\"otzky--Koecher principle.
\subsection{Hilbert-Eisenstein series}
We now introduce the Hilbert theta series and Hilbert-Eisenstein series. Holomorphic HMF are generated by Hilbert--Eisenstein series of parallel, even integral weight $k > 2$. To each non-zero fractional ideal $\mathfrak{m}$ of $F$, for $k \in 2 \mathbb{Z}_+$we associate a function $E(\tau_n, \mathfrak{m}, k)$ given by
\begin{align}
\label{eq:hilbeis}
      E(\tau_n, \mathfrak{m}, k) = \text{Nm}(\mathfrak{m})^k \sum_{ (c,d)\in (\mathfrak{m}\times \mathfrak{m}) - \left\{ 0 \right\}/\mathcal{O}^*_F } \left( \prod_{j = 1}^n \left( \sigma_j(c) \tau_j + \sigma_j(d) \right)  \right)^{(-k)} ~.
  \end{align} 
  \eqref{eq:hilbeis} is a Hilbert modular form of parallel weight $k$ on $\Gamma_F$ and admits the following Fourier expansion: 
  \begin{align}
       E(\tau_n, \mathfrak{m}, k) = \zeta(\mathfrak m,k) + \frac{1}{\sqrt{D_F}} \left( \frac{(-2 \pi i)^k}{\Gamma(k)} \right)^n  \sum_{\substack{\nu \in \mathfrak d_F^{-1}}} \sigma_{k-1}(\nu; \mathfrak m) e^{2 \pi i \Tr(\nu \tau_n)}~,
  \end{align}
  where $D_F$ is the discriminant of $F$ in $\mathbb{Z}$, and $\mathfrak d_F^{-1}$ denotes the inverse of the different ideal of $F$. It is straightforward to see that for $F = \mathbb{Q}, \ \mathfrak m = \mathbb{Z}$, we recover the classical Eisenstein series (such as the ones derived in \eqref{eq:eisensteinexample}). 
  \subsection{Lattice theta functions and Hilbert modular forms}
  It is a canonized statement that the theta function of a positive definite lattice of signature $(p,0)$ is an elliptic theta function of weight $p/2$. The lattice theta function can also be expressed in terms of a quadratic form where the norm of the vector expressed as the value of a quadratic form $Q$. This provides an equivalency between quadratic forms and lattices $\Lambda^{p,0} \subset \mathbb{R}^p$. \\[0.2cm]
  Consider now for a positive integer $p$, $\Lambda_F^{p,0} \in F^p$ is a lattice of rank $p$ in $F$ and a quadratic form $Q: F^p \rightarrow F$. The lattice/quadratic form is even-integral in $F$ if $\forall \alpha, \beta \in \Lambda_F$, $Q(\alpha, \beta): = Q(\alpha+\beta) - Q(\alpha) - Q(\beta) \in \mathfrak m_F^{-1}$. Given such an even, integral lattice with $p \in 2 \mathbb{Z}$ with $Q^{-1}$ also even integral and $\det Q$ being a square in $F- \left\{ 0 \right\} $, the function
  \begin{align}
      \Theta(\tau_p; Q) := \sum_{\nu \in \mathcal O_F^p} e^{2 \pi i \Tr (Q(\nu) \tau_p)} 
  \end{align}
  is a Hilbert modular form of weight $p/2$ in $\Gamma_F$. 
  \begin{remark}It is also possible to recover a Hilbert--Eisenstein series from a Hilbert-theta function in the sense of Siegel--Weil (see \cite{Hidasw} and references therein).  Siegel-Weil theorems over the classical modular group (and subgroups thereof) are of interest in string theory (see \cite{Kachru:2017zur} from the perspective of BPS attractors, \cite{Maloney:2020nni}(and works that follow) from the perspective of holography, and \cite{Ashwinkumar:2021kav} from the perspective of topological invariants). The Hilbert-Eisenstein series obtained via
\end{remark}
\providecommand{\href}[2]{#2}\begingroup\raggedright\endgroup

\end{document}